\documentclass[a4paper,11pt]{article}
\pdfoutput=1 % if your are submitting a pdflatex (i.e. if you have
\usepackage{jheppub} % for details on the use of the package, please
\usepackage{import}

\usepackage{lmodern}
\usepackage{mathtools}

\DeclarePairedDelimiter\floor{\lfloor}{\rfloor}
\usepackage{tikz}
\usepackage{steinmetz}
\usepackage[T1]{fontenc} % if needed
\graphicspath{{img/}} % set images path
\usepackage{tikz}
\makeatletter
\gdef\@fpheader{}
\makeatother

\title{\boldmath{ \texorpdfstring{$c=1$}{TEXT}, \texorpdfstring{$R=1$}{TEXT} and \texorpdfstring{$N\gg 1$}{TEXT}: ZZ instantons in 2D String Theory and Matrix Integrals}}

\author[a]{Rishabh Kaushik}
\affiliation{International Center for Theoretical Sciences(ICTS-TIFR),
Shivakote, Hesaraghatta Hobli, Bengaluru North 560089, India.}

\emailAdd{rishabh.kaushik@icts.res.in}

\abstract{We explore the non-perturbative aspects of $c=1$ string with compactified Euclidean time, its $0+0$ dimensional matrix model duals (at self-dual radius), and $0+1$ dimensional Matrix Quantum Mechanics (free fermion) description. We calculate the instanton normalizations, disk two-point function, and annulus one-point function in worldsheet formalism using string field theory insights. We further match them with the corresponding predictions from the matrix model descriptions. We also have some results and remarks regarding the multi ZZ instanton normalizations and the general $(1,n)$ ZZ instanton normalization for $c=1$ string at both self-dual and generic radius.}

\begin{document}
\maketitle
\raggedbottom
\section{Overview and Summary of Results}
There has been significant development in the study of non-perturbative effects in string theories after the regularization procedure was devised in \cite{Sen_202183}, \cite{Eniceicu_2022} and \cite{Sen_2021}. It started with the attempt by Sen in \cite{Sen:2019qqg} to fix the undetermined constants in the analysis of Balthazar Rodriguez and Yin in \cite{Balthazar:2019rnh}. Non-perturbative effects corresponding to D-instantons in string theory come from the worldsheets having boundaries with some specific boundary conditions on different worldsheet fields. These effects are constituted by different pieces like the annulus partition function, disk two-point function, annulus one-point function, disk with two holes, etc. Out of these, the annulus partition function $\sim g_s^0$, which means that the contribution from this piece can be factored out in the form of \textit{exponentiated annulus} and appears as the normalization of the instanton amplitude. However, the different pieces discussed above are \textit{typically divergent} in worldsheet CFT calculations as the worldsheet undergoes degeneration(pinching of the Riemann surface for example). Hence, proper regularization is needed which involves the insights from string field theory as shown in \cite{Sen_202183}, \cite{Eniceicu_2022}, and \cite{Sen_2021}. \cite{Sen_202183} mainly dealt with the regularizing annulus one-point function and disk two-point function in $c=1$ string theory with non-compact boson. On the other hand, \cite{Sen_2021} dealt with the regularization of normalization of instanton effects i.e. exponentiated annulus in the same theory.

Following \cite{Sen_202183}, \cite{Eniceicu_2022} and \cite{Sen_2021}, there has been a significant amount of work in the direction of instanton calculations in string theory and matching them with the corresponding effects in the dual matrix models. \cite{Eniceicu_2022Multi_instantons} discussed instantons in minimal string theory and showed matching with the eigenvalue instanton effects in the two-matrix model. Similarly,  \cite{chakrabhavi2024normalizationzzinstantonamplitudes} did it for the type 0B minimal superstring and the unitary matrix models. \cite{alexandrov2023instantons} discussed it for the $c=1$ string with compactified(at radius $R$) boson and the dual Matrix Quantum Mechanics(MQM) description. Another work on similar lines is \cite{Chakravarty_2023} which discusses the type 0B superstring with free boson matter and the dual MQM description.

All of the examples mentioned above involve double-scaled matrix models i.e. away from the 't Hooft limit. There are very few examples of these dualities which involve matrix models in the 't Hooft limit. 't Hooft limit is relevant in order to make connections with AdS/CFT correspondence while studying these toy models as emphasized in \cite{gopakumar2023derivingsimplestgaugestringduality,Mazenc_2023}. 

One such example is the $c=1$ string at self-dual radius and the two-matrix model which was first discussed in \cite{Bonora_1995} and has been reiterated recently in relation to the Feynman graph dualities and the topological strings in \cite{gopakumar2023derivingsimplestgaugestringduality,Mazenc_2023}. We will call this two-matrix model as the KM Model throughout this article(the two matrices involved are $K$ and $M$). Another matrix model, originally proposed in \cite{Dijkgraaf_1993} and later corrected in \cite{Imbimbo_1995}, also exhibits duality with the $c=1$ string at self-dual radius. It was derived using the $W_{\infty}$ constraints obtained in the free fermion description of the $c=1$ string. We will call it Imbimbo Mukhi(IM) Model. It was argued in \cite{gopakumar2023derivingsimplestgaugestringduality} that these two matrix models roughly exhibit Feynman graph duality at the perturbative level using the integrating in and out procedure. We will explore the non-perturbative aspects of this duality between these matrix models and the $c=1$ string at self-dual radius in this article. We will also explore some details of the non-perturbative aspects of the $c=1$ string at generic radius $R$ and the dual MQM.

The $c=1$ string we will be working with is comprised of the Liouville theory with $c=25$ ($b=1, Q=1$) coupled to a free boson. The action looks like,
\begin{align}
    \frac{1}{\pi}\int d^2z\left(\partial \phi\bar{\partial}\phi+\pi \mu e^{2\phi}+\partial X\bar{\partial}X\right)
\end{align}
where $\phi$ and $X$ are respectively the Liouville field and the free boson. There will be a $bc$ ghost action as well. We further have the boson compactified, i.e.
\begin{align}
    X\sim X+2\pi R
\end{align}
and self-dual radius corresponds to the $R=1$ case. This characterizes the so-called \textit{$c=1$ string at self-dual radius}. 

\noindent{}We summarize our main results as follows, 
\begin{itemize}
    \item We first show how the total free energy expression we have from MQM(or free fermion) description of the $c=1$ string at self-dual radius \eqref{freeenergyatR} (setting $R=1$) can arise from the two matrix model descriptions (the KM and the IM matrix model description) in \eqref{freeenergyatR=1} and \eqref{IMfreeeng}.
    \item We further look at the non-perturbative part of the above-mentioned free energy and particularly, the one-instanton sector ($\sim e^{-2\pi \mu}$). We see a matching of the matrix model prediction for the instanton normalization \eqref{one_inst_1} and the worldsheet string theory calculation of the normalization aided with the string field theory insights in \eqref{eq: norm_ws_result}. The main difference from \cite{alexandrov2023instantons} comes from the string theory side of the calculation, where we see two extra bosonic zero modes \eqref{eq: zero_modes} because of the specialization to the self-dual radius. We see that the boundary states corresponding to the bosons are now parametrized by $SU(2)$ (equivalently, $S^3$) instead of a circle $S^1$ in $R\neq 1$ case. This plays a crucial role in the calculation.
    \item Next, we look at the disk two-point function and the annulus one-point functions as well. We use the method devised in \cite{Eniceicu_2022} to first get the matrix model predictions in \eqref{ann_one_pt_MM} and \eqref{disk_2_pt_MM}. We then carry out the worldsheet calculation(following \cite{ Sen_202183,Eniceicu_2022}) again aided by the string field theory insights (\eqref{eq: ann_1_pt_ws_result} and \eqref{eq: disk_2_pt_ws_result}) and see the perfect matching. We again witness the extra zero modes, appearing at self-dual radius, playing a crucial role in this matching.
    \item We look at a particular contribution to the two instanton amplitude in $R\neq 1$ case. Particularly, we explain the term proportional to $e^{-2\pi \mu(1+R)}$ in the partition function trans-series in \eqref{Rneq1_trans_series}. This is explained by the absence of the contribution from the mixed (boson) boundary condition annuli in both $R=1$ as well as $R\neq 1$ case as shown in \eqref{eq: mixed_bc_result} and the discussion thereafter. We also show that annuli with mixed (boson) boundary conditions don't seem to contribute even for more general Liouville boundary conditions in \eqref{eq: gen_alpha_result1} and \eqref{eq: gen_alpha_result2} for both $R=1$ and $R\neq 1$ (last paragraph of section \ref{general_liouvill_bc}).
    \item We also study the two $(1,1)$ ZZ instanton normalization and show the triviality of the cross annulus stretched between the two instantons in \eqref{eq: 2_inst_(1,1)_result} using the unitary prescription of \cite{Sen:2020ruy}. It was devised to deal with the singularities coming at some specific separation between the two instantons. We show a matching with the matrix model result in \eqref{Rneq1_trans_series}. We further note that the lorentzian prescription of \cite{balthazar2019multiinstanton}, on the other hand, leads to an extra imaginary piece which has the same functional form as the normalization of a single $(1,2)$ ZZ instanton \eqref{eq: two_inst_lorentz} and another term occurring at the same order in the trans series \eqref{Rneq1_trans_series} obtained from MQM. 
\end{itemize}
Finally, we also have some observations and remarks about general $(1,n)$ ZZ boundary conditions. \cite{balthazar2019multiinstanton} studied the non-compact boson case and showed that only the $(1,n)$ ZZ instantons seem to contribute when matching with the MQM results. We see something similar here for the compactified case (for both $R=1$ and $R\neq 1$). For $R\neq 1$, we see a matching of terms(functional form) in non-perturbative free energy from MQM \eqref{npfreenergyatR} and the string theory calculation in \eqref{eq: (1,n)_ws_result} and \eqref{eq: (1,n)_ws_result_neumann} up to an overall normalization. For $R=1$ case, we see a similar matching (\eqref{npfreeenergyatR=1} and \eqref{eq: (1,n)_ws_R=1}) again up to an overall normalization. We further note some subtleties in these cases. We study the two instanton case for $(1,n)$ and $(1,n')$ boundary conditions and see how these can possibly contribute a normalization proportional to that of a single $(1,n+n')$ ZZ instanton in \eqref{eq: 2_(1,n)_inst_main_result}. We further have some comments about the multi $(1,n)$ ZZ instantons in section \ref{subsec: multi_(1,n)_ZZ}. 

The discussion will be organized as follows. In section \ref{sec_2}, we will start by reviewing the Matrix Quantum Mechanics(MQM) or free fermion results for the free energy of $c=1$ string at general radius $R$. Further subsections will discuss the free energy calculation in the KM Matrix model and the Imbimbo Mukhi(IM) Matrix Model to show how the non-perturbative free energy arises in those cases. Finally, there will be a discussion on the matrix model predictions for the relevant quantities like exponentiated annulus, disk two-point function, and annulus one-point function. Next, section \ref{sec_3} will discuss the exponentiated annulus in different cases, and in section \ref{sec_4}, we will discuss the disk two-point function, and annulus one-point function using the string field theory regularization. We will show the matching with the matrix model predictions made in section \ref{sec_2}. Section \ref{subsec: general(1,n)bc} has some results and observations related to general $(1,n)$ ZZ boundary condition at both the self-dual radius and the generic radius.
\section{Exact Free Energies from Matrix Models}\label{sec_2}
In this section, we will calculate the free energies from the matrix model sides. We will start with describing a mapping of parameters and operators between the Matrix Models and the $c=1$ string theory results from the free fermion description. With this mapping of parameters, we will calculate the free energy from the matrix model and try to match it with the free energy expression from the free fermion description reviewed in the following section \ref{ReviewMQM}.

\subsection{Review: MQM/free fermion results}\label{ReviewMQM}
We will list some results from the free fermion description of the $c=1$ string theory at radius $R$. We choose to work in the double scaling limit of the MQM which amounts to taking the continuum limit of the discretization provided by the Feynman diagrams of the matrix quantum mechanics. Working with the gauged MQM further leads to the restriction to singlet sector and makes it possible for the system to have a free fermion description which leads to various useful results, for example, the free energy in the compactified case is given by \cite{Alexandrov_2003, klebanov2003string}, 
\begin{align}\label{freeenergyatR}
    \mathcal{F}_R(\mu)=-\frac{1}{4}\int_{\frac{1}{\Lambda}}^{\infty}\frac{dt}{t}\frac{e^{i\mu t}}{\sinh\frac{t}{2}\sinh\frac{t}{2R}}.
\end{align}
This is complete free energy i.e. it has perturbative as well as non-perturbative parts. $\text{Re}(\mathcal{F}_R(\mu))$ is the perturbative part and $\text{Im}(\mathcal{F}_R(\mu))$ is the non-perturbative part. The cutoff $\Lambda$ is introduced to account for the non-universal contributions that are ignored in this double scaling limit. We will be specifically interested in the non-perturbative part of free energy. It can be evaluated by closing the contour in the upper half of the complex $t$-plane. Then the integral simply reduces to a sum of residues at the two types of simple poles on the positive imaginary axes at $t=2in\pi, 2in\pi R$ for $n=\mathbb{Z}_+$ \cite{alexandrov2023instantons, Alexandrov_2005} (detailed calculation in appendix \ref{appendix_np_freeenergy}), 
\begin{align}\label{npfreenergyatR}
    \mathcal{F}_{\text{np},R}(\mu)\equiv i\text{Im}\mathcal{F}_R(\mu)=i\sum_{n=1}^{\infty}\frac{1}{4n(-1)^n}\frac{e^{-2\pi n\mu}}{\sin\frac{\pi n}{R}}+i\sum_{n=1}^{\infty}\frac{1}{4n(-1)^n}\frac{e^{-2\pi n \mu R}}{\sin\pi Rn} \ .
\end{align}
Another way of deriving the non-perturbative contribution to free energy is using the language of the chiral Matrix Quantum Mechanics (MQM) as done in \cite{alexandrov2023instantons, Alexandrov:2004cg}. It simply comes from the scattering phase relating the incoming left-moving fermions to the outgoing right-moving fermions in the fermionic sea filling the right side of the harmonic oscillator \cite{alexandrov2023instantons, Alexandrov:2004cg}. 
\begin{align}
    e^{i\phi_0(E)}=\frac{e^{-\frac{\pi}{2}(E-\frac{i}{2})}}{\sqrt{2\pi}}\Gamma\left(\frac{1}{2}+iE\right)
\end{align}
This is equal to the reflection coefficient $R(E)$ appearing in \cite{Balthazar:2019rnh} by complex conjugation upto some overall phase, which doesn't affect the non-perturbative part of the free energy as we will see below.
It has been argued in \cite{Balthazar:2019rnh, Alexandrov_2005,alexandrov2023instantons} that the imaginary part of the phase $\phi_0(E)$ (or the overall factor of $(1+e^{2\pi E})^{-\frac{1}{2}}$ in $R(E)$ in \cite{Balthazar:2019rnh}) is responsible for the non-perturbative effects i.e. the leaking of the fermions through the inverted harmonic oscillator potential. 
\begin{align}\label{eq: Im_phi}
    \text{Im}\phi_0(E)=\frac{1}{2}\log(1+e^{2\pi E})
\end{align}
The free energy is related to this scattering phase as follows \cite{Alexandrov_2005, alexandrov2023instantons}, 
\begin{align}
    2\sin{\frac{\partial_{\mu}}{2R}}\mathcal{F}_{R}(\mu)=\phi_0(-\mu)
\end{align}
Using \eqref{eq: Im_phi} on RHS, we can deduce $\mathcal{F}_{\text{np},R}(\mu)$ as stated in \eqref{npfreenergyatR}.
The two types of terms (from the two types of poles) in the \eqref{npfreenergyatR} have the interpretation of coming respectively from the some D instantons and D0 branes.
It was shown in \cite{alexandrov2023instantons} using the worldsheet calculation aided by the string field theory insights that these non-perturbative terms appearing here indeed correspond to ZZ instantons in the $c=1$ string theory with free boson compactified at radius $R$. ZZ instanton corresponds to the contributions due to ZZ boundary conditions in the Liouville direction along with some specific boundary conditions in the boson direction. In particular, the exponentiated annulus calculation with $(1,1)$ ZZ boundary condition in the Liouville direction and Dirichlet boundary condition in the free boson direction gave the result, 
\begin{align}\label{DDBC_result}
    \frac{i}{4\sin\frac{\pi}{R}}
\end{align}
which is exactly the normalization of the $n=1$ term in the first sum in \eqref{npfreenergyatR}. A similar result(with the same derivation) holds for the case when we have Neumann boundary condition in the free boson direction, 
\begin{align}\label{NBC_result}
    \frac{i}{4\sin{\pi R}}
\end{align}
which is the normalization of the  $n=1$ term in the second sum in \eqref{npfreenergyatR}.

\subsection{The Two-Matrix model (or the KM Matrix Model)}
We are interested in a matrix model with two hermitian matrices $K$ and $M$ \cite{gopakumar2023derivingsimplestgaugestringduality,Mazenc_2023} given as follows,
\begin{align}
	Z=\frac{1}{\text{Vol(U(N))}}\int dM dK\exp\bigg(-\frac{1}{2g_s}(c_1\text{Tr}K^2+c_2\text{Tr}M^2+2c_3\text{Tr}(KM))\bigg) \ .
\end{align}
It is supposed to describe the $c=1$ string at self-dual radius. We will treat it like a formal matrix integral for now, and later, we will discuss the convergence. The propagators for it can be evaluated as follows (see, for example, \cite{lando2003graphs}), 
\begin{align}
	\langle K_{ij}K_{kl}\rangle=\frac{\delta_{il}\delta_{jk}c_2g_s}{c_1c_2-c_3^2}, \qquad \langle M_{ij}M_{kl}\rangle=\frac{\delta_{il}\delta_{jk}c_1g_s}{c_1c_2-c_3^2}, \qquad \langle K_{ij}M_{kl}\rangle=-\frac{\delta_{il}\delta_{jk}c_3g_s}{c_1c_2-c_3^2} \ .
\end{align}
We are interested in the case when the only non-vanishing propagator is the mixed propagator, i.e., $c_1,c_2=0$ and $c_3=1$ as it was argued in \cite{Mazenc_2023}. The reason for this choice is that the correlators of gauge-invariant operators Tr$M^n$ and Tr$K^n$ can be described in terms of three permutations and consequently be written as a sum over branched coverings of $\mathbb{P}^1$ (see \cite{Mazenc_2023} for more details). This interpretation of correlators as a sum over branched coverings plays a crucial role in relating this matrix model to topological strings. Furthermore, this matrix model exhibits graph duality at the level of Feynman diagrams upon integrating in and out procedure of \cite{gopakumar2023derivingsimplestgaugestringduality, Brown:2010af} as shown in \cite{Mazenc_2023}. With the above choice of measure and the identification\footnote{The factors of $\Gamma(1-|l|)$ are divergent whenever $l$ is an integer(which is the case here because $R=1$ leads to integer momenta). We should really be thinking of the vertex operators $T(l)$ getting renormalized at $R=1$ and this divergent factor getting absorbed in the definition of the vertex operator itself.}
\begin{align}\label{identification}
	\text{Tr}(M^l)\leftrightarrow\frac{-i|l|\Gamma(|l|)T(l)}{\Gamma(1-|l|)}, \quad \text{Tr}(K^l)\leftrightarrow\frac{-i|l|\Gamma(|l|)T(-l)}{\Gamma(1-|l|)}, \quad g_s=i, \text{ and } N=-i\mu,
\end{align} 
with the tachyons $T(l)$ ($l$ is the tachyon momentum) and chemical potential $\mu$ in free fermion formulation \cite{klebanov2003string}, we have matches with the tachyon correlators at the self-dual radius ($R=1$ in the notation of \cite{klebanov2003string}) \cite{Mazenc_2023,Bonora_1995}. Note that we have the freedom to rescale $g_s$ in the above mapping of parameters. We just need to rescale the mapping of operators accordingly to compensate. With this mapping of parameters, we can evaluate the free energy as follows, 
\begin{align}
    F(\mu)=\ln Z\bigg|_{g_s=i, N=-i\mu} \ .
\end{align}
We can evaluate $Z$ using the orthogonal polynomial techniques by first reducing it to an eigenvalue integral, 
\begin{align}
    Z=\frac{(2\pi g_s)^{N^2}(c_1c_2-c_3^2)^{\frac{N^2}{2}}}{\text{Vol(U(N))}}\bigg|_{(c_1,c_2=0,c_3=1)}=(2\pi)^{\frac{N(N-1)}{2}}G_2(N+1)\left(-g_s^2\right)^{\frac{N^2}{2}} \ .
\end{align}
Since we have already argued that $g_s$ can be scaled by a compensating scaling of the mapping of operators, the $g_s$ dependence can be ignored (or we can just set it to $i$) as it will just add an additive constant to the free energy. Hence, we have the following free energy,
\begin{align}
    F(\mu)=&\frac{i\mu(i\mu+1)}{2}\ln(2\pi) +\ln G_2(1-i\mu)\nonumber\\
    =&i\mu\int_0^{\infty}\frac{dt}{t^2}-\frac{\mu^2}{2}\bigg(\ln 2\pi+\int_0^{\infty}dt\frac{e^{-t}}{t}\bigg)+\mathcal{F}(\mu)-\mathcal{F}(0)\label{eq_int_step_freen_1}\\
    \text{where, }\mathcal{F}(\mu)\equiv \mathcal{F}_{R=1}(\mu)=&-\int_{0}^{\infty}\frac{dt}{t}\frac{e^{i\mu t}}{4\sinh^2 \frac{t}{2}}\label{freeenergyatR=1}
\end{align}
where we have used the integral form of the Barnes gamma function, 
\begin{align}\label{barnes_gamma_exp}
    \ln G_2(1+z)=\frac{z}{2}\ln2\pi+\int_0^{\infty}\frac{dt}{t}\bigg(\frac{1-e^{-zt}}{4\sinh^2{\frac{t}{2}}}+\frac{z^2}{2}e^{-t}-\frac{z}{t}\bigg) \ .
\end{align}
An important thing to note is that the integral term $\int_0^{\infty}\frac{dt}{t}\frac{e^{-zt}}{4\sinh^2{\frac{t}{2}}}$ is divergent from $t\rightarrow 0$ limit and the other terms appearing in the integral act as a regulator to this divergence ultimately making the function finite. This can be related to the cutoff $\Lambda$ in the integral \eqref{freeenergyatR}. 

Another way to argue the existence of non-perturbative effects is by looking at the Taylor series expansion of the $\log{G_2(1+z)}$ and then taking the Borel transform and consequently the Borel resummation. Borel resummation picks up residues at the poles as the perturbation parameter ($1/N$ in this case) is analytically continued in the complex plane. This was exactly the procedure followed in \cite{Pasquetti_2010}. The results and calculations that we will show here using \eqref{freeenergyatR=1} are similar to the discussion in \cite{Pasquetti_2010}.

$\mathcal{F}(\mu)$ is exactly the free energy obtained from the free fermion formulation in \eqref{freeenergyatR}. Hence, we see that with the correct mapping of parameters, the matrix integral is capable of producing not only perturbative free energy and correlators but also non-perturbative free energy. This is a general phenomenon that non-perturbative effects are encountered when we move the coupling in the complex plane and in this case, we have moved $(N,g_s)\rightarrow (-i\mu, i)$.

\subsection{Imbimbo Mukhi Model}
In this section, we will discuss the Imbimbo Mukhi (IM) matrix model that was shown to capture the $c=1$ string at self-dual radius in \cite{Imbimbo_1995}. It certainly does capture the perturbative results because it is designed to satisfy the $W_{\infty}$ constraints followed by the double-scaled MQM partition function calculated for the $c=1$ string theory at self-dual radius \cite{Dijkgraaf_1993}.
It was classified as an F-type model in  \cite{Mazenc_2023,gopakumar2023derivingsimplestgaugestringduality} i.e. it is related to the KM matrix via graph duality as shown in \cite{gopakumar2023derivingsimplestgaugestringduality}. Hence, $\mu$ has to appear explicitly in the matrix integral instead of $N=-i\mu$ identification/analytic continuation needed in the two-matrix model case.  The IM matrix integral looks like, 
\begin{align}\label{IM model}
    &Z_{IM}(\{t_k,\bar{t}_k\})=(\text{det}(A))^{\nu}\int dM\exp\bigg(-\nu \text{Tr}( MA)-(Q-\nu)\text{Tr}\ln M-\nu\sum_{k=1}^{\infty}\bar{t}_k\text{Tr}M^k\bigg)\nonumber\\
    &\text{where, }\nu:=-i\mu\text{ and } t_k:=\frac{\text{Tr}A^{-k}}{\nu k}
\end{align}
where $M,A$ are $Q\times Q$ matrices and $A$ can be thought of as a source matrix. The tachyon correlators can be extracted from this matrix model by the use of $W_{\infty}$ constraints \cite{Imbimbo_1995}. For the case of a stationary background in which we are interested, we have to set all $t_k, \bar{t}_k$s to zero. In this case, it can be simply noticed that if we set $\bar{t}_k$s to zero, $t_{k}$ (and hence, $A$) dependence drops out automatically. This is a reflection of the conservation of tachyon momentum on the string side or the MQM side. Hence, the problem comes down to the free energy calculation of the matrix integral,
\begin{align}
Z_{IM}(\{t_k,0\})=&(\text{det}(A))^{\nu}\int dM\exp\bigg(-\nu \text{Tr}(MA)-(Q-\nu)\text{Tr}\ln M\bigg)\nonumber\\
=&\int dM\exp(-\nu \text{Tr}M-(Q-\nu)\text{Tr}\ln M)
\end{align}
where we have done the redefinition, $M\rightarrow \Tilde{M}=MA^{-1}$, to get rid of the $A$ dependence. Also, the large $Q$ limit is implicit from the beginning to ensure the independence of all the $t_k$s. This integral is a Penner-type matrix integral. This is exactly calculated for real $\nu$ using the orthogonal polynomials and then, we can analytically continue $\nu=-i\mu$ in the end. This analytic continuation will be the reason for the non-perturbative effects as we don't usually expect non-perturbative effects in Penner-like matrix integrals. We need the orthogonal polynomials for the measure $\frac{dm}{2\pi}m^{\nu-Q}e^{-\nu m}$ to evaluate the integral. This is exactly the measure for the Laguerre polynomials which satisfy the following orthogonality relation, 
\begin{align}
    \int \frac{dx}{2\pi}x^{\alpha}e^{-x}L_n(x)L_m(x)=\frac{\Gamma(n+\alpha+1)}{2\pi}n!\delta_{nm}
\end{align}
which can be easily checked by using the definition $L_n:=x^{-\alpha}e^x(-\partial_x)^n(x^{n+\alpha}e^{-x})$. We put $\alpha=\nu-Q$ and $x=\nu m$ to get, 
\begin{align}
    \int \frac{dm}{2\pi}m^{\nu-Q}e^{-\nu m}l_n(m)l_m(m)=h_n\delta_{nm}\text{ where, }h_n=\frac{\Gamma(\nu-Q+1+n)\Gamma(n+1)}{2\pi\nu^{\nu-Q+1+2n}}
\end{align}
In terms of these $h_n$'s, the matrix integral takes a simple form (see, for example, \cite{marino2005chern}), 
\begin{align}
    \frac{Z_{IM}(\{t_k,0\})}{\text{Vol(U(Q))}}=&\prod_{n=0}^{Q-1}h_n=\frac{G_2(\nu+1)G_2(Q+1)}{(2\pi)^Q\nu^{Q\nu}G_2(\nu-Q+1)}\label{Imb_Part_funt}
\end{align}
We will now put $\nu=-i\mu$ and immediately see a factor of $G_2(1-i\mu)$, which indeed has the same corrections of type $e^{-2\pi\mu}$ upon taking a logarithm\footnote{ Note that we are taking partition function to be the gauge-fixed matrix model as we have divided by the $U(Q)$ group volume. This makes sense because only the correlation functions of the gauge invariant operators like traces and determinants are considered in \cite{Imbimbo_1995,Dijkgraaf_1993} to get the tachyon correlators.}, 
\begin{align}\label{IMfreeeng}
    \log\bigg( \frac{Z_{IM}(\{t_k,0\})}{\text{Vol(U($Q$))}}\bigg)=&\log{G_2(1-i\mu)}+\log{G_2(1+Q)}-\log{G_2(-i\mu-Q+1)}\nonumber\\
    &-Q\log{2\pi}+i\mu Q\log(-i\mu)
\end{align}
The first term is $Q$ independent and it exactly gives the free energy term we get from the MQM \eqref{freeenergyatR} (setting $R=1$) as we saw in \eqref{eq_int_step_freen_1}. The imaginary part of that gives us the non-perturbative effects. However, extra $Q$ dependent terms may have some non-trivial $\mu$ dependent imaginary part. The KM model was shown to reduce to a product of IM Model and Penner Model in \cite{gopakumar2023derivingsimplestgaugestringduality} by a field redefinition. So, the extra terms appearing here may be thought of as compensating for some decoupled Penner integral. We highlight in appendix \ref{app: KMIMrel} that the decoupling which was shown at a formal level in \cite{gopakumar2023derivingsimplestgaugestringduality} works up to a non-perturbative contribution in $\mu^{-1}$.  

Another important observation is that the required term comes from the factor of $G_2(1-i\mu)$ in \eqref{Imb_Part_funt} which has a completely different origin here as compared to the two matrix integral. In the case of the KM Model, the origin was that we chose to work with the gauged matrix model, and that led to the factor of $\text{Vol(U}(N))^{-1}$ which gave us the $G_{2}(1-i\mu)$ upon using the mapping $N=-i\mu$. Here, we see this factor coming automatically. 
\subsection{Non-perturbative Free Energy}
Non-perturbative part of the free energy \eqref{freeenergyatR=1} can again be evaluated by the same procedure of closing the contour in the upper half of the complex $t$-plane. But this time, the integral will amount to calculating residues at the double poles on the imaginary axes as the two types of poles that appear in \eqref{npfreenergyatR} merge to give double poles. The details are again discussed in the appendix \ref{appendix_np_freeenergy}, and here, we simply state the result, 
\begin{align}\label{npfreeenergyatR=1}
    \mathcal{F}_{\text{np}}(\mu)=-i\sum_{n=1}^{\infty} e^{-2 \pi  \mu  n} \bigg(\frac{\mu }{2n}+\frac{1}{4\pi n^2}\bigg)
\end{align}
Hence, we see that there are only one type of terms in $R=1$ case. Note that $\mathcal{F}_{\text{np}}(\mu)$ can also be obtained as the $R\rightarrow 1$ limit of the \eqref{npfreenergyatR} i.e. $\mathcal{F}_{\text{np}}(\mu)=\lim_{R\rightarrow 1}\mathcal{F}_{\text{np},R}(\mu)$. We have shown this in detail in the appendix \ref{appendix_np_freeenergy}. The figure \ref{Commutingdiag} indicates that \textit{``extracting non-perturbative effects from the integral" and ``taking $R\rightarrow 1$ limit" commutes in this case which may not be true in general}.
\begin{figure}[ht!]
    \centering

\tikzset{every picture/.style={line width=0.75pt}} %set default line width to 0.75pt        

\begin{tikzpicture}[x=0.75pt,y=0.75pt,yscale=-1,xscale=1]
%uncomment if require: \path (0,189); %set diagram left start at 0, and has height of 189

%Straight Lines [id:da7655512320593572] 
\draw    (291.14,38) -- (379.59,38) ;
\draw [shift={(381.59,38)}, rotate = 180] [color={rgb, 255:red, 0; green, 0; blue, 0 }  ][line width=0.75]    (10.93,-3.29) .. controls (6.95,-1.4) and (3.31,-0.3) .. (0,0) .. controls (3.31,0.3) and (6.95,1.4) .. (10.93,3.29)   ;
%Straight Lines [id:da723273677301332] 
\draw    (291.14,151.9) -- (379.59,151.9) ;
\draw [shift={(381.59,151.9)}, rotate = 180] [color={rgb, 255:red, 0; green, 0; blue, 0 }  ][line width=0.75]    (10.93,-3.29) .. controls (6.95,-1.4) and (3.31,-0.3) .. (0,0) .. controls (3.31,0.3) and (6.95,1.4) .. (10.93,3.29)   ;
%Straight Lines [id:da39727525605292935] 
\draw    (413.65,57.81) -- (413.65,133.81) ;
\draw [shift={(413.65,135.81)}, rotate = 270] [color={rgb, 255:red, 0; green, 0; blue, 0 }  ][line width=0.75]    (10.93,-3.29) .. controls (6.95,-1.4) and (3.31,-0.3) .. (0,0) .. controls (3.31,0.3) and (6.95,1.4) .. (10.93,3.29)   ;
%Straight Lines [id:da613738963871213] 
\draw    (253.35,55.33) -- (253.35,131.33) ;
\draw [shift={(253.35,133.33)}, rotate = 270] [color={rgb, 255:red, 0; green, 0; blue, 0 }  ][line width=0.75]    (10.93,-3.29) .. controls (6.95,-1.4) and (3.31,-0.3) .. (0,0) .. controls (3.31,0.3) and (6.95,1.4) .. (10.93,3.29)   ;

% Text Node
\draw (236.08,25.69) node [anchor=north west][inner sep=0.75pt]    {$\mathcal{F}_{R}( \mu )$};
% Text Node
\draw (395.72,29.16) node [anchor=north west][inner sep=0.75pt]    {$\mathcal{F}( \mu )$};
% Text Node
\draw (308.21,16.78) node [anchor=north west][inner sep=0.75pt]    {$R\rightarrow 1$};
% Text Node
\draw (224.57,142.3) node [anchor=north west][inner sep=0.75pt]    {$\mathcal{F}_{\text{np} ,R}( \mu )$};
% Text Node
\draw (393.23,143.54) node [anchor=north west][inner sep=0.75pt]    {$\mathcal{F}_{\text{np}}( \mu )$};
% Text Node
\draw (308.21,130.69) node [anchor=north west][inner sep=0.75pt]    {$R\rightarrow 1$};
% Text Node
\draw (256.57,81) node [anchor=north west][inner sep=0.75pt]   [align=left] {{\small Evaluation of Im(Integral)}};

\end{tikzpicture}

    \caption{Extracting the imaginary parts of the free energies and $R\rightarrow 1$ limits commute}
    \label{Commutingdiag}
\end{figure}

An important thing to emphasize here is that the non-perturbative free energy above does not have an obvious interpretation as the eigenvalue instantons in the matrix integrals. There was some work in giving these non-perturbative effects an interpretation in terms of the eigenvalue instantons using the multi-sheeted nature of the effective potential in \cite{Pasquetti_2010,Matsuo_2006}. This behaviour is similar to the saddle point approximation in $\Gamma(z)$.

\subsection{Matrix Model Predictions}\label{subsec: MM_predictions}
We have the result for the full non-perturbative partition function from the matrix model side (collectively MQM and both matrix models),
\begin{align}
    \mathcal{Z}=&\exp(\mathcal{F}_{\text{p}}+\mathcal{F}_{\text{np}})=\exp(\mathcal{F}_{\text{p}})\exp\bigg(-i\sum_{n=1}^{\infty}e^{-2\pi \mu n}\bigg(\frac{\mu}{2n}+\frac{1}{4\pi n^2}\bigg)\bigg)\nonumber\\
    =&\mathcal{Z}^{(0)}\bigg(1-ie^{-2\pi \mu }\bigg(\frac{\mu}{2}+\frac{1}{4\pi}\bigg)+\mathcal{O}(e^{-4\pi \mu})\bigg)\label{eq: trans_series_R=1_part_func}
\end{align}
where $\mathcal{Z}^{(0)}:=e^{\mathcal{F}_{\text{p}}}$ is the perturbative or zero-instanton sector contribution to the partition function. Focusing on the one-instanton contribution to the partition function (using $g_s=(2\pi \mu)^{-1}$ identification because of the KPZ scaling law \cite{alexandrov2023instantons}),
\begin{align}
    \frac{\mathcal{Z}^{(1)}}{\mathcal{Z}^{(0)}}=&-ie^{-2\pi \mu}\bigg(\frac{\mu}{2}+\frac{1}{4\pi}\bigg)=-\frac{i\mu}{2}\exp\bigg(-2\pi \mu+\ln\bigg(1+\frac{1}{2\pi \mu}\bigg)\bigg)\label{one_inst_1}\\
    =&-\frac{i}{2}\exp\bigg(-2\pi \mu+\ln \mu+\ln\bigg(1+\frac{1}{2\pi \mu}\bigg)\bigg)\label{one_inst_2}\\
    =&-\frac{i}{2}\exp\bigg(-g_s^{-1}-\ln g_s-\ln 2\pi+\ln(1+g_s)\bigg).\label{one_inst_3}
\end{align}
From \eqref{one_inst_1}, we can simply read off the expected result for the exponentiated annulus with boundary conditions which is $-i\mu/2$.
Following the general method described in \cite{Eniceicu_2022}, we can predict the ratio of the annulus one-point function to the disk one-point function (for zero-momentum insertions), $g$, using \eqref{one_inst_2},
\begin{align}\label{ann_one_pt_MM}
    g_s g =\frac{\frac{\partial (\ln \mu)}{\partial \mu }}{\frac{\partial (-2\pi \mu)}{\partial \mu}}=-(2\pi \mu)^{-1}\quad \Rightarrow \quad  g=-1 \ .
\end{align}
The result for $g$ differs from that in \cite{Eniceicu_2022} since $\ln\left(\frac{\mathcal{Z}^{(1)}}{\mathcal{Z}^{(0)}}\right)$ has the term $-\ln g_s$ \eqref{one_inst_3} in this case instead of $\frac{1}{2}\ln g_s$ in the minimal string case discussed in \cite{Eniceicu_2022}. The ratio of the disc two-point function to disk one-point function squared, $f$ can be similarly determined, 
\begin{align}\label{disk_2_pt_MM}
    g_s f=\frac{\frac{\partial^2 (-2\pi \mu)}{\partial \mu^2}}{\bigg(\frac{\partial(-2\pi \mu )}{\partial \mu}\bigg)^2}\quad \Rightarrow \quad f=0 \ .
\end{align}
Also, the two instanton($\sim e^{-4\pi \mu}$) contribution to the trans-series \eqref{eq: trans_series_R=1_part_func} will be 
\begin{align}\label{eq: 2_inst_R=1_term_trans}
    \frac{\mathcal{Z}^{(2)}}{\mathcal{Z}^{(0)}}=-\frac{i\mu e^{-4 \pi  \mu }}{4}\left(1+\frac{1}{4\pi \mu}\right)-\frac{\mu^2e^{-4 \pi  \mu }}{8}\left(1+\frac{1}{\pi \mu}+\frac{1}{4\pi^2\mu^2}\right)\ .
\end{align}

We will note similar results for the general $R$ case as well. The partition function can be written as follows using \eqref{npfreenergyatR},
\begin{align}
    \mathcal{Z}_R=&\exp\bigg(\mathcal{F}_{\text{p},R}-\frac{i}{4}\bigg(\frac{e^{-2\pi \mu}}{\sin\frac{\pi }{R}}+\frac{e^{-2\pi \mu R}}{\sin\pi R}\bigg)+\frac{i}{8}\bigg(\frac{e^{-4\pi\mu}}{\sin\frac{2\pi }{R}}+\frac{e^{-4\pi  \mu R}}{\sin 2\pi R}\bigg)+\mathcal{O}(e^{-2\pi \mu(a+bR)})\bigg)\nonumber\\
    =&\mathcal{Z}_{R}^{(0)}\bigg\{1+\frac{\mathcal{Z}_R^{(1)}}{\mathcal{Z}_R^{(0)}}+\frac{\mathcal{Z}^{(2)}_R}{\mathcal{Z}^{(0)}_R}+\mathcal{O}(e^{-2\pi \mu(a+bR)})\bigg\}\\
    \text{where }&\frac{\mathcal{Z}_R^{(1)}}{\mathcal{Z}_R^{(0)}}=-\frac{i}{4}\left(\frac{e^{-2\pi \mu}}{\sin\frac{\pi }{R}}+\frac{e^{-2\pi \mu R}}{\sin\pi R}\right)\text{ and }\\
    &\frac{\mathcal{Z}_R^{(2)}}{\mathcal{Z}_R^{(0)}}=\left(\frac{i}{8\sin\frac{2\pi }{R}}-\frac{1}{32\sin^2\frac{\pi}{R}}\right)e^{-4\pi\mu}+\left(\frac{i}{8\sin 2\pi R}-\frac{1}{32\sin^2\pi R}\right)e^{-4\pi \mu R}\nonumber\\
    &\qquad \qquad-\frac{e^{-2\pi\mu(1+R)}}{16\sin \frac{\pi}{R}\sin\pi R} \ .\label{Rneq1_trans_series}
\end{align}
Here, we have kept terms up to two D-Instanton or D0-brane contributions($a+b=2$). Consider the following term in the two-instanton sector, $\frac{\mathcal{Z}_R^{(2)}}{\mathcal{Z}_R^{(0)}}$, corresponding to one D-instanton and one wrapped D0-brane, 
\begin{align}\label{2_inst_mixed_contri}
    -\frac{e^{-2\pi \mu(1+R)}}{16\sin \frac{\pi}{R}\sin \pi R}=\frac{i}{4\sin \frac{\pi}{R}}\frac{i}{4\sin \pi R}e^{-2\pi \mu-2\pi \mu R}
\end{align}
The first factor on the RHS corresponds to the exponentiated annulus with Dirichlet boundary condition (result quoted in \eqref{DDBC_result}) on both boundaries, whereas the second factor corresponds to the Neumann boundary condition on both boundaries \eqref{NBC_result}. Hence, we expect that the exponentiated annulus for the mixed boundary condition (Neumann on one boundary and Dirichlet on the other boundary) should be just $1$ or the annulus partition function should vanish (upon appropriate integration over the moduli). If we picture the free energy \eqref{freeenergyatR} as the sum of connected diagrams, this statement is just that there is no diagram connecting a D instanton (Dirichlet) and a D0 brane (Neumann).

\section{String Theory Calculation of Instanton Normalizations}\label{sec_3}
In this section, we will carry out the calculation of the instanton normalizations in the $c=1$ string theory at self-dual radius as well as at a generic radius using the worldsheet CFT formulation aided with string field theory insights. We will mainly follow the methods developed in \cite{Sen_2021,Sen_202183} and used in \cite{alexandrov2023instantons}. The main difference in $R=1$ case will come from the fact that there will be a family of boundary conditions parametrized by $g\in SU(2)$ as opposed to just Dirichlet or Neumann in generic $R$ case \cite{Recknagel_Schomerus_2013,Recknagel_1999,Gaberdiel_2002} parametrized respectively by the position on compactification circle or the dual circle (see appendix \ref{app_freeboson_review}). This will lead to three bosonic zero modes instead of the one bosonic zero mode in the $R\neq 1$ case. These zero modes, along with the other zero modes, lead to divergences in the worldsheet CFT calculations and need to be properly interpreted as a path integral using the string field theory \cite{Sen_2021}.

Instanton Normalization is controlled by the \textit{Exponentiated Annulus} in string theory. Since the annulus diagram in string theory has Euler characteristic $\chi=2-2g-n=0$ (because $g=0, n=2$ for the annulus\footnote{Here, $g$ is the genus and not the element of SU(2).
$n$ is the number of boundaries.}), it scales as $g_s^{-\chi}=g_s^0$ with the string coupling $g_s$. Hence, the exponentiated annulus factors out and appears as the normalization of the instanton amplitudes. We will first work out the case of $(1,1)$ ZZ boundary condition on both the boundaries of the annulus for the Liouville direction. However, we will be considering two cases for the boson direction, the same boundary condition, and different boundary conditions on the two boundaries of the annulus. Later, we will look at the general $(m,n)$ Liouville boundary conditions and the generic radius $R$ cases. 
\begin{figure}[ht!]
    \centering

\tikzset{every picture/.style={line width=0.75pt}} %set default line width to 0.75pt        

\begin{tikzpicture}[x=0.75pt,y=0.75pt,yscale=-0.91,xscale=0.91]
%uncomment if require: \path (0,294); %set diagram left start at 0, and has height of 294

%Shape: Path Data [id:dp5140994606885412] 
\draw  [fill={rgb, 255:red, 80; green, 227; blue, 194 }  ,fill opacity=0.3 ] (348.72,22.5) .. controls (378.39,22.5) and (402.45,46.15) .. (402.45,75.33) .. controls (402.45,104.5) and (378.39,128.15) .. (348.72,128.15) .. controls (319.05,128.15) and (295,104.5) .. (295,75.33) .. controls (295,46.15) and (319.05,22.5) .. (348.72,22.5) -- cycle (317.56,75.33) .. controls (317.56,92.25) and (331.51,105.97) .. (348.72,105.97) .. controls (365.93,105.97) and (379.88,92.25) .. (379.88,75.33) .. controls (379.88,58.4) and (365.93,44.69) .. (348.72,44.69) .. controls (331.51,44.69) and (317.56,58.4) .. (317.56,75.33) -- cycle ;
%Shape: Path Data [id:dp6390141239002456] 
\draw  [fill={rgb, 255:red, 80; green, 227; blue, 194 }  ,fill opacity=0.3 ] (178.72,152.5) .. controls (208.39,152.5) and (232.45,176.15) .. (232.45,205.33) .. controls (232.45,234.5) and (208.39,258.15) .. (178.72,258.15) .. controls (149.05,258.15) and (125,234.5) .. (125,205.33) .. controls (125,176.15) and (149.05,152.5) .. (178.72,152.5) -- cycle (147.56,205.33) .. controls (147.56,222.25) and (161.51,235.97) .. (178.72,235.97) .. controls (195.93,235.97) and (209.88,222.25) .. (209.88,205.33) .. controls (209.88,188.4) and (195.93,174.69) .. (178.72,174.69) .. controls (161.51,174.69) and (147.56,188.4) .. (147.56,205.33) -- cycle ;
%Shape: Path Data [id:dp10702856113230719] 
\draw  [fill={rgb, 255:red, 80; green, 227; blue, 194 }  ,fill opacity=0.3 ] (348.72,152.25) .. controls (378.39,152.25) and (402.45,175.9) .. (402.45,205.08) .. controls (402.45,234.25) and (378.39,257.9) .. (348.72,257.9) .. controls (319.05,257.9) and (295,234.25) .. (295,205.08) .. controls (295,175.9) and (319.05,152.25) .. (348.72,152.25) -- cycle (317.56,205.08) .. controls (317.56,222) and (331.51,235.71) .. (348.72,235.71) .. controls (365.93,235.71) and (379.88,222) .. (379.88,205.08) .. controls (379.88,188.15) and (365.93,174.44) .. (348.72,174.44) .. controls (331.51,174.44) and (317.56,188.15) .. (317.56,205.08) -- cycle ;
%Shape: Path Data [id:dp5100923965302295] 
\draw  [fill={rgb, 255:red, 80; green, 227; blue, 194 }  ,fill opacity=0.3 ] (531.72,152.25) .. controls (561.39,152.25) and (585.45,175.9) .. (585.45,205.08) .. controls (585.45,234.25) and (561.39,257.9) .. (531.72,257.9) .. controls (502.05,257.9) and (478,234.25) .. (478,205.08) .. controls (478,175.9) and (502.05,152.25) .. (531.72,152.25) -- cycle (500.56,205.08) .. controls (500.56,222) and (514.51,235.71) .. (531.72,235.71) .. controls (548.93,235.71) and (562.88,222) .. (562.88,205.08) .. controls (562.88,188.15) and (548.93,174.44) .. (531.72,174.44) .. controls (514.51,174.44) and (500.56,188.15) .. (500.56,205.08) -- cycle ;

% Text Node
\draw (333.37,81.91) node [anchor=north west][inner sep=0.75pt]  [font=\scriptsize]  {$( m,n)$};
% Text Node
\draw (37.28,41.81) node [anchor=north west][inner sep=0.75pt]  [font=\scriptsize]  {$R=1:$};
% Text Node
\draw (36.28,178.31) node [anchor=north west][inner sep=0.75pt]  [font=\scriptsize]  {$R\neq 1:$};
% Text Node
\draw (391.37,112.41) node [anchor=north west][inner sep=0.75pt]  [font=\scriptsize]  {$( m',n')$};
% Text Node
\draw (159.37,192.91) node [anchor=north west][inner sep=0.75pt]  [font=\scriptsize]  {$( D,x_{1})$};
% Text Node
\draw (233.87,221.41) node [anchor=north west][inner sep=0.75pt]  [font=\scriptsize]  {$( D,x_{2})$};
% Text Node
\draw (329.87,187.16) node [anchor=north west][inner sep=0.75pt]  [font=\scriptsize]  {$\left( N,\tilde{x}_{1}\right)$};
% Text Node
\draw (402.87,216.16) node [anchor=north west][inner sep=0.75pt]  [font=\scriptsize]  {$\left( N,\tilde{x}_{2}\right)$};
% Text Node
\draw (512.37,190.66) node [anchor=north west][inner sep=0.75pt]  [font=\scriptsize]  {$( D,x_{0})$};
% Text Node
\draw (588.87,211.16) node [anchor=north west][inner sep=0.75pt]  [font=\scriptsize]  {$\left( N,\tilde{x}_{0}\right)$};
% Text Node
\draw (162.37,212.41) node [anchor=north west][inner sep=0.75pt]  [font=\scriptsize]  {$( m,n)$};
% Text Node
\draw (218.37,246.91) node [anchor=north west][inner sep=0.75pt]  [font=\scriptsize]  {$( m',n')$};
% Text Node
\draw (332.87,212.41) node [anchor=north west][inner sep=0.75pt]  [font=\scriptsize]  {$( m,n)$};
% Text Node
\draw (389.37,248.91) node [anchor=north west][inner sep=0.75pt]  [font=\scriptsize]  {$( m',n')$};
% Text Node
\draw (516.37,210.16) node [anchor=north west][inner sep=0.75pt]  [font=\scriptsize]  {$( m,n)$};
% Text Node
\draw (576.37,244.16) node [anchor=north west][inner sep=0.75pt]  [font=\scriptsize]  {$( m',n')$};
% Text Node
\draw (343.87,60.91) node [anchor=north west][inner sep=0.75pt]  [font=\scriptsize]  {$g_{1}$};
% Text Node
\draw (408.37,89.41) node [anchor=north west][inner sep=0.75pt]  [font=\scriptsize]  {$g_{2}$};

\end{tikzpicture}

    \caption{The continuum of possible boundary conditions parametrized by $g\in$ SU(2) in $R=1$ case reduces to only two possibilities Dirichlet ($D$) and Neumann ($N$) in the case of $R\neq 1$. $x_0,x_1$ and $x_2$ denote the collective coordinates on the compactification circle of radius $R$ whereas $\tilde{x}_0, \tilde{x}_1$ and $\tilde{x}_2$ denote the coordinate on the dual circle of radius $\frac{1}{R}$.}
    \label{different_BC}
\end{figure}
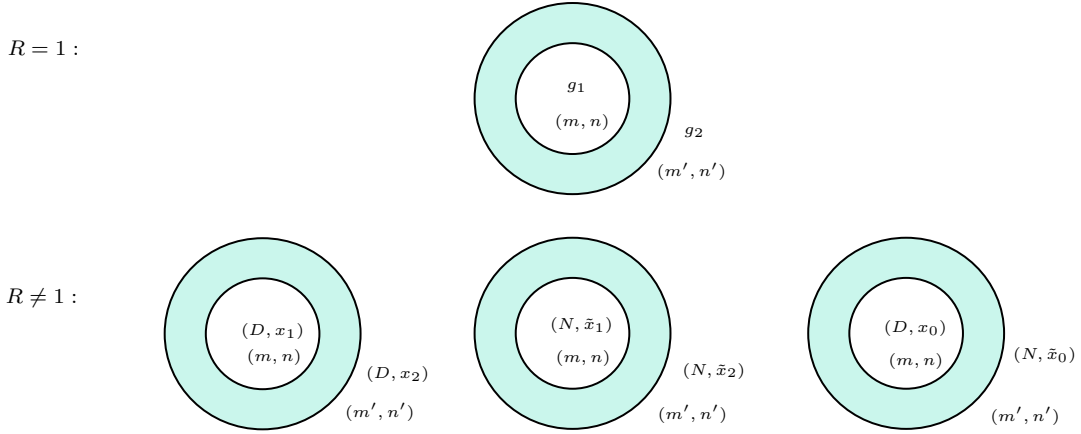
\subsection{Single $(1,1)$ ZZ Instanton at Self-Dual $R$}
In this section, we will evaluate the exponentiated annulus using the worldsheet CFT \cite{alexandrov2023instantons,balthazar2019c1}. It is given by the following, 
\begin{align}\label{expannulus}
    \exp\bigg(\int_0^{\infty} \frac{dt}{2t}Z(t)\bigg)
\end{align}
where $Z(t)$ is the cylinder/annulus partition function which is just a product of Liouville \cite{balthazar2019multiinstanton,ZZoriginal}, free boson and the ghost cylinder partition functions with appropriate boundary conditions i.e.
\begin{align}
     &Z(t)=Z_{\varphi,(1,1)}(t)Z_{X,g}(t)Z_{\text{gh}}(t)\\
    Z_{\varphi,(1,1)}(t)=\frac{v^{-1}-1}{\eta(it)}, \quad &Z_{X,g}(t)=\frac{1}{\eta(it)}\sum_{j\in \mathbb{Z}}v^{j^2}, \quad Z_{\text{gh}}(t)=(\eta(it))^2\quad(v:=e^{-2\pi t})\label{cylpartfuncs}
\end{align}
where, the $Z_{X,g}(t)$ is evaluated for the same boundary conditions (parametrized by $g\in$ SU(2)) on both the boundaries of the annulus (figure \ref{different_BC}). $Z_{\varphi,(1,1)}(t)$ corresponds to the $(1,1)$ ZZ boundary condition on both the boundaries. Hence, this annulus corresponds to the case $(m,n)=(m',n')=(1,1)$ and $g_1=g_2=g$ in $R=1$ case in figure \ref{different_BC}. It is clearly visible that $Z(t)$ will be a sum over all integers, and the terms corresponding to $j=0,\pm1$ will lead to the divergences we were anticipating. For the $R\neq 1$ case discussed in \cite{alexandrov2023instantons}, it was just the $j=0$ term. This signifies that there are two extra zero modes now compared to the $R\neq 1$ case. Separating the convergent and the divergent terms, we get,
\begin{align}
    \underbrace{\exp\bigg(\int_0^{\infty}\frac{dt}{2t}(e^{2\pi t}-1+2(1-e^{-2\pi t}))\bigg)}_{\text{Divergent}}\underbrace{\exp\bigg(\sum_{\substack{n\in \mathbb{Z}\\n\neq 0,\pm 1}}\int_0^{\infty}\frac{dt}{2t}(e^{-2\pi t(n^2-1)}-e^{-2\pi tn^2})\bigg)}_{\text{Convergent}}
\end{align}
The basic formula to keep in mind is the following interpretation of the integral for $h_1,h_2\neq 0$, 
\begin{align}\label{SFTinput1}
    \exp\bigg(\int_0^{\infty}\frac{dt}{2t}(e^{-2\pi h_1 t}-e^{-2\pi h_2 t})\bigg)\equiv \bigg(\frac{h_2}{h_1}\bigg)^{\frac{1}{2}}
\end{align}
This comes from the path integral interpretation of the exponentiated annulus in the string field theory \cite{Sen_2021}. It is an identity whenever $h_1,h_2> 0$ but string field theoretic path integral representation tells us that we can analytically continue and use it for the $h_1,h_2<0$ as well (as we will see in \eqref{normalizationcontri2}). This amounts to using the steepest descent contour to evaluate the Gaussian integral (see \eqref{eq: path_int_inter_exp_ann}). The convergent factor can be simply evaluated,
\begin{align}
    \exp\bigg(\sum_{\substack{n\in \mathbb{Z}\\n\neq 0,\pm 1}}\int_0^{\infty}\frac{dt}{2t}(e^{-2\pi t(n^2-1)}-e^{-2\pi tn^2})\bigg)=\prod_{n=2}^{\infty}\frac{n^2}{n^2-1}=2 \ .\label{eq: normalizationcontri_4}
\end{align}
The first factor has serious divergences because it corresponds to the cases $h_1$ or $h_2=0$ (which cannot be simply treated by analytic continuation). We have to again rely on the path integral interpretation and look at the origin of these zero modes.
These typically come from problems in gauge fixing and the zero modes in the string theory. It is similar to the case studied in \cite{alexandrov2023instantons} and the only difference is that there are three bosonic zero-modes instead of one bosonic zero-mode, 
\begin{align}\label{eq: zero_modes}
    c\partial X, \qquad ce^{2iX}, \text{ and } \qquad  ce^{-2iX}.
\end{align}
Hence, this divergent factor has to be interpreted in path integral language as follows,
\begin{align}
    &\exp\bigg(\int_0^{\infty}\frac{dt}{2t}(e^{2\pi t}+3-2-2e^{-2\pi t})\bigg)=\exp\bigg\{\int_0^{\infty}\frac{dt}{2t}\bigg(\sum_{i=1}^4e^{-2\pi h_i^bt}-\sum_{i=1}^22e^{-2\pi h_i^ft}\bigg) \bigg\}\nonumber\\
    &\qquad=\int \left(\prod_{i=1}^4\frac{db_i}{\sqrt{2\pi}}\right)df_1d\tilde{f_1}df_2d\tilde{f}_2\exp\bigg(-\sum_{i=1}^4\frac{h_i^bb_i^2}{2}+h_1^f\tilde{f}_1f_1+h_2^f\tilde{f}_2f_2 \bigg)\label{eq: path_int_inter_exp_ann}
\end{align}
where $b_i$ are bosonic variables  and $f_i, \tilde{f}_i$s are fermionic variables. Here, we have $h_{2}^b,h_{3}^b,h_{4}^b, $ $h_1^f=0$, $h_1^b=-1$ and $h_2^f=1$. We enlist the origin of these different modes, 
\begin{itemize}
    \item $h_2^b,h_3^b$ and $h_4^b$ denote the conformal weight of the three bosonic zero modes corresponding to the D-instanton collective coordinate, which in this case, corresponds to $g\in $SU(2). Hence, the integral over these modes will be just proportional to the SU(2) volume. For exact normalization, we pick the result from \cite{Sen_2021,alexandrov2023instantons} which was just $db=dx/\sqrt{2}\pi g_o$ and $dx$ goes over $(0,2\pi R)$. Specializing to the current case, we have $R=1$ and three zero modes, hence, $db_i=dx_i/\sqrt{2}\pi g_o$ for $i=2,3,4$ which uses the fact that all directions on the manifold $S_3$ of SU(2) are identical(hence, the same normalization). Integration now has to be done over the SU(2) group manifold i.e. $S^3$, 
    \begin{align}\label{normalizationcontri1}
        \int \frac{db_2}{\sqrt{2\pi}}\frac{db_3}{\sqrt{2\pi}}\frac{db_4}{\sqrt{2\pi}}=\frac{1}{(\sqrt{2}\pi g_o)^3}\int_{S^3} \frac{dx_2}{\sqrt{2\pi}}\frac{dx_3}{\sqrt{2\pi}}\frac{dx_4}{\sqrt{2\pi}} =\frac{1}{4 g_o^3\pi^{\frac{5}{2}}} \ .
    \end{align}
    Here, we have taken the radius of the $S^3$ to be $1$ i.e. same as the radius $R$ of the circle corresponding to the zero mode we had earlier(in $R\neq 1$ case). 
    \item For $h_1^b=-1$, we will just treat it like an analytic continuation of the Gaussian integral, which just amounts to taking the steepest descent contour in path integral, giving us simply a factor of $\sqrt{\frac{2\pi}{-1}}$. Hence, we get the following, 
    \begin{align}\label{normalizationcontri2}
        \int \frac{db_1}{\sqrt{2\pi}}e^{\frac{b_1^2}{2}}=\sqrt{-1} \ .
    \end{align}
    \item Next, the fermionic zero modes $f_1,\tilde{f}_1$ have their origin in the gauge fixing ghosts which, as argued in \cite{Sen_2021} and \cite{Sencern}, doesn't make sense since the corresponding gauge symmetry is the rigid U(1) rotation of open string at its ends. The way to deal with this is to use the gauge unfixed form of the integral, which is just a bosonic Gaussian (Nakanishi Lautrup) integral divided by the U(1) group volume. The normalizations were again determined in \cite{Sen_2021}, so we will just directly use the result here.
    \begin{align}\label{normalizationcontri3}
        \int df_1d \tilde{f}_1\rightarrow \frac{\int d\phi e^{\phi^2}}{\int d\theta}=\frac{\sqrt{-\pi}}{2\pi i /g_o}
    \end{align}
where $i/g_o$ is the normalization relating the $\theta$ with the rigid U(1) transformation parameter determined in \cite{Sen_2021}.
\end{itemize}
Using \eqref{eq: normalizationcontri_4}, \eqref{normalizationcontri1}, \eqref{normalizationcontri2} and \eqref{normalizationcontri3}, we get the following expression for the normalization of D-instanton,
\begin{align}\label{eq: norm_ws_result}
    \mathcal{N}=\zeta \times \frac{1}{4g_o^3\pi^{\frac{5}{2}}}\times\sqrt{-1}\times \frac{\sqrt{-\pi}}{2\pi i/g_o}\times 2= \frac{i\zeta }{4g_o^2\pi^3}=i\zeta\mu
\end{align}
where we have used $g_o^2=g_s/2\pi^2$ and $g_s=1/2\pi \mu$. And finally, it was argued in \cite{Sen_2021} that only half the steepest descent contour contributes for this particular case because of the cubic nature of the tachyon effective potential. This just means that $\zeta=1/2$ and hence, this normalization exactly matches the normalization of the leading term in \eqref{npfreeenergyatR=1} or equivalently in \eqref{one_inst_1} up to an overall sign.
\subsection{Two $(1,1)$ ZZ Instantons}
The case of different boundary conditions on both boundaries is surprisingly simple. Such diagrams appear in the case of multi-instanton contributions to the partition function. For $R\neq 1$, the simplification comes from the fact that the corresponding annulus partition function (with Dirichlet and Neumann boundary conditions on the two boundaries) is independent of the radius $R$ \eqref{DNpartfunc}. Hence, we will just discuss the $R=1$ case, and the result will be applicable directly to the generic $R$ case. 

\subsubsection{Self-Dual $R$}
For $R=1$ case, the annulus partition function for generic boundary condition $g_1\in $ SU(2) on one boundary and $g_2\in$ SU(2) on the other is given by \cite{Gaberdiel_2002, Recknagel_Schomerus_2013}, 
\begin{align}\label{mixedpartfunc}
    Z_{g_1,g_2;X}(t)=\frac{1}{\eta(it)}\sum_{n\in \mathbb{Z}}v^{(n-\frac{\alpha}{2\pi})^2}
\end{align}
where, $\alpha$ is such that $\hat{g}=\text{diag}(e^{i\alpha},e^{-i\alpha})$ is the Jordan normal form of $g_1g_2^{-1}$. Clearly, $\alpha=2l\pi$ ($l\in \mathbb{Z}$) case reduces to the result \eqref{cylpartfuncs} used in previous section. This gives the following result for the exponentiated annulus(for generic $\alpha$), 
\begin{align}
    \exp\bigg(\int \frac{dt}{2t}Z_{ \varphi,(1,1)}(t)Z_{\text{gh}}(t)Z_{g_1,g_2;X}(t)\bigg)=&\exp\bigg(\int_0^{\infty}\frac{dt}{2t}(e^{-2\pi t((n-\frac{\alpha}{2\pi})^2-1)}-e^{-2\pi t(n-\frac{\alpha}{2\pi})^2})\bigg)\nonumber\\
    =&\prod_{n\in \mathbb{Z}}\bigg(\frac{(n-\frac{\alpha}{2\pi})^2}{(n-\frac{\alpha}{2\pi})^2-1} \bigg)^{\frac{1}{2}}\nonumber\\
    =&\prod_{n\in \mathbb{Z}}\bigg(\frac{(n-\frac{\alpha}{2\pi})^2}{(n-\frac{\alpha}{2\pi}-1)(n-\frac{\alpha}{2\pi}+1)} \bigg)^{\frac{1}{2}}=1\label{eq: mixed_bc_result}
\end{align}
where we have used \eqref{SFTinput1}. Here, we have considered the annulus with $(m,n)=(m',n')=(1,1)$ while keeping $g_1,g_2$ general in the $R=1$ case in figure \ref{different_BC}. Hence, using the string field theory input \eqref{SFTinput1}, we see that the contribution from the exponentiated annulus with mixed boundary condition is trivial.

An important consequence of this is that the instanton normalization corresponding to the two $(1,1)$ ZZ instanton cases will be simply a product of the normalization of the individual $(1,1)$ ZZ instanton. The normalization for the two-instanton will look like, 
\begin{align}
    \frac{1}{2}\int \frac{d^3x_1}{(2\pi)^{\frac{3}{2}}}\int \frac{d^3x_2}{(2\pi)^{\frac{3}{2}}}\exp\left(\int \frac{dt}{2t}Z_1(t)\right)\exp\left(\int \frac{dt}{2t}Z_2(t)\right)\underbrace{\exp\left(\int \frac{dt}{t}Z_{12}(t)\right)}_{=1}=-\frac{\mu^2}{8}
\end{align}
where $Z_1(t)$ and $Z_2(t)$ are the annulus partition functions corresponding to the same boundary conditions on both the boundaries whereas $Z_{12}(t)$ corresponds to the annulus with mixed boundary conditions which we just discussed. Since the last factor is $1$, the normalization becomes the product of the individual $(1,1)$ ZZ instanton normalization \eqref{eq: norm_ws_result}. Such a term is present there in the two instanton contribution \eqref{eq: 2_inst_R=1_term_trans} to the trans-series \eqref{eq: trans_series_R=1_part_func}. The overall $1/2$ factor outside accounts for the exchange symmetry of the two instantons. We will see that such factorization is true for more general Liouville boundary conditions as well in section \ref{general_liouvill_bc}. Hence, this feature will be present at all orders in the trans-series.
\subsubsection{Generic $R$ case with Mixed Boson Boundary Conditions}
For the generic $R$ case, where there is a possibility for the mixed boson boundary condition, i.e., Neumann on one boundary and Dirichlet on the other (last annulus in $R\neq 1$ case in figure \ref{different_BC}), just corresponds to the special case of $\alpha=\pi/2$ in the above calculation (shown in detail in appendix \ref{app_freeboson_review}, equation \eqref{eq: app_equiv_R_1}). Since we have shown the trivial nature of the exponentiated annulus for the general $\alpha$, it holds for $\alpha=\pi/2$ as well. This clearly explains the absence of terms $\sim e^{-2\pi \mu(1+R)}$ in the expression for non-perturbative free energy \eqref{npfreenergyatR} at general radius $R$. This was expected from looking explicitly at the coefficient of the term $e^{-2\pi \mu(1+R)}$ in two instanton contribution in \eqref{2_inst_mixed_contri} as it is simply a product of two exponentiated annulus on two different $(1,1)$ ZZ instantons corresponding to Dirichlet \eqref{DDBC_result} and Neumann \eqref{NBC_result} boundary conditions in the boson. For clarity,
\begin{align}\label{eq: mixed_instanton(1,1)}
    \int_0^{2\pi R} dx_0\int_0^{\frac{2\pi}{R}}d\tilde{x}_0\exp\left(\int \frac{dt}{2t}Z_1(t)\right)\exp\left(\int \frac{dt}{2t}Z_2(t)\right)\underbrace{\exp\left(\int \frac{dt}{t}Z_3(t)\right)}_{=1}
\end{align}
where, $Z_1(t)$ ($Z_2(t)$) has the free boson partition function \eqref{Dirichlepartfunc} (\eqref{Neumannpartfunc}) with $x_0=x_0'$ ($\tilde{x}_0=\tilde{x}_0'$). $Z_3(t)$ has the free boson partition function for the mixed boundary condition \eqref{DNpartfunc}. Since exponentiated annulus corresponding to $Z_3(t)$ is $1$, the integral just becomes a product of two integrals as anticipated in section \ref{subsec: MM_predictions}.

\subsubsection{Generic $R$ case with Same Boson Boundary Conditions}\label{subsubsec: 2(1,1)R>1}
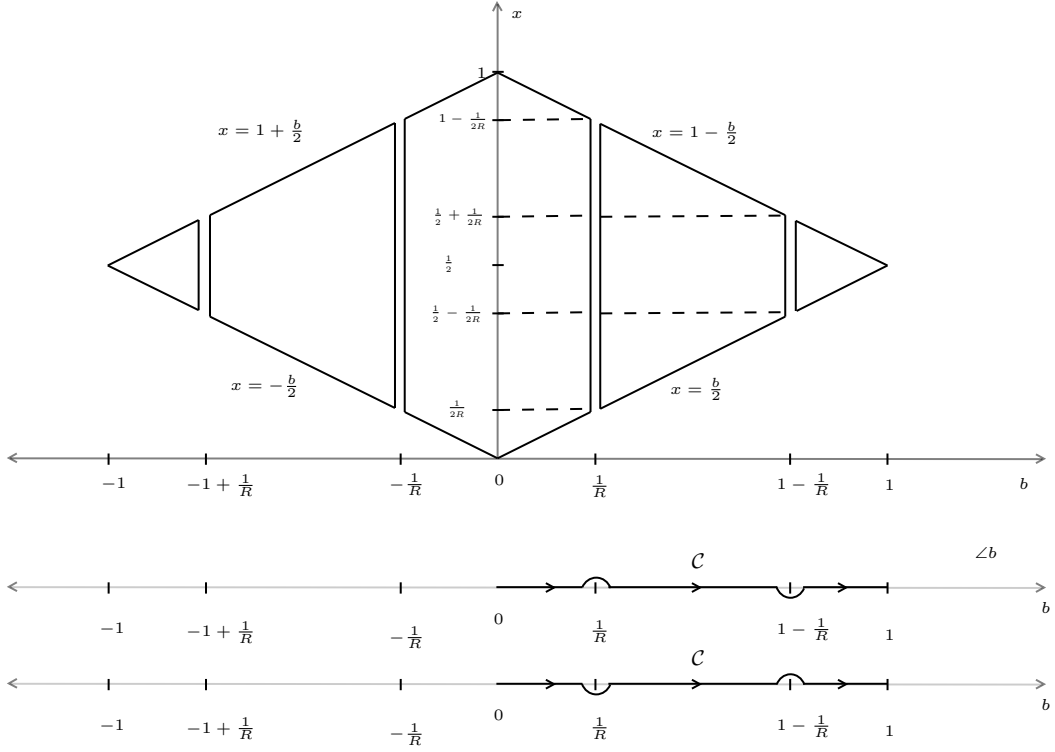
\begin{figure}
    \centering

\tikzset{every picture/.style={line width=0.75pt}} %set default line width to 0.75pt        

\begin{tikzpicture}[x=0.75pt,y=0.75pt,yscale=-0.81,xscale=0.81]
%uncomment if require: \path (0,481); %set diagram left start at 0, and has height of 481

%Straight Lines [id:da0387932157546802] 
\draw [color={rgb, 255:red, 0; green, 0; blue, 0 }  ,draw opacity=0.5 ]   (10,290) -- (646,290.5) ;
%Straight Lines [id:da24365482339975508] 
\draw    (250.5,287) -- (250.5,294) ;
%Straight Lines [id:da5965252738597622] 
\draw    (490.5,287) -- (490.5,294) ;
%Straight Lines [id:da6918383947518807] 
\draw    (370.5,287) -- (370.5,294) ;
%Straight Lines [id:da6267327712124071] 
\draw    (130.5,287) -- (130.5,294) ;
%Straight Lines [id:da9587658513855246] 
\draw    (70,170.25) -- (126,198) ;
%Straight Lines [id:da09720162203709726] 
\draw    (70,170.25) -- (126,142) ;
%Straight Lines [id:da6550833332010273] 
\draw    (126,142) -- (126,198) ;
%Straight Lines [id:da9725443458470053] 
\draw    (133,139) -- (133,202) ;
%Straight Lines [id:da1681420969417271] 
\draw    (133,139) -- (247,81.75) ;
%Straight Lines [id:da8921886020143659] 
\draw    (133,202) -- (247,258.75) ;
%Straight Lines [id:da27671140885613155] 
\draw    (247,81.75) -- (247,258.75) ;
%Straight Lines [id:da6645023863174135] 
\draw    (253,79.25) -- (253,261.25) ;
%Straight Lines [id:da9566513412523212] 
\draw    (253,261.25) -- (310.25,290) ;
%Straight Lines [id:da6198825539531674] 
\draw    (253,79.25) -- (310.25,50.5) ;
%Straight Lines [id:da5830539460670625] 
\draw    (310.25,50.5) -- (367.5,79.25) ;
%Straight Lines [id:da06628204650089886] 
\draw    (310.25,290) -- (367.5,261.25) ;
%Straight Lines [id:da42410396465590416] 
\draw    (367.5,79.25) -- (367.5,261.25) ;
%Straight Lines [id:da7369544211778296] 
\draw    (373.5,82.25) -- (373.5,259.25) ;
%Straight Lines [id:da20454763080905747] 
\draw    (373.5,259.25) -- (487.5,202) ;
%Straight Lines [id:da0880115770043588] 
\draw    (373.5,82.25) -- (487.5,139) ;
%Straight Lines [id:da9681520261838741] 
\draw    (487.5,139) -- (487.5,202) ;
%Straight Lines [id:da07852339244943995] 
\draw    (494,142.5) -- (494,198.5) ;
%Straight Lines [id:da9694832693542961] 
\draw    (494,142.5) -- (550.5,170.25) ;
%Straight Lines [id:da00290285636650478] 
\draw    (494.5,198.5) -- (550.5,170.25) ;
%Straight Lines [id:da09147023404665644] 
\draw [color={rgb, 255:red, 0; green, 0; blue, 0 }  ,draw opacity=0.5 ]   (310.25,8) -- (310.25,290) ;
%Straight Lines [id:da6222733895300434] 
\draw    (307,170) -- (314,170) ;
%Straight Lines [id:da12312681583467944] 
\draw    (307,260) -- (314,260) ;
%Straight Lines [id:da35878835388893937] 
\draw    (307,80) -- (314,80) ;
\draw  [color={rgb, 255:red, 0; green, 0; blue, 0 }  ,draw opacity=0.5 ] (642,287.5) -- (647,290.25) -- (642,293) ;
\draw  [color={rgb, 255:red, 0; green, 0; blue, 0 }  ,draw opacity=0.5 ] (14.04,292.46) -- (9,289.79) -- (13.96,286.96) ;
\draw  [color={rgb, 255:red, 0; green, 0; blue, 0 }  ,draw opacity=0.5 ] (307.35,12.35) -- (309.9,7.25) -- (312.84,12.14) ;
%Straight Lines [id:da24655318247652414] 
\draw    (70.5,287) -- (70.5,294) ;
%Straight Lines [id:da277612638059205] 
\draw    (550.5,287) -- (550.5,294) ;
%Straight Lines [id:da20520631062366768] 
\draw    (307,200) -- (314,200) ;
%Straight Lines [id:da5945871973801624] 
\draw    (307,140) -- (314,140) ;
%Straight Lines [id:da9411960077938313] 
\draw    (307,50) -- (314,50) ;
%Straight Lines [id:da3176547557032112] 
\draw  [dash pattern={on 4.5pt off 4.5pt}]  (311,80) -- (367.5,79.25) ;
%Straight Lines [id:da12626391739728882] 
\draw  [dash pattern={on 4.5pt off 4.5pt}]  (310.5,140) -- (367,139.25) ;
%Straight Lines [id:da8217501208947107] 
\draw  [dash pattern={on 4.5pt off 4.5pt}]  (310.5,200) -- (367,199.25) ;
%Straight Lines [id:da0816902997416713] 
\draw  [dash pattern={on 4.5pt off 4.5pt}]  (310.5,260) -- (367,259.25) ;
%Straight Lines [id:da8909801405983839] 
\draw  [dash pattern={on 4.5pt off 4.5pt}]  (373.5,200) -- (488,199.25) ;
%Straight Lines [id:da24539253775687309] 
\draw  [dash pattern={on 4.5pt off 4.5pt}]  (373.5,140) -- (488,139.25) ;
%Straight Lines [id:da37391537683614984] 
\draw [color={rgb, 255:red, 0; green, 0; blue, 0 }  ,draw opacity=0.2 ]   (10.04,370) -- (646.04,370.5) ;
%Straight Lines [id:da7242942862240618] 
\draw    (250.54,367) -- (250.54,374) ;
%Straight Lines [id:da2122702377466108] 
\draw    (490.54,367) -- (490.54,374) ;
%Straight Lines [id:da5591142178133559] 
\draw    (370.54,367) -- (370.54,374) ;
%Straight Lines [id:da01941069846553667] 
\draw    (130.54,367) -- (130.54,374) ;
\draw  [color={rgb, 255:red, 0; green, 0; blue, 0 }  ,draw opacity=0.5 ] (14.08,372.46) -- (9.04,369.79) -- (14,366.96) ;
%Straight Lines [id:da17930942448303955] 
\draw    (70.54,367) -- (70.54,374) ;
%Straight Lines [id:da9323498696685459] 
\draw    (550.54,367) -- (550.54,374) ;
\draw  [color={rgb, 255:red, 0; green, 0; blue, 0 }  ,draw opacity=0.5 ] (642,367.5) -- (647,370.25) -- (642,373) ;
%Straight Lines [id:da5501350834027783] 
\draw    (309.5,370) -- (362.5,370) ;
%Straight Lines [id:da10381072325712548] 
\draw    (378.5,370) -- (482.5,370) ;
%Straight Lines [id:da4430971029950477] 
\draw    (498.5,370) -- (550.5,370) ;
%Shape: Arc [id:dp23738368949325062] 
\draw  [draw opacity=0] (362.45,370.08) .. controls (363.94,366.55) and (367.23,364.1) .. (371.05,364.1) .. controls (375.05,364.1) and (378.47,366.79) .. (379.85,370.61) -- (371.05,374.35) -- cycle ; \draw   (362.45,370.08) .. controls (363.94,366.55) and (367.23,364.1) .. (371.05,364.1) .. controls (375.05,364.1) and (378.47,366.79) .. (379.85,370.61) ;  
%Shape: Arc [id:dp13761525076359016] 
\draw  [draw opacity=0] (499.08,370.76) .. controls (497.54,374.3) and (494.25,376.7) .. (490.5,376.59) .. controls (486.57,376.48) and (483.3,373.64) .. (482.09,369.75) -- (490.8,366.35) -- cycle ; \draw   (499.08,370.76) .. controls (497.54,374.3) and (494.25,376.7) .. (490.5,376.59) .. controls (486.57,376.48) and (483.3,373.64) .. (482.09,369.75) ;  
\draw  [color={rgb, 255:red, 0; green, 0; blue, 0 }  ,draw opacity=1 ] (340,367.5) -- (345,370.25) -- (340,373) ;
\draw  [color={rgb, 255:red, 0; green, 0; blue, 0 }  ,draw opacity=1 ] (430,367.5) -- (435,370.25) -- (430,373) ;
\draw  [color={rgb, 255:red, 0; green, 0; blue, 0 }  ,draw opacity=1 ] (520,367.5) -- (525,370.25) -- (520,373) ;
%Straight Lines [id:da5978605054751511] 
\draw [color={rgb, 255:red, 0; green, 0; blue, 0 }  ,draw opacity=0.2 ]   (10.04,430) -- (646.04,430.5) ;
%Straight Lines [id:da002496631055468601] 
\draw    (250.54,427) -- (250.54,434) ;
%Straight Lines [id:da1227718027135547] 
\draw    (490.54,427) -- (490.54,434) ;
%Straight Lines [id:da05133272098155062] 
\draw    (370.54,427) -- (370.54,434) ;
%Straight Lines [id:da29546872988068684] 
\draw    (130.54,427) -- (130.54,434) ;
\draw  [color={rgb, 255:red, 0; green, 0; blue, 0 }  ,draw opacity=0.5 ] (14.08,432.46) -- (9.04,429.79) -- (14,426.96) ;
%Straight Lines [id:da6704914226372631] 
\draw    (70.54,427) -- (70.54,434) ;
%Straight Lines [id:da4586049956130489] 
\draw    (550.54,427) -- (550.54,434) ;
\draw  [color={rgb, 255:red, 0; green, 0; blue, 0 }  ,draw opacity=0.5 ] (642,427.5) -- (647,430.25) -- (642,433) ;
%Straight Lines [id:da3687504161044297] 
\draw    (309.5,430) -- (362.5,430) ;
%Straight Lines [id:da025906653612542074] 
\draw    (378.5,430) -- (482.5,430) ;
%Straight Lines [id:da17561608003368012] 
\draw    (498.5,430) -- (550.5,430) ;
%Shape: Arc [id:dp27138917854820876] 
\draw  [draw opacity=0] (379.64,430.62) .. controls (378.14,434.15) and (374.85,436.6) .. (371.04,436.6) .. controls (367.03,436.59) and (363.61,433.89) .. (362.24,430.08) -- (371.05,426.35) -- cycle ; \draw   (379.64,430.62) .. controls (378.14,434.15) and (374.85,436.6) .. (371.04,436.6) .. controls (367.03,436.59) and (363.61,433.89) .. (362.24,430.08) ;  
%Shape: Arc [id:dp4182784569431339] 
\draw  [draw opacity=0] (482.28,430.4) .. controls (483.62,426.78) and (486.77,424.2) .. (490.52,424.1) .. controls (494.45,423.99) and (497.87,426.64) .. (499.3,430.46) -- (490.8,434.35) -- cycle ; \draw   (482.28,430.4) .. controls (483.62,426.78) and (486.77,424.2) .. (490.52,424.1) .. controls (494.45,423.99) and (497.87,426.64) .. (499.3,430.46) ;  
\draw  [color={rgb, 255:red, 0; green, 0; blue, 0 }  ,draw opacity=1 ] (340,427.5) -- (345,430.25) -- (340,433) ;
\draw  [color={rgb, 255:red, 0; green, 0; blue, 0 }  ,draw opacity=1 ] (430,427.5) -- (435,430.25) -- (430,433) ;
\draw  [color={rgb, 255:red, 0; green, 0; blue, 0 }  ,draw opacity=1 ] (520,427.5) -- (525,430.25) -- (520,433) ;

% Text Node
\draw (317,10.4) node [anchor=north west][inner sep=0.75pt]  [font=\tiny]  {$x$};
% Text Node
\draw (630,300.4) node [anchor=north west][inner sep=0.75pt]  [font=\tiny]  {$b$};
% Text Node
\draw (421,73.4) node [anchor=north west][inner sep=0.75pt]  [font=\scriptsize]  {$x=1-\frac{b}{2}$};
% Text Node
\draw (432.5,234.03) node [anchor=north west][inner sep=0.75pt]  [font=\scriptsize]  {$x=\frac{b}{2}$};
% Text Node
\draw (144,236.4) node [anchor=north west][inner sep=0.75pt]  [font=\scriptsize]  {$x=-\frac{b}{2}$};
% Text Node
\draw (136,77.9) node [anchor=north west][inner sep=0.75pt]  [font=\scriptsize]  {$x=1+\frac{b}{2}$};
% Text Node
\draw (480.5,297.9) node [anchor=north west][inner sep=0.75pt]  [font=\fontsize{0.59em}{0.71em}\selectfont]  {$1-\frac{1}{R}$};
% Text Node
\draw (365,298.9) node [anchor=north west][inner sep=0.75pt]  [font=\fontsize{0.59em}{0.71em}\selectfont]  {$\frac{1}{R}$};
% Text Node
\draw (241.5,297.9) node [anchor=north west][inner sep=0.75pt]  [font=\fontsize{0.59em}{0.71em}\selectfont]  {$-\frac{1}{R}$};
% Text Node
\draw (116.5,298.4) node [anchor=north west][inner sep=0.75pt]  [font=\fontsize{0.59em}{0.71em}\selectfont]  {$-1+\frac{1}{R}$};
% Text Node
\draw (64,300.4) node [anchor=north west][inner sep=0.75pt]  [font=\fontsize{0.59em}{0.71em}\selectfont]  {$-1$};
% Text Node
\draw (547,301.4) node [anchor=north west][inner sep=0.75pt]  [font=\fontsize{0.59em}{0.71em}\selectfont]  {$1$};
% Text Node
\draw (306.5,298.4) node [anchor=north west][inner sep=0.75pt]  [font=\fontsize{0.59em}{0.71em}\selectfont]  {$0$};
% Text Node
\draw (282,252.4) node [anchor=north west][inner sep=0.75pt]  [font=\tiny]  {$\frac{1}{2R}$};
% Text Node
\draw (264.5,192.4) node [anchor=north west][inner sep=0.75pt]  [font=\tiny]  {$\frac{1}{2} -\frac{1}{2R}$};
% Text Node
\draw (286.5,162.4) node [anchor=north west][inner sep=0.75pt]  [font=\tiny]  {$\frac{1}{2}$};
% Text Node
\draw (265.5,131.4) node [anchor=north west][inner sep=0.75pt]  [font=\tiny]  {$\frac{1}{2} +\frac{1}{2R}$};
% Text Node
\draw (267,73.4) node [anchor=north west][inner sep=0.75pt]  [font=\tiny]  {$1-\frac{1}{2R}$};
% Text Node
\draw (295.5,45.9) node [anchor=north west][inner sep=0.75pt]  [font=\tiny]  {$1$};
% Text Node
\draw (643.54,377.4) node [anchor=north west][inner sep=0.75pt]  [font=\tiny]  {$b$};
% Text Node
\draw (480.54,387.9) node [anchor=north west][inner sep=0.75pt]  [font=\tiny]  {$1-\frac{1}{R}$};
% Text Node
\draw (365.04,388.9) node [anchor=north west][inner sep=0.75pt]  [font=\fontsize{0.59em}{0.71em}\selectfont]  {$\frac{1}{R}$};
% Text Node
\draw (240.54,385.9) node [anchor=north west][inner sep=0.75pt]  [font=\fontsize{0.59em}{0.71em}\selectfont]  {$-\frac{1}{R}$};
% Text Node
\draw (114.54,385.4) node [anchor=north west][inner sep=0.75pt]  [font=\fontsize{0.59em}{0.71em}\selectfont]  {$-1+\frac{1}{R}$};
% Text Node
\draw (63.04,390.4) node [anchor=north west][inner sep=0.75pt]  [font=\fontsize{0.59em}{0.71em}\selectfont]  {$-1$};
% Text Node
\draw (547.04,394.4) node [anchor=north west][inner sep=0.75pt]  [font=\fontsize{0.59em}{0.71em}\selectfont]  {$1$};
% Text Node
\draw (306.04,384.4) node [anchor=north west][inner sep=0.75pt]  [font=\fontsize{0.59em}{0.71em}\selectfont]  {$0$};
% Text Node
\draw (428,347.4) node [anchor=north west][inner sep=0.75pt]  [font=\scriptsize]  {$\mathcal{C}$};
% Text Node
\draw (643.54,437.4) node [anchor=north west][inner sep=0.75pt]  [font=\tiny]  {$b$};
% Text Node
\draw (480.54,447.9) node [anchor=north west][inner sep=0.75pt]  [font=\fontsize{0.59em}{0.71em}\selectfont]  {$1-\frac{1}{R}$};
% Text Node
\draw (365.04,448.9) node [anchor=north west][inner sep=0.75pt]  [font=\fontsize{0.59em}{0.71em}\selectfont]  {$\frac{1}{R}$};
% Text Node
\draw (240.54,446.9) node [anchor=north west][inner sep=0.75pt]  [font=\fontsize{0.59em}{0.71em}\selectfont]  {$-\frac{1}{R}$};
% Text Node
\draw (116.54,449.4) node [anchor=north west][inner sep=0.75pt]  [font=\fontsize{0.59em}{0.71em}\selectfont]  {$-1+\frac{1}{R}$};
% Text Node
\draw (63.04,450.4) node [anchor=north west][inner sep=0.75pt]  [font=\fontsize{0.59em}{0.71em}\selectfont]  {$-1$};
% Text Node
\draw (547.04,454.4) node [anchor=north west][inner sep=0.75pt]  [font=\fontsize{0.59em}{0.71em}\selectfont]  {$1$};
% Text Node
\draw (306.04,444.4) node [anchor=north west][inner sep=0.75pt]  [font=\fontsize{0.59em}{0.71em}\selectfont]  {$0$};
% Text Node
\draw (428,407.4) node [anchor=north west][inner sep=0.75pt]  [font=\scriptsize]  {$\mathcal{C}$};

\end{tikzpicture}

    \caption{Integration Domain $D$ of the integral in the $x-b$ plane and the contours in the complex $b$ plane used to calculate the principal value integral.}
    \label{fig:two_inst_int_domain}
\end{figure}

We now consider the two $(1,1)$ ZZ instanton configuration and try to calculate the normalization. On the free boson, we have Dirichlet boundary condition such that the collective coordinates values are $x_1$ and $x_2$. We will have to consider the three types of annuli, one with both boundaries on $x_1$, one with both boundaries on $x_2$, and finally, the one stretched between $x_1$ and $x_2$. The corresponding annulus partition functions will be respectively,
\begin{align}\label{eq: two_inst_ann_part_func}
    Z_1(v)=&(v^{-1}-1)\sum_{j\in \mathbb{Z}}v^{j^2R^2}, \quad Z_2(v)=(v^{-1}-1)\sum_{j\in \mathbb{Z}}v^{j^2R^2}\text{ and }\nonumber\\
    Z_3(v)=&(v^{-1}-1)\sum_{j\in \mathbb{Z}}v^{\left(jR+\frac{x_1-x_2}{2\pi}\right)^2}\ .
\end{align}
$Z_3(v)$ corresponds to setting $(m,n)=(m',n')=(1,1)$ in the first annulus in $R\neq 1$ case of figure \ref{different_BC}. $Z_{1}(v)$ and $Z_{2}(v)$ correspond to further setting $x_1=x_2$.
Normalization will be simply given by following in this case, 
\begin{align}
    \frac{\tilde{\zeta}}{(\sqrt{2}\pi g_o)^2}\int_0^{2\pi R} \frac{dx_1}{\sqrt{2\pi}}\int_{0}^{2\pi R}\frac{dx_2}{\sqrt{2\pi}}\exp\left(\int\frac{dt}{2t}Z_1(t)+\int\frac{dt}{2t}Z_2(t)+\int\frac{dt}{t}Z_3(t)\right)
\end{align}
where $\tilde{\zeta}$ is a constant factor to be fixed later. The overall factors of $(\sqrt{2}\pi g_o)^{-1}$ are the same as the one we saw in \eqref{normalizationcontri1}. The first two exponents will give the same factors as calculated in \cite{alexandrov2023instantons} using the replacement rule \eqref{SFTinput1}. The factor is constituted of the convergent part and divergent part(after regularization), which are respectively,
\begin{align}
    \prod_{j\neq 0}\left(\frac{j^2R^2}{j^2R^2-1}\right)^{\frac{1}{2}}=\prod_{j=1}^{\infty}\frac{j^2R^2}{j^2R^2-1}=\frac{\frac{\pi}{R}}{\sin{\frac{\pi}{R}}} \quad \text{and}\quad \frac{\sqrt{-\pi}}{\frac{2\pi i}{g_o}}.
\end{align}
The third exponent leads to the following using the \eqref{SFTinput1}, 
\begin{align}\label{eq: cross_cyl_result_1}
    \prod_{j\in \mathbb{Z}}\frac{(jR+\frac{x_1-x_2}{2\pi})^2}{(jR+\frac{x_1-x_2}{2\pi})^2-1} \ .
\end{align}
Hence, the integral of interest is as follows, 
\begin{align}
    \frac{\tilde{\zeta}}{16\pi^2R^2\sin^2{\frac{\pi}{R}}}\int_0^{2\pi R}dx_1\int_0^{2\pi R}dx_2\prod_{j\in \mathbb{Z}}\frac{(jR+\frac{x_1-x_2}{2\pi})^2}{(jR+\frac{x_1-x_2}{2\pi})^2-1}\ .
\end{align}
We can make a change of variable $x_{1,2}\rightarrow \frac{x_{1,2}}{2\pi R}$,
\begin{align}
    \frac{\tilde{\zeta}}{4\sin^2{\frac{\pi}{R}}}\int_{0}^{1}\int_{0}^{1}dx_1dx_2\prod_{j\in \mathbb{Z}}\left(1-\frac{1}{R^2(j+b)^2}\right)^{-1}\label{eq: initial_int_R}
\end{align}
where $b:=x_1-x_2$ in terms of new variables.
For $R=1$, the integrand simplifies and becomes $1$ which we already saw in \eqref{eq: mixed_bc_result}. The integrand can be simplified as follows by first separating the $j=0$ term and then evaluating products of $j$th and $-j$th term such that only the product over positive integers remains, 
\begin{align}\label{eq: cross_cyl_intermediate_step_1}
    \frac{\prod_{j\in \mathbb{Z}}(j+b)^2}{\prod_{j\in \mathbb{Z}}\left(j+b-\frac{1}{R}\right)\prod_{j\in \mathbb{Z}}\left(j+b+\frac{1}{R}\right)}
    =&\frac{b^2\prod_{j=1}^{\infty}\left(1-\frac{b^2}{j^2}\right)\prod_{j= 1}^{\infty}\left(1-\frac{b^2}{j^2}\right)}{\left(b^2-\frac{1}{R^2}\right)\prod_{j=1}^{\infty}\left(1-\frac{(b-\frac{1}{R})^2}{j^2}\right)\prod_{j=1}^{\infty}\left(1-\frac{(b+\frac{1}{R})^2}{j^2}\right)} \ .
\end{align}
Now, using the fact that $\prod_{j=1}^{\infty}(1-\frac{x^2}{j^2})=\frac{\sin{\pi x}}{\pi x}$, we get, 
\begin{align}\label{eq: cross_cyl_intermediate_step_2}
    \prod_{j\in \mathbb{Z}}\left(1-\frac{1}{R^2(j+b)^2}\right)^{-1}=\frac{\sin^2{\pi b}}{\sin{\pi(b-\frac{1}{R})}\sin{\pi(b+\frac{1}{R})}}\ .
\end{align}
Hence, the integral takes a simple form,
\begin{align}\label{eq: fin_form_norm}
     &\frac{\tilde{\zeta}}{4\sin^2{\frac{\pi}{R}}}\int_{0}^{1}\int_{0}^{1}dx_1dx_2f(b)\quad\text{where }f(b):=\frac{\sin^2{\pi b}}{\sin{\pi(b-\frac{1}{R})}\sin{\pi(b+\frac{1}{R})}} \ .
\end{align}
We can make a change of variable $x_1=x+ \frac{b}{2}$, $x_2=x- \frac{b}{2}$. The unit square in $(x_1,x_2)$ plane will be mapped to the rhombus enclosed by $b=2x, b=-2x, b=2-2x, b=2x-2$ in the $(b,x)$ plane (see figure \ref{fig:two_inst_int_domain}).
This will be the integration domain $D$.
We get, 
\begin{align}
     \frac{\tilde{\zeta}}{4\sin^2{\frac{\pi}{R}}}\int_{D}dx dbf(b)
\end{align}

Since $f(b)$ have singularities in the domain(at $b=\pm R^{-1},\pm(1-R^{-1})$), we will be employing the unitary prescription devised in \cite{Sen:2020ruy} to deal with these singularities. Unitary prescription is simply that we consider the principal value of the integral.
\begin{align}
    P\int_D db dx f(b)=&2\left\{\int_0^{1}db\frac{1}{2}\left(\int_{\frac{b+i\epsilon}{2}}^{1-\frac{b+i\epsilon}{2}}dxf(b+i\epsilon)+\int_{\frac{b-i\epsilon}{2}}^{1-\frac{b-i\epsilon}{2}}dxf(b-i\epsilon)\right)\right\}\nonumber\\
    =&\int_0^{1}db\left((1-b-i\epsilon)f(b+i\epsilon)+(1-b+i\epsilon)f(b-i\epsilon)\right)\nonumber\\
    =&2P\int_0^{1}db(1-b)f(b)\label{eq: PV_1D_integral}
\end{align}
where we have used the $b\rightarrow -b$ symmetry of the integrand and restricted our attention to the right half of the domain $D$ (see figure \ref{fig:two_inst_int_domain}) by introducing an overall factor of $2$ in the first step. In the second step, we have carried out the $x$ integral. In the last line \eqref{eq: PV_1D_integral}, the principal value gets rid of the poles at $b=R^{-1}$ and $b=1-R^{-1}$. 

We can carry out this integral by considering the integral over the contour $\mathcal{C}$ (see figure \ref{fig:two_inst_int_domain}). 
\begin{align}
\int_{\mathcal{C}}db(1-b)f(b)=\frac{1}{2}\pm i\frac{\tan{\frac{\pi}{R}}}{2} \ .
\end{align}
Here, $\pm$ corresponds to the two choices of the contour highlighted in figure \ref{fig:two_inst_int_domain}. The real part is simply the principal value integral and the imaginary part is the contribution coming from the residues at the poles $b=R^{-1}, 1-R^{-1}$. Hence,
\begin{align}
    P\int_D db dx f(b)=2P\int_0^1db(1-b)f(b)=2\times \frac{1}{2}=1 \ .
\end{align}
Hence, we see that \textit{the principal value of the integral is independent of $R$ and simply equals $1$.} All the $R$ dependence goes in the half-residues at the poles (through the small semi-circular contours in figure \ref{fig:two_inst_int_domain}) which are, by definition, removed in the principal value prescription and are purely imaginary. 

 \textit{This simply implies that the contribution from the exponentiated cross-annulus (annulus stretched between two different points) is trivial if we consider the unitary prescription of \cite{Sen:2020ruy}.} To put this in perspective, we are saying that
 \begin{align}
     \frac{1}{4\pi^2R^2}\int_0^{2\pi R} dx_1\int_0^{2\pi R} dx_2 \exp\left(\int_0^{\infty}\frac{dt}{t}Z_3(t)\right)=1
 \end{align}
 with the unitary prescription.
 This is similar to what we saw in the $R=1$ case. The only difference is that the trivial nature of the cross-annulus is reflected at the level of the integrand itself in the $R=1$ case, whereas here, we observe it after we evaluate the integral with a particular prescription.
Hence, the total normalization \eqref{eq: fin_form_norm} takes the form, 
\begin{align}
    \frac{\tilde{\zeta}}{4\sin^2{\frac{\pi}{R}}} \ .
\end{align}
Now, we argue that $\tilde{\zeta}$ will be constituted of the two factors of $\frac{1}{2}$, one for each instanton because only half the steepest descent contour contributes due to the cubic nature of the effective tachyon potential \cite{Sen_2021, Sen:2020ruy}. Another $1/2$ factor will enter due to the exchange symmetry of two $(1,1)$ ZZ instantons (the same $1/2!$ factor that comes from the exponentiation of free energy i.e. the exchange of disconnected pieces in the diagram).  Hence, the normalization becomes
\begin{align}\label{eq: 2_inst_(1,1)_result}
    \frac{1}{32\sin^2{\frac{\pi}{R}}},
\end{align}
which matches with the term in $\frac{\mathcal{Z}_R^{(2)}}{\mathcal{Z}_R^{(0)}}$ \eqref{Rneq1_trans_series}.

An important thing to note here is that if we had chosen the Lorentzian prescription of \cite{balthazar2019multiinstanton} instead of the unitary prescription of the \cite{Sen:2020ruy}, we would have had multiple choices of the contours around the residues. For the particular choice highlighted in figure \ref{fig:two_inst_int_domain}, we would get the extra imaginary contribution to the exponentiated cross annulus leading to the normalization getting corrected, 
\begin{align}\label{eq: two_inst_lorentz}
    \frac{1}{32 \sin^2{\frac{\pi}{R}}}\left(1\pm i\tan{\frac{\pi}{R}}\right)=\frac{1}{32\sin^2{\frac{\pi}{R}}}\pm \frac{i}{16\sin{\frac{2\pi}{R}}} \ .
\end{align}
The second term here is proportional to another term that appears in the same order($\sim e^{-4\pi \mu}$) in the matrix model result \eqref{Rneq1_trans_series}. We will see in the section \ref{subsec: general(1,n)bc} that a single $(1,2)$ ZZ instanton normalization can also produce such a term. We will also see a generalization of the discussion here to more general Liouville boundary conditions in section \ref{subsubsec: two_inst(1,n)}.

\subsection{General Liouville Boundary Conditions at Self-Dual $R$}\label{general_liouvill_bc}
In this section, we will generalize the previous results and try to see the general boundary conditions in the Liouville direction as well. The annulus partition function for the Liouville with ZZ boundary condition of type $(m,n)$ on one boundary and $(m',n')$ on the other boundary is given by \cite{balthazar2019multiinstanton, ZZoriginal},
\begin{align}\label{eq: gen_liouville_part_func}
    Z_{\varphi;(m,n),(m'n')}(t)=\frac{1}{\eta(it)}\sum_{p=0}^{\text{min}(m,m')-1}\quad\sum_{l=0}^{\text{min}(n,n')-1}\bigg(&v^{-\frac{(m+m'+n+n'-2p-2l-2)^2}{4}}\nonumber\\
    &\hspace{30 pt}-v^{-\frac{(m+m'-n-n'-2p+2l)^2}{4}}\bigg) \ .
\end{align}
Hence, the complete general annulus partition function takes the form after including the ghost and free boson contribution ($R=1$ annulus in figure \ref{different_BC}),
\begin{align}
    Z(t)=\sum_{p=0}^{\text{min}(m,m')-1} \quad \sum_{l=0}^{\text{min}(n,n')-1}\sum_{j\in \mathbb{Z}}\bigg(&v^{\left(j-\frac{\alpha}{2\pi}\right)^2-\frac{(m+m'+n+n'-2p-2l-2)^2}{4}}\nonumber\\
    &\hspace{40 pt}-v^{\left(j-\frac{\alpha}{2\pi}\right)^2-\frac{(m+m'-n-n'-2p+2l)^2}{4}}\bigg) \ .
\end{align}
We can now redefine the summing variable as, $p\rightarrow m+m'-2p$ and $l\rightarrow n+n'-2l$, which gives us the following simple formula, 
\begin{align}\label{eq: (m,n)_liov_part_func}
    Z(t)=\sum_{p=|m-m'|+2,2}^{m+m'}\quad\sum_{l=|n-n'|+2,2}^{n+n'} \quad\sum_{j\in \mathbb{Z}}\left(v^{\left(j-\frac{\alpha}{2\pi}\right)^2-\frac{(p+l-2)^2}{4}}-v^{\left(j-\frac{\alpha}{2\pi}\right)^2-\frac{(p-l)^2}{4}}\right),
\end{align}
where $,2$ in the lower limit of the sum over $p$ and $l$ denotes that these indices run in steps of two. To calculate the annulus contribution to the amplitude, we want to calculate the following integral,
\begin{align}
    \int_{0}^{\infty}\frac{dt}{2t}Z(t)=&\int_0^{\infty}\frac{dt}{2t}\sum_{p=|m-m'|+2,2}^{m+m'}\quad\sum_{l=|n-n'|+2,2}^{n+n'}\quad\sum_{j\in \mathbb{Z}}\big(v^{\left(j-\frac{\alpha}{2\pi}\right)^2-\frac{(p+l-2)^2}{4}}-v^{\left(j-\frac{\alpha}{2\pi}\right)^2-\frac{(p-l)^2}{4}}\big)\nonumber\\
    =&\frac{1}{2}\sum_{p=|m-m'|+2,2}^{m+m'}\quad\sum_{l=|n-n'|+2,2}^{n+n'}\log\left(\prod_{j\in \mathbb{Z}}\frac{\left(2j-\frac{\alpha}{\pi}+p-l\right)\left(2j-\frac{\alpha}{\pi}-p+l\right)}{\left(2j-\frac{\alpha}{\pi}+p+l\right)\left(2j-\frac{\alpha}{\pi}-p-l\right)}\right), 
\end{align}
where we have again used the SFT replacement rule \eqref{SFTinput1} which can be done for general $\alpha\in (0,\pi)$ because there are no zero modes in that case and further simplified a bit by using the $j$ translation invariance of the summand. 

It is not very difficult to see that it will vanish for the generic $\alpha$. The reason is that we can translate in $j$ as $j$ runs over $\mathbb{Z}$. Note that $p\pm l$ stays even or odd throughout the sum because the even-ness and odd-ness of $p\pm l$ are fixed once we have a particular $m,n,m',n'$. Hence, we have two cases at hand, $p\pm l$ being odd or even. We see that it reduces to the following in these two cases using the $j$ shift-invariance of the summand, 
\begin{align}
    &\text{Even }p\pm l: \quad \frac{1}{2}\sum_{p=|m-m'|+2,2}^{m+m'}\quad\sum_{l=|n-n'|+2,2}^{n+n'}\log\left(\prod_{j\in \mathbb{Z}}\frac{\left(2j-\frac{\alpha}{\pi}\right)\left(2j-\frac{\alpha}{\pi}\right)}{\left(2j-\frac{\alpha}{\pi}\right)\left(2j-\frac{\alpha}{\pi}\right)}\right)=0,\label{eq: gen_alpha_result1}\\
    &\text{Odd }p\pm l: \quad \frac{1}{2}\sum_{p=|m-m'|+2,2}^{m+m'}\quad\sum_{l=|n-n'|+2,2}^{n+n'}\log\left(\prod_{j\in \mathbb{Z}}\frac{\left(2j-\frac{\alpha}{\pi}+1\right)\left(2j-\frac{\alpha}{\pi}+1\right)}{\left(2j-\frac{\alpha}{\pi}+1\right)\left(2j-\frac{\alpha}{\pi}+1\right)}\right)=0 \ .\label{eq: gen_alpha_result2}
\end{align}
In the $\alpha=0$ case, we will have a lot of zero modes for a general $(m,n),(m',n')$ boundary condition on the Liouville field. This will require us to look at all the zero-mode states and deal with them individually which is beyond the scope of this article. However, we will comment on one special case in the next section. But we now at least know that whatever the complete worldsheet result for the instanton normalizations is, \textit{it won't involve any contribution from the annuli corresponding to the $\alpha\neq 0$}. 

Note that this will also work for the $R\neq 1$ case when there is general Liouville boundary condition and the mixed boson boundary condition i.e. the last annulus in $R\neq 1$ case in figure \ref{different_BC} with general $(m,n)$ and $(m',n')$. We just have to set $\alpha=\frac{\pi}{2}$. This generalizes the result \eqref{eq: mixed_instanton(1,1)}. Hence, we also verify the absence of terms $\sim e^{-2n\pi \mu-2m \pi \mu R}$ in the free energy \eqref{npfreenergyatR}. Another way of saying this is that \textit{the normalization corresponding to the general two instanton configurations involving the Dirichlet boundary condition in boson and the Neumann boundary condition on the boson will factorize into the individual exponentiated annuli with the same boson boundary conditions on both boundaries irrespective of the Liouville boundary conditions.}
\section{\texorpdfstring{$(1,n)$}{TEXT} ZZ Boundary Conditions}\label{subsec: general(1,n)bc}
\subsection{Single \texorpdfstring{$(1,n)$}{TEXT} ZZ Instanton}
Here, we highlight some of the observations related to the exponentiated annulus calculation for the $(1,n)$ ZZ boundary condition, first for the case of generic radius $R$ and then for the $R=1$ case. 
\subsubsection{Generic $R$}
First of all, we note that the action of a general $(m,n)$ ZZ instanton is given by $mn/g_s$ (or $2\pi mn\mu$) \cite{balthazar2019multiinstanton, Balthazar:2019rnh, McGreevy:2003kb}. Hence, the $(1,n)$ ZZ instanton will have the action $2\pi n\mu$ and hence will appear at the order $\sim e^{-2\pi n \mu}$ in the non-perturbative free energy. The association of $e^{-2\pi n \mu}$ suppressed effects in the MQM with $(1,n)$ ZZ branes in $c=1$ string goes back to \cite{Alexandrov:2003un, Alexandrov:2004ip} where the sine-Liouville deformation of string theory and MQM led to the comparison between the two. Consider the general $(1,n)$ ZZ boundary on both the boundaries of the annulus, the Liouville partition function will look like (using \eqref{eq: gen_liouville_part_func}),
\begin{align}\label{eq: (1,n)part_func}
    Z_{\varphi;(1,n),(1,n)}(t)=\frac{1}{\eta(it)}\sum_{l=0}^{n-1}\left(v^{-(n-l)^2}-v^{-(n-l-1)^2}\right)=\frac{v^{-n^2}-1}{\eta(it)}\ ,
\end{align}
which gives the total annulus partition function for the generic radius case as follows, 
\begin{align}\label{eq: (1,n)tot_part_func_R}
    Z(t)=\sum_{j\in \mathbb{Z}}\left(v^{j^2R^2-n^2}-v^{j^2R^2}\right)=v^{-n^2}+1-2+2\sum_{j=1}^{\infty}\left(v^{j^2R^2-n^2}-v^{j^2R^2}\right) \ .
\end{align}
We can try to calculate the exponentiated annulus in a manner similar to how \eqref{DDBC_result} was calculated in \cite{alexandrov2023instantons}. There are two fermionic zero modes and one bosonic zero mode in this case. Fermionic zero modes are the same as those discussed in \eqref{normalizationcontri3} and will have the same remedy. There will be just one bosonic zero mode $c\partial X$ and will lead to measure and prefactor similar to \eqref{normalizationcontri1} but integrated over a circle. The tachyons can again be dealt with by doing the analytic continuation as in \eqref{normalizationcontri2}. Combining all these gives, 
\begin{align}\label{eq: (1,n)_ws_result}
   \zeta \times\frac{\sqrt{-\pi}}{2\pi i/g_o}\times \frac{1}{\sqrt{2}\pi g_o}\int_{0}^{2\pi R}\frac{dx}{\sqrt{2\pi}}\times\frac{1}{\sqrt{-n^2}}\times\prod_{j=1}^{\infty}\frac{j^2R^2}{j^2R^2-n^2}
   =-\frac{i\zeta}{2\sin{\frac{n\pi}{R}}} \ .
\end{align}
The factor $\zeta$ is supposed to carry the information of the steepest descent contours associated with the tachyons. There are more tachyons in this case compared to the $(1,1)$ case which only had one and $\zeta$ was fixed to be $1/2$ by studying the effective tachyon potential in \cite{Sen_2021}. Nevertheless, it has the same functional form as the coefficient of the $n$th term in \eqref{npfreenergyatR} (it will match exactly if $\zeta=(2n)^{-1}$). A similar calculation can also be done for the case of the Neumann boundary condition\footnote{All the statements we have in this article to describe the matching of Dirichlet boundary condition on free boson and the terms in the first sum in \eqref{npfreenergyatR} can be readily generalized to describe the matching of Neumann boundary condition on free boson and the terms in the second sum in \eqref{npfreenergyatR}.} on the boson, the only difference will be that $R\rightarrow R^{-1}$ in the partition function, 
\begin{align}
    Z(t)=\sum_{j\in \mathbb{Z}}\left(v^{\frac
    {j^2}{R^2}-n^2}-v^{\frac{j^2}{R^2}}\right).
\end{align}
$\exp(\int \frac{dt}{2t}Z(t))$ gives upon regularization,
\begin{align}\label{eq: (1,n)_ws_result_neumann}
    \zeta \times\frac{\sqrt{-\pi}}{2\pi i/g_o}\times \frac{1}{\sqrt{2}\pi g_o}\int_{0}^{\frac{2\pi}{R}}\frac{d\tilde{x}}{\sqrt{2\pi}}\times\frac{1}{\sqrt{-n^2}}\times\prod_{j=1}^{\infty}\frac{\frac{j^2}{R^2}}{\frac{j^2}{R^2}-n^2}
   =-\frac{i\zeta}{2\sin{n\pi R}}
\end{align}
which is again proportional to the coefficient of the $n$th term in the second sum of the \eqref{npfreenergyatR}. It will again match if $\zeta=(2n)^{-1}$. This matching of the functional form of normalizations is a strong indication that the MQM answer for non-perturbative free energy captures more general ZZ instantons. 
\subsubsection{Self-Dual $R$}
We can now carry out a similar analysis for the $R=1$ case. We have already seen in \eqref{eq: gen_alpha_result1} and \eqref{eq: gen_alpha_result2} that for $R=1$ and $\alpha\neq 0$ case, we have trivial exponentiated annulus. For $\alpha=0$ case, the corresponding partition function will be simply (putting $R=1$ in \eqref{eq: (1,n)part_func}), 
\begin{align}
    Z(t)=&\sum_{j\in \mathbb{Z}}\left(v^{j^2-n^2}-v^{j^2}\right)\nonumber\\
    =&v^{-n^2}-1+2\sum_{j=1}^n\left(v^{j^2-n^2}-v^{j^2}\right)+2\sum_{j=n+1}^{\infty}\left(v^{j^2-n^2}-v^{j^2}\right) \ .
\end{align}
The naive computation of the exponentiated annulus, $\exp\left(\int\frac{dt}{2t}Z(t)\right)$, gives (using the \eqref{normalizationcontri1} and \eqref{normalizationcontri3}), 
\begin{align}\label{eq: (1,n)_ws_R=1}
    \zeta\times \frac{1}{4 g_o^3\pi^{\frac{5}{2}}}\times \frac{\sqrt{-\pi}}{2\pi i/g_o}\times\frac{1}{(-n^2)^{\frac{1}{2}}} \times\tilde{\prod}_{j=1}^{\infty}\frac{j^2}{j^2-n^2}=-i\zeta(-1)^{n-1}n\mu
\end{align}
where $\ \tilde{} \ $ in the product means we exclude any zeros occurring in the numerator or denominator. This is off by a factor of $n^{-2}$ when compared to the leading coefficient of the $n$th term in the sum in \eqref{npfreeenergyatR=1} (if we take $\zeta=1/2$). However, we should note that this result is naive because as we pointed out earlier, there are a lot more zero modes and tachyons in this case which we have completely ignored. Perhaps, if we interpret those zero modes appropriately and try to study their effect using some string field theory insight, we will be able to produce the correct normalization. Still, matching the scaling of coupling ($\sim \mu$) in this case is reassuring.

\subsection{Two \texorpdfstring{$(1,n)$}{TEXT} ZZ Instantons}\label{subsubsec: two_inst(1,n)}
In this section, we will be generalizing the calculation of section \ref{subsubsec: 2(1,1)R>1}, to the case of one $(1,n)$ instanton and one $(1,n')$ instanton. The analog of $Z_1(v)$ and $Z_2(v)$ in \eqref{eq: two_inst_ann_part_func} will be \eqref{eq: (1,n)tot_part_func_R} with $n$ and $n'$. The $Z_3(v)$ in \eqref{eq: two_inst_ann_part_func} will generalize to the following, 
\begin{align}
    Z_3(v)=\sum_{j\in \mathbb{Z}}\left(v^{\left(jR+\frac{x_1-x_2}{2\pi}\right)^2-\frac{(n+n')^2}{4}}-v^{\left(jR+\frac{x_1-x_2}{2\pi}\right)^2-\frac{(n-n')^2}{4}}\right)
\end{align}
where we have used the \eqref{eq: (m,n)_liov_part_func} for the $m,m'=1$. With this, we can easily follow through and get the generalization of \eqref{eq: cross_cyl_result_1} using \eqref{SFTinput1} which can be further simplified in a fashion similar to \eqref{eq: cross_cyl_intermediate_step_1} \eqref{eq: cross_cyl_intermediate_step_2},
\begin{align}
    \prod_{j\in \mathbb{Z}}\frac{\left(jR+\frac{x_1-x_2}{2\pi}-\frac{|n-n'|}{2}\right)\left(jR+\frac{x_1-x_2}{2\pi}+\frac{|n-n'|}{2}\right)}{\left(jR+\frac{x_1-x_2}{2\pi}-\frac{n+n'}{2}\right)\left(jR+\frac{x_1-x_2}{2\pi}+\frac{n+n'}{2}\right)}=\frac{\sin{\pi\left(b-\frac{|n-n'|}{2R}\right)}\sin{\pi\left(b+\frac{|n-n'|}{2R}\right)}}{\sin{\pi\left(b-\frac{n+n'}{2R}\right)}\sin{\pi\left(b+\frac{n+n'}{2R}\right)}}\label{eq: 2_(1,n)_inst_cross_exp_ann}
\end{align}
where $b=\frac{x_1-x_2}{2\pi R}$. Using the \eqref{eq: (1,n)_ws_result}, we can get the analog of the \eqref{eq: fin_form_norm}, 
\begin{align}\label{eq: 2_(1,n)_inst_main_int}
    \frac{\tilde{\zeta}}{4\sin{\frac{n\pi}{R}}\sin{\frac{n'\pi}{R}}}\int_{0}^1dx_1\int_0^1dx_2f(b) \text{ where }f(b)=\frac{\sin{\pi\left(b-\frac{|n-n'|}{2R}\right)}\sin{\pi\left(b+\frac{|n-n'|}{2R}\right)}}{\sin{\pi\left(b-\frac{n+n'}{2R}\right)}\sin{\pi\left(b+\frac{n+n'}{2R}\right)}}
\end{align}
where we have done the rescaling $x_{1,2}\rightarrow \frac{x_{1,2}}{2\pi R}$ and consequently $b=x_1-x_2$ in terms of the new variables. Here, $\tilde{\zeta}$ carries the ambiguities of the $\zeta$ in \eqref{eq: (1,n)_ws_result} as well. Again there are simple poles of the integrand at $b=\pm \left(\frac{n+n'}{2R}-\floor*{\frac{n+n'}{2R}}\right),\pm \left(1-\frac{n+n'}{2R}+\floor*{\frac{n+n'}{2R}}\right)$. We can again change the variables to $x_1=x+\frac{b}{2}, x_2=x-\frac{b}{2}$ and then use the $b\rightarrow -b$ symmetry of the integrand $f(b)$ to restrict our attention to just the right-half of the domain $D$ in figure \ref{fig:two_inst_int_domain}. It is very easy to see that again the principal value integral (using the unitary prescription of \cite{Sen:2020ruy}),
\begin{align}
    P\int_D dbdxf(b)=2P\int_0^1db(1-b)f(b)=2\times\frac{1}{2}=1.
\end{align}
\textit{This again (indirectly) implies the trivial nature of the cross-annulus stretched between $(1,n)$ and $(1,n')$ ZZ instantons if we use the unitary prescription of} \cite{Sen:2020ruy} \textit{, and hence, that normalization of the $(1,n)$ and $(1,n')$ ZZ instanton configuration is simply the product of the individual normalizations.} However, we again point out that if we consider the particular contour $\mathcal{C}$ in figure \ref{fig:two_inst_int_domain}, we get the following instead,
\begin{align}\label{eq: 2_(1,n)_inst_main_result}
    \frac{\tilde{\zeta}}{4\sin{\frac
    {n\pi}{R}}\sin{\frac
    {n'\pi}{R}}}\left(1\pm i\frac{2\sin{\frac{\pi n}{R}}\sin{\frac{\pi n'}{R}}}{\sin{\pi\frac{(n+n')}{R}}}\right)=\tilde{\zeta}\left(\frac{1}{4\sin{\frac
    {n\pi}{R}}\sin{\frac
    {n'\pi}{R}}}\pm \frac{i}{2\sin{\pi\frac{(n+n')}{R}}}\right)
\end{align}
where, the imaginary part comes from the residue contributions from the poles at the $b= \frac{n+n'}{2R}-\floor*{\frac{n+n'}{2R}}, 1-\frac{n+n'}{2R}+\floor*{\frac{n+n'}{2R}}$. It reduces to \eqref{eq: two_inst_lorentz} upon setting $n=n'=1$. Hence, we can see that with this changed prescription, this configuration can produce a piece proportional to the normalization of the single $(1,n+n')$ ZZ instanton. Hence, the normalization of the $k$th term in the first sum of \eqref{npfreenergyatR} can, in principle, get contributions from the two-instanton configuration made of $(1,k_1)$ and $(1,k_2)$ ZZ instanton $\forall k_1,k_2$ satisfying $k_1+k_2=k$ along with the single $(1,k)$ ZZ instanton normalization depending on the choice of prescription and upto the ambiguities of $\zeta, \tilde{\zeta}$.

\subsection{Multi \texorpdfstring{$(1,n)$}{TEXT} ZZ Instantons}\label{subsec: multi_(1,n)_ZZ}
The case of previous section can in principle be generalized to the multi-instanton case of $(1,n_1),(1,n_2),...,(1,n_l)$ instantons. Such configurations will appear at the order $\sim e^{-2\pi\mu(n_1+...+n_l)}$. The normalization will be constructed of the annuli on the same instantons and the annuli on two different instantons (cross-annuli). It will have the structure \cite{balthazar2019multiinstanton,Sen:2020ruy}, 
\begin{align}\label{eq: gen_multi_inst_norm}
    \tilde{\zeta}'\int \prod_{i=1}^l\frac{dx_i}{\sqrt{2\pi}}\exp\left(\sum_{i=1}^l\int\frac{dt}{2t}Z_i(v)+\sum_{\substack{i<j,\\i,j=1}}^l\int \frac{dt}{t}Z_{ij}(v)\right)
\end{align}
where $Z_i(v)$ is the annulus partition function for the annulus having $(1,n_i)$ boundary conditions on both boundaries. $Z_{ij}(v)$ is the annulus partition function corresponding to annulus having $(1,n_i)$ boundary condition on one boundary and $(1,n_j)$ boundary condition on the other. The first term in the exponent can be regularized as before, and will lead to $x_i$ independent contributions. The second term will lead to the $^lC_2$ factors like $f(b)$ in \eqref{eq: 2_(1,n)_inst_cross_exp_ann}. Hence, we can directly write the analog of \eqref{eq: 2_(1,n)_inst_main_int} after rescaling $x_i\rightarrow \frac{x_i}{2\pi R}$ such that $x_i\in (0,1]$. 
\begin{align}\label{eq: multi_inst_norm_formula}
    \frac{\tilde{\zeta}'}{2^l\prod_{i=1}^l\sin{\frac{n_i\pi}{R}}}\int \prod_{i=1}^ldx_i\prod_{\substack{i<j,\\ i,j=1}}^l\frac{\sin{\pi\left(x_{ij}-\frac{|n_i-n_j|}{2R}\right)}\sin{\pi\left(x_{ij}+\frac{|n_i-n_j|}{2R}\right)}}{\sin{\pi\left(x_{ij}-\frac{n_i+n_j}{2R}\right)}\sin{\pi\left(x_{ij}+\frac{n_i+n_j}{2R}\right)}}
\end{align}
where $x_{ij}:=x_i-x_j$. We have evaluated the above integral for the case of $l=3$ numerically with the unitary prescription of \cite{Sen:2020ruy}. We notice that the integral with this prescription carries a non-trivial dependence on both $R$ and the $\{n_i\}_{i=1,2,3}$ as opposed to the $l=2$ case studied in the previous section. However, the residues can also lead to some real contributions because of the multi-dimensional nature of the integral. A detailed analysis of these integrals will be the subject
of future work \cite{Alexandrov:2025pzs}.

For $R=1$, it is simply clear by the discussion in section \ref{general_liouvill_bc}, that the integrand will be equal to $1$ and the normalization \eqref{eq: gen_multi_inst_norm}\footnote{To be precise, the analog of it, because the integration of collective coordinates will be over $S^3$ (SU(2) group manifold) in $R=1$ case.} becoming a product of individual normalizations.

\section{Disc Two-Point and Annulus One-Point Function}\label{sec_4}
In this section, we will match the matrix model predictions \eqref{disk_2_pt_MM} and \eqref{ann_one_pt_MM} with the corresponding worldsheet results for the disk two-point function and the annulus one-point function. We will follow the notation and conventions of the \cite{Eniceicu_2022}. First, note that we have $b=1$ in this case, as discussed earlier. Hence, the equation of motion for the Liouville field will look like,
\begin{align}
    \frac{\partial \bar{\partial}\phi(z, \bar{z})}{\pi \mu}=e^{2\phi(z, \bar{z})}=:V(z, \bar{z}) \ .
\end{align}
Next, we use the following expansions on the Upper Half Plane (UHP) with ZZ boundary condition on the boundary i.e. the real line, 
\begin{align}
    \partial \phi(z, \bar{z})=-\frac{2}{z-\bar{z}}+O(z-\bar{z}), \quad \bar{\partial} \phi(z, \bar{z})=\frac{2}{z-\bar{z}}+O(z-\bar{z}) \ .
\end{align}
We can simply read off one-point function in UHP from the above expansion,
\begin{align}
    \langle V(z,\bar{z})\rangle_{\text{UHP}}=\frac{2}{\pi \mu |z-\bar{z}|^2}
\end{align}
Using above and $\langle c(z_1)c(z_2)c(z_3)\rangle=z_{12}z_{23}z_{31}$, we can calculate the disk one point function with the closed string insertion as follows, 
\begin{align}\label{Disk_1_pt_zeromom}
    A_{\text{disk}}(\psi_c)=\frac{1}{4g_s}\langle (\partial c-\bar{\partial}\bar{c})c\bar{c}V\rangle_{\text{UHP}}=(\pi \mu g_s)^{-1}=2 \ .
\end{align}
For the ratio of the disk two-point function to the disk one-point function, the calculation will be exactly the same as the corresponding calculation in \cite{Eniceicu_2022} for the minimal string case. Hence, we can simply borrow the result and specialize to $b=1$ and $Q=2$ case, 
\begin{align}\label{eq: disk_2_pt_ws_result}
    \boxed{g_sf=g_s\bigg(\frac{2b}{Q}-1\bigg)\bigg|_{b=1,Q=2}=0 \ .}
\end{align}
which exactly matches the prediction using the matrix model result of free energy in \eqref{disk_2_pt_MM}. The reason this works is because we are calculating the correlator for the zero momentum insertions ($e^{2\phi}$) i.e. there is no free boson field ($X$) appearing in the insertions. Hence, the answer just corresponds to the one in the minimal string case with the appropriate value of parameter $b$.

We can now proceed with the annulus one-point function calculation. Consider the annulus partition function for the same boundary condition on both boundaries \eqref{expannulus} ($v=e^{-2\pi t}$),
\begin{align}\label{ann_part_func_exp}
    Z(t)\equiv Z(v)=(v^{-1}-1)\sum_{n\in \mathbb{Z}}v^{n^2}\quad \xrightarrow{v \rightarrow \text{ small}}\quad  v^{-1}+1+O(v)
\end{align}
which will be useful for us later. We are mainly interested in the ratio of the annulus one-point function and disk one-point function. We will follow the calculation in \cite{Eniceicu_2022} and this ratio is given by,
\begin{align}\label{ann_1pt_derivation}
 &g_s g =\int_0^1 dv \int_{0}^{\frac{1}{4}}dx F(v,x)=\int_0^1 dv \int_{0}^{\frac{1}{4}}dx \partial_xG(v,x)\\
 &\text{where, } F(v,x)=2g_s\text{Tr}(V(w, \bar{w})b_0c_0v^{L_0-1}), \quad G(v,x)=\frac{g_s}{8\pi^2}\text{Tr}(\partial_x\phi(w, \bar{w})b_0c_0v^{L_0-1})\nonumber\\
 &\text{ and }w=2\pi(x+iy)
\end{align}
where, we have used the overall normalization $2g_s$ fixed in the appendix of \cite{Eniceicu_2022} using the arguments similar to the case of critical bosonic string theory in section 6.4 in \cite{Polchinski:1998rq}. This integral experiences the divergences from the small $x$ and small $v$ regions of the integration domain. Hence, we will note the corresponding behaviours of $F(v,x)$ and $G(v,x)$ as it will be useful for us later \cite{Eniceicu_2022}, 
\begin{align}
    G(v,x)& = 
    \begin{cases}
       -\frac{g_s}{4\pi^2 x}\frac{Z(v)}{v},&  \text{small }x\\
       -\frac{g_s}{2\pi}\cot{2\pi x}(v^{-2}+v^{-1}+O(1)),&  \text{small }v
    \end{cases}\label{G(v,x)_limits} \\
     F(v,x)&=  \begin{cases}
          \frac{g_s}{4\pi^2 x^2}\frac{Z(v)}{v},&  \text{small }x\\
       \frac{g_s}{\sin^22\pi x}(v^{-2}+v^{-1}+O(1)),&  \text{small }v \ .
         \end{cases}\label{F(v,x)limits}
\end{align}
Main difference from \cite{Eniceicu_2022} here is that the $Z(v)$ is \eqref{ann_part_func_exp} which reflects at the level of small $v$ expansion as well. Now, we will calculate the different contributions to $g$.
\begin{itemize}
    \item We will first consider the worldsheet contribution $g_{\text{ws}}=g^{(a)}+g^{(b)}+g^{(c)}+g^{(d)}$, where superscript denotes the contributions coming from the four different regions of the integration domain or moduli space (see figure \ref{C_annulus_moduli}). We have reviewed the details in appendix \ref{appendix_ws_calc}. $g^{(d)}$ can be reduced to an integral over the boundary because of the total derivative integrand. Contributions from boundaries at $x=\frac{1}{4}$ and $v=1$ vanish as argued in \cite{Eniceicu_2022} and hence, we have only the contributions from the rest of the two boundaries at the interface of the region $(d)$ with $(c)$ and $(b)$
    \begin{align}
        g^{(d)}=g^{(b)-(d)}+g^{(c)-(d)} \ .
    \end{align}
    Further $g^{(b)}$ cancels exactly $g^{(b)-(d)}$ as shown in \cite{Sen_202183} and \cite{Eniceicu_2022}, we are just left with, 
    \begin{align}\label{worldsheetcontribution}
        g_{\text{ws}}=g^{(a)}+g^{(c)}+g^{(c)-(d)} \ .
    \end{align}
    We will now list down the results for these different contributions (detailed calculation in appendix \ref{appendix_ws_calc}), 
        \begin{align}
            g^{(a)}=&\frac{\tilde{\lambda} \alpha^2}{2\pi}, \\
            g^{(c)}=&-\frac{\alpha^2\tilde{\lambda}^2}{4}+\frac{\alpha^2\tilde{\lambda}}{2\pi}-\frac{\alpha^2}{8},\\
            g^{(c)-(d)}=&\frac{\alpha^2\tilde{\lambda}^2}{4}-\frac{\alpha^2\tilde{\lambda}}{\pi}+\frac{\alpha^2}{8}+\frac{9\tilde{\lambda}}{4\pi}-1-\frac{2\tilde{\lambda}^2}{\pi}\int_{(2\tilde{\lambda})^{-1}}^1d\beta\frac{(f(\beta))^2}{\beta^2+1},\label{gcdresult}
        \end{align} 
    where $\alpha$ and $\tilde{\lambda}$ are two large constants and $f(\beta)$ is a function that appears in the construction of string field theory vertices used for the computation. The final result is expected to be independent of the choice of $\alpha, \tilde{\lambda}$ and $f(\beta)$. Note that $g^{(a)}$ and $g^{(c)}$ are same as the results from \cite{Eniceicu_2022} whereas $g^{(c)-(d)}$ has changed by $3\tilde{\lambda}/\pi-3/2$. The reason for this change is again the presence of extra bosonic zero modes compared to non-compactified case of \cite{Sen_202183} and the minimal case of \cite{Eniceicu_2022}. This is reflected in the partition function \eqref{ann_part_func_exp} as well which ultimately leads to the above result \eqref{gcdresult} as shown in appendix \ref{appendix_ws_calc}.
    Putting all these results in \eqref{worldsheetcontribution}, we get the total worldsheet contribution,
    \begin{align}
        g_{\text{ws}}=\frac{9\tilde{\lambda}}{4\pi}-1-\frac{2\tilde{\lambda}^2}{\pi}\int_{(2\tilde{\lambda})^{-1}}^1d\beta\frac{(f(\beta))^2}{\beta^2+1} \ .
    \end{align}
    \begin{figure}
        \centering

\tikzset{every picture/.style={line width=0.75pt}} %set default line width to 0.75pt        

\begin{tikzpicture}[x=0.75pt,y=0.75pt,yscale=-1,xscale=1]
%uncomment if require: \path (0,284); %set diagram left start at 0, and has height of 284

%Shape: Arc [id:dp4809145067535663] 
\draw  [draw opacity=0] (67.88,234.81) .. controls (70.39,224.98) and (79.27,217.72) .. (89.85,217.72) .. controls (102.37,217.72) and (112.52,227.9) .. (112.52,240.45) .. controls (112.52,240.5) and (112.52,240.55) .. (112.52,240.61) -- (89.85,240.45) -- cycle ; \draw   (67.88,234.81) .. controls (70.39,224.98) and (79.27,217.72) .. (89.85,217.72) .. controls (102.37,217.72) and (112.52,227.9) .. (112.52,240.45) .. controls (112.52,240.5) and (112.52,240.55) .. (112.52,240.61) ;  
%Shape: Arc [id:dp39134893542465576] 
\draw  [draw opacity=0] (112.34,243.34) .. controls (110.92,254.53) and (101.39,263.18) .. (89.85,263.18) .. controls (80.15,263.18) and (71.87,257.08) .. (68.64,248.48) -- (89.85,240.45) -- cycle ; \draw   (112.34,243.34) .. controls (110.92,254.53) and (101.39,263.18) .. (89.85,263.18) .. controls (80.15,263.18) and (71.87,257.08) .. (68.64,248.48) ;  
%Curve Lines [id:da46706464580904217] 
\draw    (49.8,235.15) .. controls (81.54,235.15) and (86.22,236.66) .. (86.07,231.36) ;
%Curve Lines [id:da5597295709606] 
\draw    (49.95,248.03) .. controls (81.69,248.03) and (87.43,247.27) .. (87.73,251.82) ;
%Shape: Ellipse [id:dp784196764384752] 
\draw   (48.29,241.59) .. controls (48.29,238.03) and (49.03,235.15) .. (49.95,235.15) .. controls (50.87,235.15) and (51.61,238.03) .. (51.61,241.59) .. controls (51.61,245.15) and (50.87,248.03) .. (49.95,248.03) .. controls (49.03,248.03) and (48.29,245.15) .. (48.29,241.59) -- cycle ;
%Straight Lines [id:da9397333443699101] 
\draw    (112.52,240.61) -- (137.91,240.6) ;
%Straight Lines [id:da4292908390057073] 
\draw    (112.34,243.35) -- (137.88,243.5) ;
%Shape: Arc [id:dp9945324221736989] 
\draw  [draw opacity=0] (137.63,240.88) .. controls (138.2,228.83) and (148.12,219.23) .. (160.28,219.23) .. controls (164.94,219.23) and (169.27,220.65) .. (172.88,223.07) -- (160.28,241.97) -- cycle ; \draw   (137.63,240.88) .. controls (138.2,228.83) and (148.12,219.23) .. (160.28,219.23) .. controls (164.94,219.23) and (169.27,220.65) .. (172.88,223.07) ;  
%Shape: Arc [id:dp24763189862428736] 
\draw  [draw opacity=0] (173.31,260.57) .. controls (169.62,263.17) and (165.13,264.7) .. (160.28,264.7) .. controls (148.12,264.7) and (138.2,255.11) .. (137.63,243.06) -- (160.28,241.97) -- cycle ; \draw   (173.31,260.57) .. controls (169.62,263.17) and (165.13,264.7) .. (160.28,264.7) .. controls (148.12,264.7) and (138.2,255.11) .. (137.63,243.06) ;  
%Shape: Arc [id:dp9110449201641928] 
\draw  [draw opacity=0] (174.31,224.12) .. controls (179.57,228.28) and (182.94,234.73) .. (182.94,241.97) .. controls (182.94,248.98) and (179.78,255.25) .. (174.8,259.42) -- (160.28,241.97) -- cycle ; \draw   (174.31,224.12) .. controls (179.57,228.28) and (182.94,234.73) .. (182.94,241.97) .. controls (182.94,248.98) and (179.78,255.25) .. (174.8,259.42) ;  
%Curve Lines [id:da9942343032567984] 
\draw    (172.85,223.05) .. controls (189.58,211.05) and (205.38,225.81) .. (206.52,240.6) .. controls (207.66,255.4) and (193.39,271.77) .. (173.29,260.59) ;
%Curve Lines [id:da09223475276323967] 
\draw    (174.29,224.1) .. controls (188.39,215.14) and (202.89,225.45) .. (204.1,240) .. controls (205.31,254.55) and (193.52,267.88) .. (174.77,259.44) ;
%Shape: Arc [id:dp4040381743367538] 
\draw  [draw opacity=0] (474.68,242.33) .. controls (477.18,232.51) and (486.06,225.24) .. (496.64,225.24) .. controls (509.16,225.24) and (519.31,235.42) .. (519.31,247.98) .. controls (519.31,248.03) and (519.31,248.08) .. (519.31,248.13) -- (496.64,247.98) -- cycle ; \draw   (474.68,242.33) .. controls (477.18,232.51) and (486.06,225.24) .. (496.64,225.24) .. controls (509.16,225.24) and (519.31,235.42) .. (519.31,247.98) .. controls (519.31,248.03) and (519.31,248.08) .. (519.31,248.13) ;  
%Shape: Arc [id:dp6374467443714043] 
\draw  [draw opacity=0] (519.13,250.87) .. controls (517.71,262.06) and (508.18,270.71) .. (496.64,270.71) .. controls (486.94,270.71) and (478.66,264.6) .. (475.43,256.01) -- (496.64,247.98) -- cycle ; \draw   (519.13,250.87) .. controls (517.71,262.06) and (508.18,270.71) .. (496.64,270.71) .. controls (486.94,270.71) and (478.66,264.6) .. (475.43,256.01) ;  
%Curve Lines [id:da9959308182397881] 
\draw    (456.59,242.67) .. controls (488.33,242.67) and (493.01,244.19) .. (492.86,238.88) ;
%Curve Lines [id:da11902633478584046] 
\draw    (456.74,255.56) .. controls (488.48,255.56) and (494.22,254.8) .. (494.52,259.34) ;
%Shape: Ellipse [id:dp06014024462087808] 
\draw   (455.08,249.11) .. controls (455.08,245.56) and (455.82,242.67) .. (456.74,242.67) .. controls (457.66,242.67) and (458.4,245.56) .. (458.4,249.11) .. controls (458.4,252.67) and (457.66,255.56) .. (456.74,255.56) .. controls (455.82,255.56) and (455.08,252.67) .. (455.08,249.11) -- cycle ;
%Straight Lines [id:da6588172585287326] 
\draw    (519.31,248.13) -- (544.7,248.13) ;
%Straight Lines [id:da017875628251730724] 
\draw    (519.13,250.87) -- (544.67,251.02) ;
%Shape: Arc [id:dp30743967134747296] 
\draw  [draw opacity=0] (544.01,247.83) .. controls (544.79,235.97) and (554.62,226.6) .. (566.63,226.6) .. controls (579.15,226.6) and (589.3,236.78) .. (589.3,249.34) .. controls (589.3,261.89) and (579.15,272.07) .. (566.63,272.07) .. controls (554.72,272.07) and (544.96,262.86) .. (544.03,251.16) -- (566.63,249.34) -- cycle ; \draw   (544.01,247.83) .. controls (544.79,235.97) and (554.62,226.6) .. (566.63,226.6) .. controls (579.15,226.6) and (589.3,236.78) .. (589.3,249.34) .. controls (589.3,261.89) and (579.15,272.07) .. (566.63,272.07) .. controls (554.72,272.07) and (544.96,262.86) .. (544.03,251.16) ;  
%Shape: Ellipse [id:dp2644761519610206] 
\draw   (554.39,249.34) .. controls (554.39,242.56) and (559.87,237.06) .. (566.63,237.06) .. controls (573.39,237.06) and (578.87,242.56) .. (578.87,249.34) .. controls (578.87,256.11) and (573.39,261.61) .. (566.63,261.61) .. controls (559.87,261.61) and (554.39,256.11) .. (554.39,249.34) -- cycle ;
%Curve Lines [id:da5362662110392704] 
\draw    (76.51,72.83) .. controls (108.25,72.83) and (112.93,74.34) .. (112.78,69.04) ;
%Curve Lines [id:da9671523485142388] 
\draw    (76.66,85.71) .. controls (108.4,85.71) and (114.14,84.95) .. (114.44,89.5) ;
%Shape: Ellipse [id:dp5939011456615424] 
\draw   (75,79.27) .. controls (75,75.71) and (75.74,72.83) .. (76.66,72.83) .. controls (77.58,72.83) and (78.32,75.71) .. (78.32,79.27) .. controls (78.32,82.82) and (77.58,85.71) .. (76.66,85.71) .. controls (75.74,85.71) and (75,82.82) .. (75,79.27) -- cycle ;
%Shape: Arc [id:dp18599388496545655] 
\draw  [draw opacity=0] (98.82,72.95) .. controls (101.67,63.66) and (110.29,56.91) .. (120.49,56.91) .. controls (125.15,56.91) and (129.49,58.32) .. (133.09,60.74) -- (120.49,79.65) -- cycle ; \draw   (98.82,72.95) .. controls (101.67,63.66) and (110.29,56.91) .. (120.49,56.91) .. controls (125.15,56.91) and (129.49,58.32) .. (133.09,60.74) ;  
%Shape: Arc [id:dp814202397205583] 
\draw  [draw opacity=0] (133.52,98.25) .. controls (129.84,100.85) and (125.34,102.38) .. (120.49,102.38) .. controls (109.98,102.38) and (101.15,95.21) .. (98.58,85.49) -- (120.49,79.65) -- cycle ; \draw   (133.52,98.25) .. controls (129.84,100.85) and (125.34,102.38) .. (120.49,102.38) .. controls (109.98,102.38) and (101.15,95.21) .. (98.58,85.49) ;  
%Shape: Arc [id:dp6041745419380213] 
\draw  [draw opacity=0] (134.53,61.79) .. controls (139.78,65.96) and (143.16,72.41) .. (143.16,79.65) .. controls (143.16,86.66) and (139.99,92.93) .. (135.01,97.1) -- (120.49,79.65) -- cycle ; \draw   (134.53,61.79) .. controls (139.78,65.96) and (143.16,72.41) .. (143.16,79.65) .. controls (143.16,86.66) and (139.99,92.93) .. (135.01,97.1) ;  
%Curve Lines [id:da0976539041062261] 
\draw    (133.07,60.73) .. controls (149.79,48.73) and (165.59,63.49) .. (166.74,78.28) .. controls (167.88,93.08) and (153.61,109.45) .. (133.5,98.26) ;
%Curve Lines [id:da12891896224097366] 
\draw    (134.5,61.77) .. controls (148.6,52.82) and (163.11,63.13) .. (164.32,77.68) .. controls (165.53,92.22) and (153.74,105.56) .. (134.99,97.12) ;
%Shape: Arc [id:dp7597641812148705] 
\draw  [draw opacity=0] (351.66,76.55) .. controls (354.19,64.59) and (365.19,55.59) .. (378.38,55.59) .. controls (393.44,55.59) and (405.65,67.33) .. (405.65,81.81) .. controls (405.65,96.3) and (393.44,108.04) .. (378.38,108.04) .. controls (365.97,108.04) and (355.5,100.07) .. (352.19,89.16) -- (378.38,81.81) -- cycle ; \draw   (351.66,76.55) .. controls (354.19,64.59) and (365.19,55.59) .. (378.38,55.59) .. controls (393.44,55.59) and (405.65,67.33) .. (405.65,81.81) .. controls (405.65,96.3) and (393.44,108.04) .. (378.38,108.04) .. controls (365.97,108.04) and (355.5,100.07) .. (352.19,89.16) ;  
%Shape: Ellipse [id:dp406544453640042] 
\draw   (368.15,81.81) .. controls (368.15,76.2) and (372.73,71.64) .. (378.38,71.64) .. controls (384.03,71.64) and (388.61,76.2) .. (388.61,81.81) .. controls (388.61,87.43) and (384.03,91.99) .. (378.38,91.99) .. controls (372.73,91.99) and (368.15,87.43) .. (368.15,81.81) -- cycle ;
%Curve Lines [id:da7273735903617309] 
\draw    (325.7,76.06) .. controls (357.44,76.06) and (362.12,77.57) .. (361.97,72.27) ;
%Curve Lines [id:da9153211708340148] 
\draw    (325.85,88.94) .. controls (357.59,88.94) and (363.33,88.18) .. (363.63,92.73) ;
%Shape: Ellipse [id:dp037983101731140057] 
\draw   (324.19,82.5) .. controls (324.19,78.94) and (324.93,76.06) .. (325.85,76.06) .. controls (326.77,76.06) and (327.51,78.94) .. (327.51,82.5) .. controls (327.51,86.06) and (326.77,88.94) .. (325.85,88.94) .. controls (324.93,88.94) and (324.19,86.06) .. (324.19,82.5) -- cycle ;
%Right Arrow [id:dp7410402611669307] 
\draw  [fill={rgb, 255:red, 208; green, 2; blue, 27 }  ,fill opacity=0.5 ] (283.53,94.16) -- (233.15,94.45) -- (233.16,97.17) -- (224,91.78) -- (233.12,86.29) -- (233.13,89.01) -- (283.51,88.72) -- cycle ;
%Right Arrow [id:dp2597247965536904] 
\draw  [fill={rgb, 255:red, 208; green, 2; blue, 27 }  ,fill opacity=0.5 ] (383.35,119.67) -- (384.3,158.09) -- (386.74,158.02) -- (382.11,168.63) -- (376.97,158.28) -- (379.41,158.22) -- (378.46,119.81) -- cycle ;
%Right Arrow [id:dp7661314300651942] 
\draw  [fill={rgb, 255:red, 208; green, 2; blue, 27 }  ,fill opacity=0.5 ] (329.57,108.87) -- (283.06,161.43) -- (285.44,163.53) -- (272.34,169.81) -- (276.99,156.05) -- (279.37,158.16) -- (325.88,105.61) -- cycle ;
%Shape: Rectangle [id:dp9309442929369403] 
\draw   (207.53,28.1) -- (466.1,28.1) -- (466.1,201.9) -- (207.53,201.9) -- cycle ;
%Straight Lines [id:da2438747903896621] 
\draw    (270.16,173.83) -- (465.87,173.57) ;
%Straight Lines [id:da3261827700481601] 
\draw    (270.16,173.83) -- (270.62,201.9) ;
%Straight Lines [id:da9759165970362289] 
\draw    (207.3,171.02) -- (270.16,173.83) ;
%Curve Lines [id:da11415599157340717] 
\draw    (211.43,28.36) .. controls (209.82,102.88) and (223.82,165.66) .. (270.16,173.83) ;
%Straight Lines [id:da6537657162960151] 
\draw    (213.5,115.9) -- (169.92,96.32) ;
\draw [shift={(168.1,95.5)}, rotate = 24.2] [color={rgb, 255:red, 0; green, 0; blue, 0 }  ][line width=0.75]    (10.93,-3.29) .. controls (6.95,-1.4) and (3.31,-0.3) .. (0,0) .. controls (3.31,0.3) and (6.95,1.4) .. (10.93,3.29)   ;
%Straight Lines [id:da424831836229584] 
\draw    (223.5,197.1) -- (206.3,226.97) ;
\draw [shift={(205.3,228.7)}, rotate = 299.94] [color={rgb, 255:red, 0; green, 0; blue, 0 }  ][line width=0.75]    (10.93,-3.29) .. controls (6.95,-1.4) and (3.31,-0.3) .. (0,0) .. controls (3.31,0.3) and (6.95,1.4) .. (10.93,3.29)   ;
%Straight Lines [id:da3462292252177077] 
\draw    (381.9,189.7) -- (460.76,234.51) ;
\draw [shift={(462.5,235.5)}, rotate = 209.61] [color={rgb, 255:red, 0; green, 0; blue, 0 }  ][line width=0.75]    (10.93,-3.29) .. controls (6.95,-1.4) and (3.31,-0.3) .. (0,0) .. controls (3.31,0.3) and (6.95,1.4) .. (10.93,3.29)   ;

% Text Node
\draw (215.82,149) node [anchor=north west][inner sep=0.75pt]  [font=\scriptsize]  {$( c)$};
% Text Node
\draw (296.78,59.18) node [anchor=north west][inner sep=0.75pt]  [font=\scriptsize]  {$( d)$};
% Text Node
\draw (193.2,115.7) node [anchor=north west][inner sep=0.75pt]  [font=\scriptsize]  {$x$};
% Text Node
\draw (336.8,204.9) node [anchor=north west][inner sep=0.75pt]  [font=\scriptsize]  {$v$};
% Text Node
\draw (197.5,202.1) node [anchor=north west][inner sep=0.75pt]  [font=\scriptsize]  {$0$};
% Text Node
\draw (467.3,201.79) node [anchor=north west][inner sep=0.75pt]  [font=\scriptsize]  {$1$};
% Text Node
\draw (190.5,15.39) node [anchor=north west][inner sep=0.75pt]  [font=\scriptsize]  {$\frac{1}{4}$};
% Text Node
\draw (244.5,79.4) node [anchor=north west][inner sep=0.75pt]  [font=\tiny]  {$\text{small} \ v$};
% Text Node
\draw (387.5,127.9) node [anchor=north west][inner sep=0.75pt]  [font=\tiny]  {$\text{small} \ x$};
% Text Node
\draw (276.58,145.03) node [anchor=north west][inner sep=0.75pt]  [font=\tiny,rotate=-312.07]  {$\text{small} \ x\ \text{and} \ v$};
% Text Node
\draw (232.51,181.28) node [anchor=north west][inner sep=0.75pt]  [font=\scriptsize,rotate=-359.42]  {$( a)$};
% Text Node
\draw (350.01,181.28) node [anchor=north west][inner sep=0.75pt]  [font=\scriptsize,rotate=-359.42]  {$( b)$};

\end{tikzpicture}

        \caption{The moduli space for the annulus with one closed string insertion. Different regions corresponds to different SFT vertices which comes from the degeneration of worldsheet near boundaries(figure reproduced from \cite{Sen_202183})}
        \label{C_annulus_moduli}
    \end{figure}
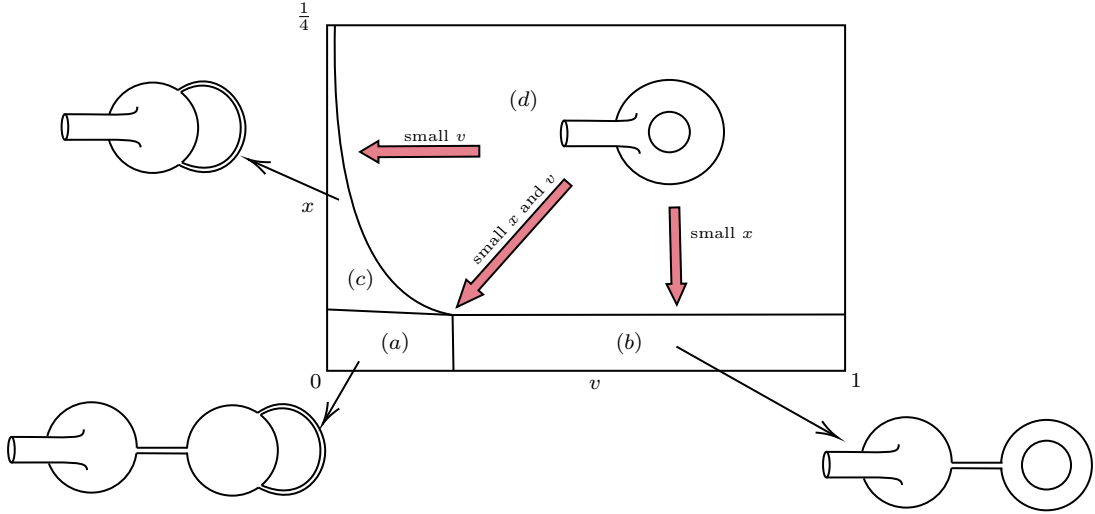
    \item Two non-worldsheet contributions, $g_{\psi}$ and $g_{\text{ghost}}$, were discussed in \cite{Sen_202183} and \cite{Eniceicu_2022} which led to following contributions, 
    \begin{align}
        g_{\psi}=\frac{\tilde{\lambda}}{4\pi}+\frac{2\tilde{\lambda}^2}{\pi}\int_{(2\tilde{\lambda})^{-1}}^{1}d\beta \frac{(f(\beta))^2}{1+\beta^2}, \quad g_{\text{ghost}}=\frac{\tilde{\lambda}}{2\pi} \ .
    \end{align}
    Here, $\psi$ is the field corresponding to the state $c_0|0\rangle$ that appears in the open string field expansion while dealing with the singular nature of the Siegel gauge. Since this state leads to a non-vanishing term in the quadratic open string field action, it can lead to contributions to the amplitudes which are invisible to the worldsheet theory. $g_{\psi}$ is exactly that contribution. $g_{\text{ghost}}$ corresponds to the contribution due to the jacobian associated with redefinition required to relate the gauge transformation parameter $\theta$ (corresponding to state $|0\rangle$) with the rigid U(1) group parameter.
    
    \item Finally we note the effect of the three bosonic zero modes corresponding to D instanton translations. It was shown in \cite{Sen_202183} that there will be a contribution to $g$ due to the jacobian corresponding to the redefinition relating the open string field corresponding to state $\alpha_{-1}c_1|0\rangle$ (with vertex operator $c\partial X$) and the collective coordinate translating the D-instanton. This contribution for the $c=1$ uncompactified case turned out to be $-\tilde{\lambda}/\pi$. But at self-dual radius, as argued earlier, we have two more zero modes i.e. we have two more open string fields corresponding to vertex operators $ce^{2iX}$ and $ce^{-2iX}$. We will have to account for the effect coming from their redefinition as well. This effect is calculated by looking at the disk amplitude with two closed string insertions and one open string insertion corresponding to the zero mode. We simply state the final result here keeping all the details in appendix \ref{Appendix_A_gjac_calc},
    \begin{align}
        g_{\text{jac}}=-\frac{3\tilde{\lambda}}{\pi} \ .
    \end{align}
    The reason that the contribution is simply three times can be traced to the fact that the three zero modes correspond to translations in $S^3$ in orthogonal directions and all directions in $S^3$ are identical. Hence, all three modes get the same Jacobian contribution upon redefinition which means the same contribution to effective action and eventually, the closed string one-point function.
\end{itemize}
Adding all the contributions, we get the following result, 
\begin{align}\label{eq: ann_1_pt_ws_result}
   \boxed{ g=g_{\text{ws}}+g_{\psi}+g_{\text{ghost}}+g_{\text{jac}}=-1 ,}
\end{align}
which is the expected result from the matrix model calculation \eqref{ann_one_pt_MM}. Again, the main difference to be noted from the minimal string case is that in \eqref{ann_part_func_exp}, we had a $+1$ instead of $-2$ in the minimal string case \cite{Eniceicu_2022}. This was again due to the extra bosonic zero modes.
\section{Discussion}
We saw different descriptions of the $c=1$ string at self-dual radius. Specifically, we saw the two matrix models (the KM model and the IM model), the MQM (or the free fermion description), and finally, the worldsheet CFT description. We saw how the total free energy expression from the MQM description can be obtained from the two matrix model descriptions. We further saw how the leading non-perturbative part of this free energy expression can be obtained from the ZZ instanton calculations in the worldsheet CFT description aided by the string field theory insights. Here, we list some of the main takeaways.
\begin{enumerate}
    \item $c=1$ string at self-dual radius has two matrix model descriptions, one of which can be classified as V-type (KM Model), and the other as F-type (IM Model)\cite{gopakumar2023derivingsimplestgaugestringduality}. V-type description requires the analytic continuation $N=-i\mu$ of the matrix dimension whereas, in the F-type description, the parameter $\mu$ is independent of the matrix dimension $Q$. The analytic continuation of the matrix dimension in the V-type description is important in order to produce the free energy expected from the MQM description correctly. The F-type model, on the other hand, requires the analytic continuation $\nu\rightarrow -i\mu$ of the parameter $\nu$. The requirement of these analytic continuations comes from the matching of matrix model correlators with the MQM results for the tachyon correlators perturbatively. Various checks we provided here indicate that ensuring perturbative matching automatically ensures matching at the non-perturbative level as well. This non-perturbative matching is seen without any double-scaling limit in the matrix integrals.
    \item The leading piece in the one-instanton sector from the string theory matches the matrix model and MQM predictions. Further, we also found that the exponentiated annulus for the mixed boundary condition (for the free boson) is just trivial in both $R=1$ and $R\neq 1$ cases. For the $R=1$ case, this simply implies that the normalization of the multi-instanton amplitude will be simply the product of individual amplitudes. For $R\neq 1$, it implies the factorization for the mixed boson boundary condition (Neumann on one boundary and Dirichlet on the other) case. We saw this for more general Liouville boundary conditions as well. Such simplifications are helpful in studying the non-perturbative effects at arbitrary order.
    
    For the $R=1$ case, we also checked that the string theory results for the disk two-point function and annulus one-point functions with zero momentum insertions match the corresponding matrix model predictions.

    \item Enhanced gauge symmetry at $R=1$ plays a crucial role in matching normalization with the string theory result. Interpretation of the translation modes living an $S^3$ ($SU(2)$ group manifold) was very important for this matching. Moreover, the radius of this $S^3$ is inherited from the compact boson radius $R$.
    \item We also explored the more general case of the $(1,n)$ ZZ instantons and how these instantons can possibly appear and completely exhaust all the terms in the matrix model free energy trans-series for both $R=1$ and $R\neq 1$ cases. Particularly for the generic $R$ case, we saw how $(1,n)$ and $(1,n')$ two instanton configuration (with a specific Lorentzian prescription) and $(1,n+n')$ single instanton configuration can lead to the normalizations proportional to the same term appearing at the order $\sim e^{-2(n+n')\pi \mu}$ in the trans series.
    \end{enumerate}
     There are many interesting future directions as well. First of all, an immediate goal is to determine from the first principles, the numerical constants  $\zeta, \tilde{\zeta}$ appearing in the \eqref{eq: (1,n)_ws_result}, \eqref{eq: (1,n)_ws_result_neumann}, \eqref{eq: (1,n)_ws_R=1}, and  \eqref{eq: 2_(1,n)_inst_main_result}. Particularly, it will be interesting to see how the choice of prescription and the ambiguity in these factors of $\zeta, \tilde{\zeta}$ can produce exactly the MQM prediction of non-perturbative free energy, \eqref{npfreenergyatR} and \eqref{npfreeenergyatR=1} and whether multiple choices are leading to the same result. Also, an analytic study of the general multi-instanton normalization \eqref{eq: multi_inst_norm_formula} can drastically improve our understanding of this duality\footnote{Work is in progress in this direction and will appear soon \cite{Alexandrov:2025pzs}.}.
     
     Partition function calculation on the sphere with three holes (disk with two holes) and a torus with one hole can be the next non-trivial check of this duality at one instanton level.  Similar calculations in superstrings, such as the Type 0B case where a dual MQM description exists \cite{Takayanagi:2003sm}, will also be a good check of this approach to account for the enhanced gauge symmetry at a self-dual radius. Studying non-perturbative effects in topological A Model string ($SL(2,\mathbb{R})_1/U(1)$ coset model) should also be possible and it will perhaps lead to some non-trivial checks of its equivalence to $c=1$ at $R=1$ shown in \cite{Mukhi:1993zb, Ashok:2005xc}.

\subsection*{Acknowledgements}
I wish to thank Ashoke Sen for essentially guiding this project through the frequent discussions, for the careful multiple readings of the drafts, and for the accompanying detailed feedback. I wish to thank my supervisor Rajesh Gopakumar for suggesting to look in this direction in the first place and later giving constant inputs. I also thank Sergey Alexandrov for collaboration on a related project which helped in improving my overall undersanding. I thank the rest of the strings group at ICTS, especially Raghu Mahajan, Omkar Shetye, and Ritwick Ghosh for some discussions on the details. I would like to thank Jyotirmoy Barman for pointing out some typographical errors. I want to thank the organizers of Les Houches School on Quantum Geometry 2024, where part of this work was presented. Research at ICTS is supported by the Department of Atomic
Energy, Government of India, under project no. RTI4001 and more broadly through the framework of
support for the basic sciences by the people of India. Finally, I also thank JHEP referee for valuable feedback which led to improved version of the draft.

\appendix

\section{String Field Theory Vertices}\label{Vertices}
In this section, we will review the construction of the vertices we will be using in our calculations. We will work in upper half plane (UHP) for all the vertices. We will use C and O to denote respectively the closed string and open string insertions. This discussion follows \cite{Sen_202183} and \cite{Eniceicu_2022}. 
\subsection*{C-O vertex} 
We consider the UHP with C and O insertions located at $z=i$ and $z=0$ respectively. We can also define the local coordinates near the open string insertion by $w$, 
\begin{align}
    w=\lambda z
\end{align}
where $\lambda$ is a large number.
\subsection*{O-O-O vertex}
We consider three O insertions at $z=0,1$ and $\infty$ on the UHP. The local coordinates near these insertions are respectively, 
\begin{align}
    w_1=\frac{2z \alpha}{2-z}, \qquad w_2=-2\alpha\frac{1-z}{1+z}, \qquad w_3=\frac{2\alpha}{1-2z}
\end{align}
where, $\alpha$ is a large number.
\subsection*{C-O-O vertex}
The C insertion is located at the $z=i$ and O insertions are at $z=\pm \beta$ $(\beta\in [0,1])$. We have restricted $\beta$ to $[0,1]$ because it covers same configurations as $\beta\geq 1$. 

It has one possible degeneration into a C-O and O-O-O vertex sewed to each other through O insertions at $z=0$ in the O-O-O UHP and O-insertion at $z=0$ in C-O UHP. String field theory Feynman rules translate to the sewing relation $ww_1=-q$ with $0\leq q\leq 1$ in terms of local coordinates near the O insertions. $q$ is the sewing parameter (or plumbing fixture variable). In the C-O UHP, the other two O insertions (at $1$ and $\infty$) in the O-O-O vertex go to $\pm q/2\tilde{\lambda}$ (where $\tilde{\lambda}:=\lambda \alpha$). Hence, this degeneration covers the $\beta\in [0,(2\tilde{\lambda})^{-1}]$ region of the moduli space i.e. when the two O insertions in the C-O-O vertex come close to each other (because $\tilde{\lambda}$ is large). 

For the rest of the moduli space, we need to choose local coordinates around the O insertions that match with the local coordinates from the $\beta \in [0,(2\tilde{\lambda})^{-1}]$ at $\beta=(2\tilde{\lambda})^{-1}$. This matching of local coordinates involves an interpolating function $f(\beta)$ which interpolates from (described in \cite{Sen_202183} and \cite{Eniceicu_2022}), 
\begin{align}
    f(\pm (2\tilde{\lambda})^{-1})=\pm \frac{4\tilde{\lambda^2}-3}{8\tilde{\lambda}^2} \text{ to } f(\pm 1)=0  \ .
\end{align}
 First condition ensures the correct behaviour at $\beta=\pm(2\tilde{\lambda})^{-1}$. Second condition is consistent with the $f(\beta)=-f(1/\beta)$. This choice is further simplified by choosing $f(-\beta)=-f(\beta)$. In terms of this function, the local coordinates near the O insertions will be, 
\begin{align}
    w_a=\alpha \tilde{\lambda}\frac{4\tilde{\lambda}^2+1}{4\tilde{\lambda}^2}\frac{z-z_a}{(1+z_az)+\tilde{\lambda}f(z_a)(z-z_a)}, \quad a=1,2
\end{align}
where $z_1=\beta$, $z_2=-\beta$ and $\beta\in [(2\tilde{\lambda})^{-1},1]$.
\subsection*{O-Annulus vertex}
We parametrize the annulus by $0\leq \text{Re} \ w\leq \pi$ and the identification $w\sim w-i\ln v$. O-insertion can be anywhere on the boundary. The moduli space is one-dimensional parametrized by $0\leq v \leq 1$. 

There is one possible degeneration for this vertex into O-O-O vertex with two open string insertions joined by a propagator. Using the UHP coordinates ($z$) of the O-O-O amplitude, sewing relation (sewing the insertions at $z=1$ and $z=\infty$) is, 
    \begin{align}
        -2\alpha\frac{1-z}{1+z}=-q\frac{1-2z}{2\alpha}, \qquad 0\leq q\leq 1 \ .
    \end{align}
    We can define coordinate $\hat{z}$ such that the above identification takes the simple form, 
    \begin{align}
        \hat{z}=u^{-1}(1-\frac{u}{2}+O(u^2))\hat{z}, \qquad u:=\frac{q}{\alpha^2}
    \end{align}
    and $\hat{z}=e^{-iw}$ will be the mapping to the annulus coordinate. \textit{Basically, the identification in the definition of annulus should be same as the identification by sewing because later is a degeneration of the former.} This just means 
    \begin{align}
        w\sim w-i\ln\left(u\left(1-\frac{u}{2}+O(u^2)\right)^{-1}\right)
    \end{align}
    which gives, 
    \begin{align}
        v=u\left(1-\frac{u}{2}\right)^{-1}+\mathcal{O}(u^2)
    \end{align} 
    and the region of moduli space covered by this degeneration is $0\leq v\leq (\alpha^2-\frac{1}{2})^{-1}$ which is simply the small $v$ region.

 The rest of the moduli space is defined as the fundamental O-Annulus vertex.
 \subsection*{C-Annulus vertex}
 We will use the same parametrization of the annulus as before but we now can place the C insertion at $2\pi x$ on $[0,\pi]$ with $x\in [0,\frac{1}{4}]$ where $w=2\pi (x+iy)$. Hence moduli space is now two dimensional parametrized by $x\in [0,\frac{1}{4}]$ and $v\in [0,1]$. This vertex has three possible degenerations coming from small $x$, small $v$ and small $x,v$ regions as discussed in \cite{Sen_202183} and \cite{Eniceicu_2022}. Hence, the moduli space can be divided into four regions as indicated in figure \ref{C_annulus_moduli}. We will now discuss the different regions. 
 
 Let's start with the region $(a)$ (small $x$ and small $v$) region corresponding to the degeneration of the C-annulus vertex into the C-O and O-O-O with two of the O insertions of the O-O-O sewed together and the remaining O insertion sewed with the O insertion on the C-O vertex. Let us further define the plumbing fixture variable $q_2$ corresponding to the sewing of two O insertions on O-O-O vertex and $q_1$ for the other sewing. Moduli parameters are related to these variables as follows,  
 \begin{align}\label{xvq_rel_(a)}
     v= \frac{q_2}{\alpha^2}\left(1-\frac{q_2}{2\alpha^2}\right)^{-1}, \qquad 2\pi x=\frac{q_1}{\tilde{\lambda}}\left(1-\frac{q_2}{\alpha^2}\right) \ .
 \end{align}
 Since $q_1$ and $q_2$ by definition vary in range $[0,1]$, we have the following region in $(x,v)$ plane, 
 \begin{align}
     0\leq v\leq \left(\alpha^2-\frac{1}{2}\right)^{-1}, \qquad 0\leq 2\pi x \leq \tilde{\lambda}^{-1}\frac{2-v}{2+v} \ .
 \end{align}

Region $(b)$ corresponds to the degeneration into O-Annulus and a C-O vertex sewed at the O-insertions using the plumbing fixture variable $q_2$. The region of the moduli space covered is given by, 
 \begin{align}
     \left(\alpha^2-\frac{1}{2}\right)^{-1}\leq v \leq 1, \qquad 0\leq 2\pi x \leq \tilde{\lambda}^{-1}(1-\alpha^{-2}) \ .
 \end{align}
Clearly, this is the small $x$ region and we have the relation with the plumbing fixture, 
\begin{align}
    2\pi x=\frac{q_1}{\tilde{\lambda}}\left(1-\alpha^{-2}\right) \ .
\end{align}

Region $(c)$ corresponds to the degeneration into a C-O-O vertex with the two O-insertions sewed with plumbing fixture $q_2$. The region is parametrized as follows. We first define a variable $u:=\frac{q_2}{\alpha^2}\left(1+\frac{1}{4\tilde{\lambda}^2}\right)^{-2}$. In terms of $u$ and $\beta$, the region $(c)$ corresponds to,
\begin{align}\label{(c)para_1}
    \frac{1}{2\tilde{\lambda}}\leq \beta \leq 1, \qquad 0\leq u \leq \alpha^{-2}\left(1+\frac{1}{4\tilde{\lambda}^2}\right)^{-2},
\end{align}
where, we have the following relation between $(v,x)$ and $(\beta,u)$, 
\begin{align}
    \label{(c)para_2}2\pi x(\beta,u)=&2\tan^{-1}\beta-\frac{u}{\beta\tilde{\lambda}^2}(1-\beta^2-2\beta\tilde{\lambda} f(\beta))\\
    \label{(c)para_3}v(\beta,u)=&\frac{u(1+\beta^2)^2}{4\beta^2\tilde{\lambda}^2}\left(1+\frac{u}{2\beta^2\tilde{\lambda}^2}\left(1-\beta^2-2\beta f(\beta)\tilde{\lambda}\right)^2\right) \ .
\end{align}

The rest of the region in moduli space is region $(d)$. Since the integrand we are interested in will be a total derivative \eqref{ann_1pt_derivation}, we will describe the boundary of the region $(d)$ (see figure \ref{C_annulus_moduli}). We have the following two boundaries at the interfaces with region $(d)$,
\begin{align}
    (b)-(d):& \quad 2\pi x=\tilde{\lambda}^{-1}(1-\alpha^{-2}), \quad \left(\alpha^2-\frac{1}{2}\right)^{-1}\leq v\leq 1, \\
    (c)-(d):& \quad \frac{1}{2\tilde{\lambda}}\leq \beta \leq 1, \quad u=\frac{1}{\alpha^2}\left(1+\frac{1}{4\tilde{\lambda}^2}\right)^{-2}\label{(c)-(d)boundary}
\end{align}
and the other two boundaries, 
\begin{align}
    \text{Top}:& \quad x=\frac{1}{4}, \quad \frac{1}{(\tilde{\lambda}\alpha)^2}\left(1+\frac{1}{4\tilde{\lambda}^2}\right)^{-2}\leq v\leq 1 \text{ and, }\\ 
    \text{Right}:& \quad v=1, \quad \frac{1}{2\pi \tilde{\lambda}}(1-\alpha^{-2})\leq x \leq \frac{1}{4}  \ .
\end{align}
Further, we also have the following replacement rules from string field theory \cite{Sen_202183}, 
\begin{align}\label{replace_rule_plmbf}
    \int_0^1\frac{dq}{q^2}\rightarrow-1, \qquad \int_0^1 \frac{dq}{q}\rightarrow 0 \ .
\end{align}
\subsection*{C-C-O vertex}
We take C insertions to be located at $i$ and $iy$ on the UHP and O insertion to be located at $x$ on the boundary $\mathbb{R}$. Again $y$ is restricted to $[0,1]$ because other configurations are equivalent to these configurations. We will now discuss different degenerations of this vertex.
\begin{itemize}
    \item C-C-O vertex can maximally degenerate into two C-O vertices (with $z$ and $\hat{z}$ UHP coordinates) and an O-O-O vertex (with $\tilde{z}$ UHP coordinates) as shown in figure \ref{fig:CCO_vertex}. The O insertions in O-O-O at $0$ and $\infty$ are respectively sewed to O insertions in C-O vertices with plumbing fixture variables $q_1$ and $q_2$,
    \begin{align}
        \tilde{\lambda}z\frac{2\tilde{z}}{2-\tilde{z}}=-q_1, \quad \tilde{\lambda}\hat{z}\frac{2}{1-2\tilde{z}}=-q_2, \quad 0\leq q_1,q_2\leq 1  \ .
    \end{align}
    Now, we can read the location of insertions in $z$ coordinates, 
    \begin{align}
        z_c^{(1)}=i, \quad z_c^{(2)}=-u_1\frac{3u_2-2i}{2(u_2+2i)}, \quad z_o=-\frac{u_1}{2}
    \end{align}
    where $u_i=q_i/\tilde{\lambda}$ for $i=1,2$. Clearly, both $z_c^{(2)}$ and $z_o$ are very close to the origin as $\tilde{\lambda}$ is large. We can now coordinate transform $z\rightarrow z'=\frac{z-a}{az+1}$ such that $z_c^{(1)}$ stays at $i$ and $z_c^{(2)}$ goes to $iy$. This fixes
    \begin{align}
        a=&\frac{u_1}{2}(1-u_2^2+\mathcal{O}(u_1^4,u_2^4,u_1^2u_2^2)),\\
        y=&u_1u_2\left(1-\frac{u_1^2}{4}-\frac{u_2^2}{4}+\mathcal{O}(u_1^4,u_2^4,u_1^2u_2^2)\right),\label{y_region_I}\\
        x=&-u_1\left(1+\frac{u_1^2}{4}-\frac{u_2^2}{2}+\mathcal{O}(u_1^4,u_2^4,u_1^2u_2^2)\right) \ .\label{x_region_I}
    \end{align}
    Now the ranges $0\leq q_1,q_2\leq 1$, tell us the region of moduli space covered by this degeneration. We will use $(u_1,u_2)$ variables while carrying out the moduli space integral. Hence, this region will be characterized by, $u_i\in [0,\tilde{\lambda}^{-1}]$ for $i=1,2$ and the corresponding replacement rules will take the form, 
    \begin{align}\label{repl_rule_reg_I}
        \int_{0}^{\tilde{\lambda}^{-1}}du_i u_i^{-2}\rightarrow -\tilde{\lambda}, \qquad \int_0^{\tilde{\lambda}^{-1}}du_iu_i^{-1}\rightarrow 0 \ .
    \end{align}
    We will call this as the region I. If we had permuted the two O insertions on the O-O-O vertex before sewing, we would have got the analog of \eqref{y_region_I} and \eqref{x_region_I} as follows, 
    \begin{align}
        y=u_1u_2\left(1-\frac{u_1^2}{4}-\frac{u_2^2}{4}+\mathcal{O}(u_1^4,u_2^4,u_1^2u_2^2)\right), \quad 
        x=u_1\left(1+\frac{u_1^2}{4}-\frac{u_2^2}{2}+\mathcal{O}(u_1^4,u_2^4,u_1^2u_2^2)\right)
    \end{align}
    where we again have the range $u_i\in [0, \tilde{\lambda}^{-1}]$ for $i=1,2$. This corresponds to the region I$'$.
\begin{figure}
    \centering

\tikzset{every picture/.style={line width=0.75pt}} %set default line width to 0.75pt        

\begin{tikzpicture}[x=0.75pt,y=0.75pt,yscale=-0.81,xscale=0.81]
%uncomment if require: \path (0,467); %set diagram left start at 0, and has height of 467

%Curve Lines [id:da5215133711615265] 
\draw    (271.01,156.83) .. controls (302.75,156.83) and (307.43,158.34) .. (307.28,153.04) ;
%Curve Lines [id:da7957460035956498] 
\draw    (271.16,164.71) .. controls (302.9,164.71) and (308.64,163.95) .. (308.94,168.5) ;
%Shape: Ellipse [id:dp42868866191072286] 
\draw   (269.16,160.6) .. controls (269.16,158.34) and (270.06,156.5) .. (271.16,156.5) .. controls (272.27,156.5) and (273.16,158.34) .. (273.16,160.6) .. controls (273.16,162.87) and (272.27,164.71) .. (271.16,164.71) .. controls (270.06,164.71) and (269.16,162.87) .. (269.16,160.6) -- cycle ;
%Shape: Arc [id:dp27826176639963784] 
\draw  [draw opacity=0] (298.61,157.02) .. controls (300.34,146.19) and (309.7,137.91) .. (320.99,137.91) .. controls (331.94,137.91) and (341.07,145.7) .. (343.2,156.05) -- (320.99,160.65) -- cycle ; \draw   (298.61,157.02) .. controls (300.34,146.19) and (309.7,137.91) .. (320.99,137.91) .. controls (331.94,137.91) and (341.07,145.7) .. (343.2,156.05) ;  
%Shape: Arc [id:dp6534879867836862] 
\draw  [draw opacity=0] (318.58,183.25) .. controls (308.32,182.16) and (300.1,174.21) .. (298.58,164.06) -- (320.99,160.65) -- cycle ; \draw   (318.58,183.25) .. controls (308.32,182.16) and (300.1,174.21) .. (298.58,164.06) ;  
%Shape: Arc [id:dp8590688983046328] 
\draw  [draw opacity=0] (343.23,165.05) .. controls (341.32,174.86) and (333.1,182.41) .. (322.98,183.29) -- (320.99,160.65) -- cycle ; \draw   (343.23,165.05) .. controls (341.32,174.86) and (333.1,182.41) .. (322.98,183.29) ;  
%Curve Lines [id:da29853787149366884] 
\draw    (366.09,164.72) .. controls (334.36,164.7) and (329.35,163.2) .. (329.5,168.5) ;
%Curve Lines [id:da9426931809784465] 
\draw    (365.95,156.83) .. controls (334.21,156.82) and (329,157.5) .. (328.17,153.03) ;
%Shape: Ellipse [id:dp433229306627128] 
\draw   (367.94,160.94) .. controls (367.94,163.21) and (367.05,165.04) .. (365.94,165.04) .. controls (364.84,165.04) and (363.94,163.2) .. (363.94,160.94) .. controls (363.95,158.67) and (364.84,156.83) .. (365.95,156.83) .. controls (367.05,156.84) and (367.95,158.67) .. (367.94,160.94) -- cycle ;
%Straight Lines [id:da2215808757704576] 
\draw    (318.58,183.25) -- (318.58,196.25) ;
%Straight Lines [id:da03595853603165167] 
\draw    (322.98,183.29) -- (322.98,196.29) ;
%Curve Lines [id:da023637630566013534] 
\draw    (412.01,95.33) .. controls (443.75,95.33) and (448.43,96.84) .. (448.28,91.54) ;
%Curve Lines [id:da45169977496909963] 
\draw    (412.16,103.21) .. controls (443.9,103.21) and (449.64,102.45) .. (449.94,107) ;
%Shape: Ellipse [id:dp09579472856739102] 
\draw   (410.16,99.1) .. controls (410.16,96.84) and (411.06,95) .. (412.16,95) .. controls (413.27,95) and (414.16,96.84) .. (414.16,99.1) .. controls (414.16,101.37) and (413.27,103.21) .. (412.16,103.21) .. controls (411.06,103.21) and (410.16,101.37) .. (410.16,99.1) -- cycle ;
%Shape: Arc [id:dp026375106485212463] 
\draw  [draw opacity=0] (439.61,95.52) .. controls (441.34,84.69) and (450.7,76.41) .. (461.99,76.41) .. controls (473.47,76.41) and (482.96,84.97) .. (484.45,96.08) -- (461.99,99.15) -- cycle ; \draw   (439.61,95.52) .. controls (441.34,84.69) and (450.7,76.41) .. (461.99,76.41) .. controls (473.47,76.41) and (482.96,84.97) .. (484.45,96.08) ;  
%Shape: Arc [id:dp73737748084078] 
\draw  [draw opacity=0] (459.58,121.75) .. controls (449.32,120.66) and (441.1,112.71) .. (439.58,102.56) -- (461.99,99.15) -- cycle ; \draw   (459.58,121.75) .. controls (449.32,120.66) and (441.1,112.71) .. (439.58,102.56) ;  
%Shape: Arc [id:dp40855882436758195] 
\draw  [draw opacity=0] (484.64,100) .. controls (484.22,111.49) and (475.29,120.8) .. (463.98,121.79) -- (461.99,99.15) -- cycle ; \draw   (484.64,100) .. controls (484.22,111.49) and (475.29,120.8) .. (463.98,121.79) ;  
%Curve Lines [id:da5052166381028689] 
\draw    (577.09,103.22) .. controls (545.36,103.2) and (540.35,101.7) .. (540.5,107) ;
%Curve Lines [id:da4442638868815505] 
\draw    (576.95,95.33) .. controls (545.21,95.32) and (540,96) .. (539.17,91.53) ;
%Shape: Ellipse [id:dp028588767977056007] 
\draw   (578.94,99.44) .. controls (578.94,101.71) and (578.05,103.54) .. (576.94,103.54) .. controls (575.84,103.54) and (574.94,101.7) .. (574.94,99.44) .. controls (574.95,97.17) and (575.84,95.33) .. (576.95,95.33) .. controls (578.05,95.34) and (578.95,97.17) .. (578.94,99.44) -- cycle ;
%Straight Lines [id:da7821182901135217] 
\draw    (459.58,121.75) -- (459.58,134.75) ;
%Straight Lines [id:da8652311647726747] 
\draw    (463.98,121.79) -- (463.98,134.79) ;
%Shape: Arc [id:dp2037352836210271] 
\draw  [draw opacity=0] (509.53,96.01) .. controls (511.06,84.94) and (520.53,76.41) .. (531.99,76.41) .. controls (542.94,76.41) and (552.07,84.2) .. (554.2,94.55) -- (531.99,99.15) -- cycle ; \draw   (509.53,96.01) .. controls (511.06,84.94) and (520.53,76.41) .. (531.99,76.41) .. controls (542.94,76.41) and (552.07,84.2) .. (554.2,94.55) ;  
%Shape: Arc [id:dp2101703695842092] 
\draw  [draw opacity=0] (554.23,103.55) .. controls (552.19,114) and (543.01,121.88) .. (531.99,121.88) .. controls (519.91,121.88) and (510.03,112.4) .. (509.36,100.45) -- (531.99,99.15) -- cycle ; \draw   (554.23,103.55) .. controls (552.19,114) and (543.01,121.88) .. (531.99,121.88) .. controls (519.91,121.88) and (510.03,112.4) .. (509.36,100.45) ;  
%Straight Lines [id:da3702235048196103] 
\draw    (484.64,100) -- (509.36,100.45) ;
%Straight Lines [id:da2121585037262652] 
\draw    (484.82,95.56) -- (509.53,96.01) ;
%Curve Lines [id:da07266855160334362] 
\draw    (58.01,94.83) .. controls (89.75,94.83) and (94.43,96.34) .. (94.28,91.04) ;
%Curve Lines [id:da9039058760561436] 
\draw    (58.16,102.71) .. controls (89.9,102.71) and (95.64,101.95) .. (95.94,106.5) ;
%Shape: Ellipse [id:dp020066373456621767] 
\draw   (56.16,98.6) .. controls (56.16,96.34) and (57.06,94.5) .. (58.16,94.5) .. controls (59.27,94.5) and (60.16,96.34) .. (60.16,98.6) .. controls (60.16,100.87) and (59.27,102.71) .. (58.16,102.71) .. controls (57.06,102.71) and (56.16,100.87) .. (56.16,98.6) -- cycle ;
%Shape: Arc [id:dp7415210453642824] 
\draw  [draw opacity=0] (85.61,95.02) .. controls (87.34,84.19) and (96.7,75.91) .. (107.99,75.91) .. controls (119.47,75.91) and (128.96,84.47) .. (130.45,95.58) -- (107.99,98.65) -- cycle ; \draw   (85.61,95.02) .. controls (87.34,84.19) and (96.7,75.91) .. (107.99,75.91) .. controls (119.47,75.91) and (128.96,84.47) .. (130.45,95.58) ;  
%Shape: Arc [id:dp033113090571936565] 
\draw  [draw opacity=0] (130.64,99.5) .. controls (130.19,111.66) and (120.22,121.38) .. (107.99,121.38) .. controls (96.65,121.38) and (87.26,113.03) .. (85.59,102.13) -- (107.99,98.65) -- cycle ; \draw   (130.64,99.5) .. controls (130.19,111.66) and (120.22,121.38) .. (107.99,121.38) .. controls (96.65,121.38) and (87.26,113.03) .. (85.59,102.13) ;  
%Curve Lines [id:da3458528936978382] 
\draw    (223.09,102.72) .. controls (191.36,102.7) and (186.35,101.2) .. (186.5,106.5) ;
%Curve Lines [id:da28793205007613354] 
\draw    (222.95,94.83) .. controls (191.21,94.82) and (186,95.5) .. (185.17,91.03) ;
%Shape: Ellipse [id:dp8426117629448235] 
\draw   (224.94,98.94) .. controls (224.94,101.21) and (224.05,103.04) .. (222.94,103.04) .. controls (221.84,103.04) and (220.94,101.2) .. (220.94,98.94) .. controls (220.95,96.67) and (221.84,94.83) .. (222.95,94.83) .. controls (224.05,94.84) and (224.95,96.67) .. (224.94,98.94) -- cycle ;
%Straight Lines [id:da9155132120355374] 
\draw    (174.58,121.25) -- (174.58,134.25) ;
%Straight Lines [id:da46464387099067905] 
\draw    (179.49,121.33) -- (179.49,134.33) ;
%Shape: Arc [id:dp9905351853195508] 
\draw  [draw opacity=0] (155.53,95.51) .. controls (157.06,84.44) and (166.53,75.91) .. (177.99,75.91) .. controls (188.94,75.91) and (198.07,83.7) .. (200.2,94.05) -- (177.99,98.65) -- cycle ; \draw   (155.53,95.51) .. controls (157.06,84.44) and (166.53,75.91) .. (177.99,75.91) .. controls (188.94,75.91) and (198.07,83.7) .. (200.2,94.05) ;  
%Shape: Arc [id:dp6059811190944087] 
\draw  [draw opacity=0] (200.23,103.05) .. controls (198.28,113.02) and (189.83,120.65) .. (179.49,121.33) -- (177.99,98.65) -- cycle ; \draw   (200.23,103.05) .. controls (198.28,113.02) and (189.83,120.65) .. (179.49,121.33) ;  
%Straight Lines [id:da7960553977033609] 
\draw    (130.64,99.5) -- (155.36,99.95) ;
%Straight Lines [id:da41687646244030874] 
\draw    (130.82,95.06) -- (155.53,95.51) ;
%Shape: Arc [id:dp32812958307761586] 
\draw  [draw opacity=0] (175.1,121.2) .. controls (164.37,119.83) and (155.98,110.93) .. (155.36,99.95) -- (177.99,98.65) -- cycle ; \draw   (175.1,121.2) .. controls (164.37,119.83) and (155.98,110.93) .. (155.36,99.95) ;  
%Curve Lines [id:da09338006489333206] 
\draw    (197.01,31.83) .. controls (228.75,31.83) and (233.43,33.34) .. (233.28,28.04) ;
%Curve Lines [id:da7472587205394001] 
\draw    (197.16,39.71) .. controls (228.9,39.71) and (234.64,38.95) .. (234.94,43.5) ;
%Shape: Ellipse [id:dp6572431330184474] 
\draw   (195.16,35.6) .. controls (195.16,33.34) and (196.06,31.5) .. (197.16,31.5) .. controls (198.27,31.5) and (199.16,33.34) .. (199.16,35.6) .. controls (199.16,37.87) and (198.27,39.71) .. (197.16,39.71) .. controls (196.06,39.71) and (195.16,37.87) .. (195.16,35.6) -- cycle ;
%Shape: Arc [id:dp6563359196459553] 
\draw  [draw opacity=0] (224.61,32.02) .. controls (226.34,21.19) and (235.7,12.91) .. (246.99,12.91) .. controls (258.47,12.91) and (267.96,21.47) .. (269.45,32.58) -- (246.99,35.65) -- cycle ; \draw   (224.61,32.02) .. controls (226.34,21.19) and (235.7,12.91) .. (246.99,12.91) .. controls (258.47,12.91) and (267.96,21.47) .. (269.45,32.58) ;  
%Shape: Arc [id:dp7875424371319957] 
\draw  [draw opacity=0] (269.64,36.5) .. controls (269.19,48.66) and (259.22,58.38) .. (246.99,58.38) .. controls (235.65,58.38) and (226.26,50.03) .. (224.59,39.13) -- (246.99,35.65) -- cycle ; \draw   (269.64,36.5) .. controls (269.19,48.66) and (259.22,58.38) .. (246.99,58.38) .. controls (235.65,58.38) and (226.26,50.03) .. (224.59,39.13) ;  
%Curve Lines [id:da9143813365467168] 
\draw    (430.09,39.22) .. controls (398.36,39.2) and (393.35,37.7) .. (393.5,43) ;
%Curve Lines [id:da001276538936824334] 
\draw    (429.95,31.33) .. controls (398.21,31.32) and (393,32) .. (392.17,27.53) ;
%Shape: Ellipse [id:dp6334481576606112] 
\draw   (430.94,35.44) .. controls (430.94,37.71) and (430.05,39.54) .. (428.94,39.54) .. controls (427.84,39.54) and (426.94,37.7) .. (426.94,35.44) .. controls (426.95,33.17) and (427.84,31.33) .. (428.95,31.33) .. controls (430.05,31.34) and (430.95,33.17) .. (430.94,35.44) -- cycle ;
%Shape: Arc [id:dp6258816729598013] 
\draw  [draw opacity=0] (362.53,32.01) .. controls (364.06,20.94) and (373.53,12.41) .. (384.99,12.41) .. controls (395.94,12.41) and (405.07,20.2) .. (407.2,30.55) -- (384.99,35.15) -- cycle ; \draw   (362.53,32.01) .. controls (364.06,20.94) and (373.53,12.41) .. (384.99,12.41) .. controls (395.94,12.41) and (405.07,20.2) .. (407.2,30.55) ;  
%Shape: Arc [id:dp33567408844621904] 
\draw  [draw opacity=0] (407.23,39.55) .. controls (405.19,50) and (396.01,57.88) .. (384.99,57.88) .. controls (372.91,57.88) and (363.03,48.4) .. (362.36,36.45) -- (384.99,35.15) -- cycle ; \draw   (407.23,39.55) .. controls (405.19,50) and (396.01,57.88) .. (384.99,57.88) .. controls (372.91,57.88) and (363.03,48.4) .. (362.36,36.45) ;  
%Shape: Arc [id:dp8257476525303555] 
\draw  [draw opacity=0] (293.38,33.03) .. controls (294.2,21.23) and (304.01,11.91) .. (315.99,11.91) .. controls (327.47,11.91) and (336.96,20.47) .. (338.45,31.58) -- (315.99,34.65) -- cycle ; \draw   (293.38,33.03) .. controls (294.2,21.23) and (304.01,11.91) .. (315.99,11.91) .. controls (327.47,11.91) and (336.96,20.47) .. (338.45,31.58) ;  
%Shape: Arc [id:dp8255694972616394] 
\draw  [draw opacity=0] (339.6,35.86) .. controls (339.18,47.17) and (330.53,56.36) .. (319.48,57.59) -- (316.94,35) -- cycle ; \draw   (339.6,35.86) .. controls (339.18,47.17) and (330.53,56.36) .. (319.48,57.59) ;  
%Straight Lines [id:da2840231682752057] 
\draw    (269.64,36.5) -- (294.36,36.95) ;
%Straight Lines [id:da06831984502387911] 
\draw    (269.45,32.58) -- (294.17,33.02) ;
%Straight Lines [id:da8769415186233838] 
\draw    (338.64,35.5) -- (363.36,35.95) ;
%Straight Lines [id:da09391415998935693] 
\draw    (338.45,31.58) -- (363.17,32.02) ;
%Shape: Arc [id:dp07433332556427552] 
\draw  [draw opacity=0] (315.56,57.69) .. controls (304.33,57.01) and (295.3,48.14) .. (294.36,36.95) -- (316.94,35) -- cycle ; \draw   (315.56,57.69) .. controls (304.33,57.01) and (295.3,48.14) .. (294.36,36.95) ;  
%Straight Lines [id:da9172393743582921] 
\draw    (315.58,57.75) -- (315.58,70.75) ;
%Straight Lines [id:da07489487347716195] 
\draw    (319.48,57.59) -- (319.48,70.59) ;
%Straight Lines [id:da2858926709161991] 
\draw    (11,250.5) -- (649.5,250) ;
%Straight Lines [id:da9971209923599389] 
\draw    (10.5,371) -- (649,370.5) ;
%Straight Lines [id:da17084794850601104] 
\draw    (10.5,430) -- (649,429.5) ;
%Straight Lines [id:da34711107052621104] 
\draw    (329.75,370.75) -- (329.75,429.75) ;
%Straight Lines [id:da02864080441187311] 
\draw    (450.25,370.75) -- (450.25,429.75) ;
%Straight Lines [id:da5848209409108025] 
\draw    (209.75,370.75) -- (209.75,429.75) ;
%Straight Lines [id:da26807387802702265] 
\draw    (329.75,429.75) -- (450.25,370.75) ;
%Straight Lines [id:da9068052326510332] 
\draw    (209.75,370.75) -- (329.75,429.75) ;
%Right Arrow [id:dp9782161946034722] 
\draw  [fill={rgb, 255:red, 208; green, 2; blue, 27 }  ,fill opacity=0.5 ] (132.81,63.62) -- (160.62,47.14) -- (159.52,45.28) -- (166.75,46) -- (163.89,52.69) -- (162.8,50.84) -- (134.99,67.32) -- cycle ;
%Right Arrow [id:dp9192897432844434] 
\draw  [fill={rgb, 255:red, 208; green, 2; blue, 27 }  ,fill opacity=0.5 ] (510.23,70.02) -- (481.61,54.98) -- (480.6,56.88) -- (477.42,50.35) -- (484.62,49.28) -- (483.62,51.18) -- (512.24,66.22) -- cycle ;
%Right Arrow [id:dp6225865202994829] 
\draw  [fill={rgb, 255:red, 208; green, 2; blue, 27 }  ,fill opacity=0.5 ] (242.69,167.77) -- (214.83,151.37) -- (213.73,153.22) -- (210.87,146.54) -- (218.12,145.82) -- (217.02,147.67) -- (244.88,164.07) -- cycle ;
%Right Arrow [id:dp4501780569519438] 
\draw  [fill={rgb, 255:red, 208; green, 2; blue, 27 }  ,fill opacity=0.5 ] (391.18,165.32) -- (419.26,149.31) -- (418.2,147.44) -- (425.41,148.28) -- (422.44,154.92) -- (421.38,153.05) -- (393.3,169.06) -- cycle ;

% Text Node
\draw (224.5,396.5) node [anchor=north west][inner sep=0.75pt]   [font=\normalsize]  {I};
% Text Node
\draw (428.5,396.5) node [anchor=north west][inner sep=0.75pt]   [font=\normalsize]  {I$'$};
% Text Node
\draw (296.5,376) node [anchor=north west][inner sep=0.75pt]   [font=\normalsize]  {II$'$};
% Text Node
\draw (352,376) node [anchor=north west][inner sep=0.75pt]   [font=\normalsize]  {II};
% Text Node
\draw (100.5,390) node [anchor=north west][inner sep=0.75pt]   [font=\normalsize]  {III};
% Text Node
\draw (544.5,390.5) node [anchor=north west][inner sep=0.75pt]   [font=\normalsize]  {III$'$};
% Text Node
\draw (323.5,288.5) node [anchor=north west][inner sep=0.75pt]   [font=\normalsize]  {IV};
% Text Node
\draw (432.5,431.9) node [anchor=north west][inner sep=0.75pt]  [font=\scriptsize]  {$x\simeq 1/\tilde{\lambda }$};
% Text Node
\draw (186.5,432.9) node [anchor=north west][inner sep=0.75pt]  [font=\scriptsize]  {$x\simeq -1/\tilde{\lambda }$};
% Text Node
\draw (314.5,434.9) node [anchor=north west][inner sep=0.75pt]  [font=\scriptsize]  {$x=0$};
% Text Node
\draw (602,342.4) node [anchor=north west][inner sep=0.75pt]  [font=\scriptsize]  {$y\simeq 1/\tilde{\lambda }^{2}$};
% Text Node
\draw (617,229.4) node [anchor=north west][inner sep=0.75pt]  [font=\scriptsize]  {$y\simeq 1$};
% Text Node
\draw (310.5,199) node [anchor=north west][inner sep=0.75pt]   [font=\normalsize]  {IV};
% Text Node
\draw (115.5,140) node [anchor=north west][inner sep=0.75pt]   [font=\normalsize]  {II and II$'$};
% Text Node
\draw (471,134.5) node [anchor=north west][inner sep=0.75pt]   [font=\normalsize]  {III and III$'$};
% Text Node
\draw (292,74) node [anchor=north west][inner sep=0.75pt]   [font=\normalsize]  {I and I$'$};

\end{tikzpicture}

    \caption{Different regions in the moduli space for the C-C-O vertex and the corresponding degenerations of the worldsheet (figure reproduced from \cite{Sen_202183})}
    \label{fig:CCO_vertex}
\end{figure}
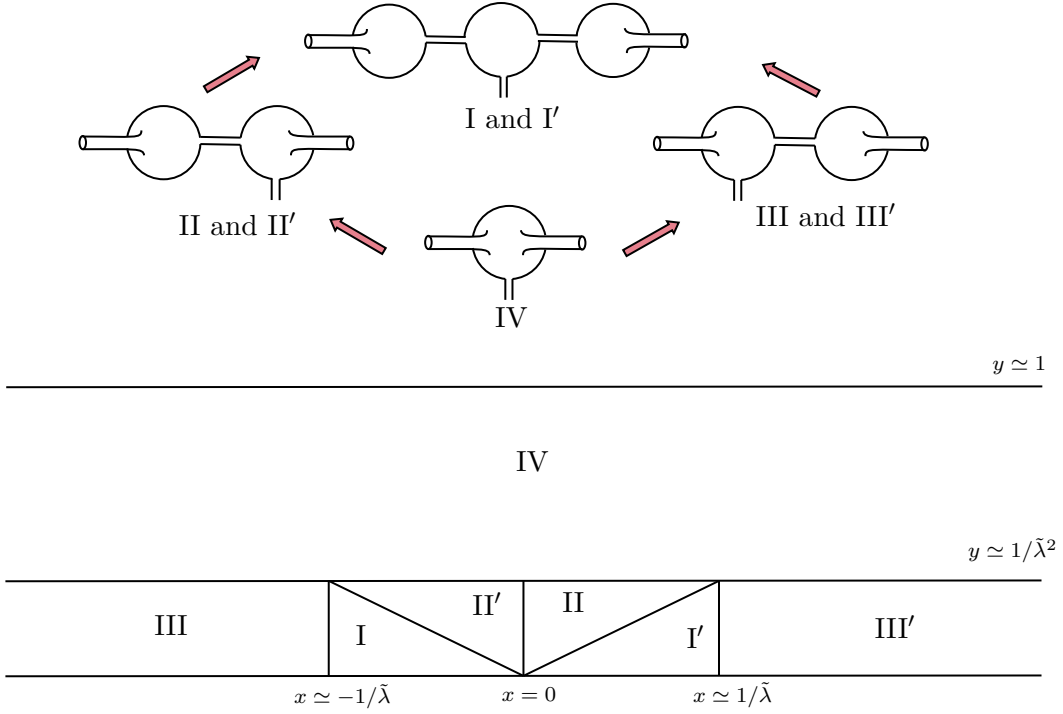
    \item Next, we can consider the degeneration into one C-O vertex ($z$ coordinate in UHP) and one C-O-O vertex ($\hat{z}$ coordinate in UHP) with O insertions at $z=0$ and $\hat{z}=-\beta$ sewed with plumbing fixture $q_1$,
    \begin{align}
        \alpha\tilde{\lambda}\frac{4\tilde{\lambda}^2+1}{4\tilde{\lambda}^2}\frac{\hat{z}+\beta}{(1-\beta\hat{z})+\tilde{\lambda}f(-\beta)(\hat{z}+\beta)}\lambda z=-q_1 \ .
    \end{align}
    This particular sewing will correspond to the region II$'$. We can also sew the O-insertion at $\hat{z}=\beta$ which will correspond to region II. The $(x,y)$ in terms of the plumbing fixture variable will be given by, \begin{align}\label{(x,y)RegionII}
        y=\frac{u}{\tilde{\lambda}}(1-u^2f(\beta)^2), \quad x=\mp\frac{u}{\tilde{\lambda}}\frac{1-\beta^2}{2\beta}\left(1-u^2f(\beta)^2+u^2f(\beta)\frac{1-\beta^2}{2\beta \tilde{\lambda}}\right)
    \end{align}
    where the upper (lower) sign corresponds to the region II$'$ (II) and 
    \begin{align}\label{u_beta_range}
        u:=\frac{4\tilde\lambda q_1}{4\tilde{\lambda}^2+1}\Rightarrow 0\leq u \leq \frac{4\tilde\lambda}{4\tilde{\lambda}^2+1}\text{ and, }(2\tilde{\lambda})^{-1}\leq \beta \leq 1\ .
    \end{align}
    Finally, the replacement rule in terms of $u$ will look like, 
    \begin{align}\label{repl_rule_regionII}
        \int_{0}^{\frac{4\tilde\lambda}{4\tilde{\lambda}^2+1}}du u^{-2}\rightarrow -\frac{4\tilde{\lambda}^2+1}{4\tilde\lambda}, \qquad \int_{0}^{\frac{4\tilde\lambda}{4\tilde{\lambda}^2+1}}du u^{-1}\rightarrow 0 \ .
    \end{align}

    \item Next degeneration corresponds to the exchange of the two closed string punctures in the previous degeneration (figure \ref{fig:CCO_vertex}). Hence, we can write the analog of \eqref{(x,y)RegionII}, 
    \begin{align}\label{(x,y)_regionIII}
        y=\frac{u}{\tilde{\lambda}}(1-u^2f(\beta)^2), \quad x=\pm\frac{2\beta}{1-\beta^2}\left(1-u^2f(\beta)\frac{1-\beta^2}{2\beta\tilde{\lambda}}\right)
    \end{align}
    where upper (lower) sign corresponds to the III$'$ (III). $u$ and $\beta$ have same range as in \eqref{u_beta_range} and replacement rules are also same as \eqref{repl_rule_regionII}.

    \item Next, we have the bulk of the moduli space characterized as follows, 
    \begin{align}
        \text{Region IV}:& -\infty \leq x \leq \infty, \quad \tilde{\lambda}^{-2}\left(1-\Delta(x)\tilde{\lambda}^{-2}\right)\leq y \leq 1\\
        \text{where  }&\Delta(x):=F(x)^2+\frac{1}{4}
    \end{align}
    and $F(x)$ is defined as follows, 
    \begin{align}
        F(x):=&f(\beta) \text{ for }0\leq |x|\leq \tilde{\lambda}^{-1}\left(1-\frac{1}{4\tilde{\lambda}^2}\right),\nonumber\\
        \text{ where }x=& \pm\frac{1}{\tilde{\lambda}}\frac{1-\beta^2}{2\beta}\left(1-\frac{\tilde{\lambda}^{-2}}{4}-\tilde{\lambda}^{-2}f(\beta)^2+\tilde{\lambda}^{-2}f(\beta)\frac{1-\beta^2}{2\beta}\right).\\
        F(x):=&f(\beta) \text{ for }\tilde{\lambda}^{-1}\left(1-\frac{1}{4\tilde{\lambda}^2}\right)\leq |x|\leq \infty,\nonumber\\
        \text{ where }x=& \pm\frac{2\beta}{1-\beta^2}\left(1-\tilde{\lambda}^{-2}f(\beta)\frac{1-\beta^2}{2\beta\tilde{\lambda}}\right) \ . 
    \end{align}
\end{itemize}

\section{\texorpdfstring{$g^{(a)}, g^{(c)}$}{TEXT} and, \texorpdfstring{$g^{(c)-(d)}$}{TEXT} calculation}\label{appendix_ws_calc}
In this section, we will evaluate the moduli space integrals for the relevant regions. Using \eqref{ann_1pt_derivation}, we have the contribution from region $(c)$,
\begin{align}
    g_sg^{(c)}=\int_{(c)} dv dxF(v,x) \ .
\end{align}
Since, region $(c)$ correspond to the small $v$, we can use the results quoted in \eqref{F(v,x)limits} to get the following,
\begin{align}
    g_sg^{(c)}=g_s\int_{(c)}\frac{dv dx}{\sin^2(2\pi x)}(v^{-2}+v^{-1}+\mathcal{O}(1)) \ . 
\end{align}
Using the parametrization of the region $(c)$ described earlier in \eqref{(c)para_1}, \eqref{(c)para_2} and \eqref{(c)para_3}, we can write this integral in terms of the $u$ and $\beta$ variables.
\begin{align}
   g^{(c)}=\int_{(2\tilde{\lambda})^{-1}}^{1}d\beta\int_0^{\alpha^{-2}(1+(4\tilde{\lambda}^2)^{-1})^{-2}}du\left(\frac{1+\beta^2}{4\pi \beta^2\tilde{\lambda}^2}+\mathcal{O}(u)\right)\frac{1}{\sin^2{(2\pi x)}}(v^{-2}+v^{-1}+\mathcal{O}(1)) \ .
\end{align}
Now, we use the definition $u=q_2\alpha^{-2}\left(1+\frac{1}{4\tilde{\lambda}^2}\right)^{-2}$ to write the integral in terms of $q_2$. Since we do not have any $\alpha$ dependence in the integration limits, we can do a large $\alpha$ expansion to get rid of the terms with negative powers of $\alpha$. Then we use the replacement rules \eqref{replace_rule_plmbf} to get,
\begin{align}
 g^{(c)}\approx-\frac{\alpha^2\tilde{\lambda}^2}{\pi}\left(1+\frac{1}{4\tilde{\lambda}^2}\right)^{2}\int_{(2\tilde{\lambda})^{-1}}^1\frac{d\beta}{1+\beta^2}\approx  -\frac{\alpha ^2 \tilde{\lambda}^2}{4}+\frac{\alpha ^2 \tilde{\lambda} }{2 \pi }-\frac{\alpha ^2}{8} 
\end{align}
where in last step we have finally done the large $\tilde{\lambda}$ expansion and ignored the negative powers of $\tilde{\lambda}$. 

Next, we go on to calculate the contribution from the region $(a)$. Since this region corresponds to the small $x$ and $v$, we have using \eqref{F(v,x)limits},  
\begin{align}
    F(v,x)\approx \frac{g_s}{4\pi^2 x^2}(v^{-2}+v^{-1}+\mathcal{O}(1)) \ .
\end{align}
Now we use the definition of $(x,v)$ in terms of the plumbing fixture variables $(q_1,q_2)$ \eqref{xvq_rel_(a)} and then use the replacement rules \eqref{replace_rule_plmbf} to get the following result,
\begin{align}
g^{(a)}=\frac{\tilde{\lambda}\alpha^2}{2\pi} \ .
\end{align}
Next, we look at $g^{(c)-(d)}$. This falls in the small $v$ region again and hence, we can use \eqref{G(v,x)_limits} to get the following, 
\begin{align}
    g^{(c)-(d)}=&-\frac{1}{2\pi}\int dv (v^{-2}+v^{-1})\cot(2\pi x)\nonumber\\
    =&\frac{1}{2\pi}\int_{(2\tilde{\lambda})^{-1}}^1 d\beta \frac{\partial}{\partial \beta}(v(\beta)^{-1}+2\ln v(\beta))\cot(2\pi x(\beta)) 
\end{align}
where we have used the fact that the boundary is defined by the \eqref{(c)-(d)boundary} and $x(\beta), v(\beta)$ are defined by setting $u=\frac{1}{\alpha^{2}}(1+\frac{1}{4\tilde{\lambda}^2})^{-2}$ in the definitions \eqref{(c)para_2} and \eqref{(c)para_3}. Carrying out the $\beta$ integral and again getting rid of the terms with negative powers of $\alpha$ and $\tilde{\lambda}$, we get, 
\begin{align}
    g^{(c)-(d)}=\frac{\alpha^2\tilde{\lambda}^2}{4}-\frac{\alpha^2\tilde{\lambda}}{\pi}+\frac{\alpha^2}{8}+\frac{9\tilde{\lambda}}{4\pi}-1-\frac{2\tilde{\lambda}^2}{\pi}\int_{(2\tilde{\lambda})^{-1}}^1d\beta\frac{(f(\beta))^2}{\beta^2+1} \ .
\end{align}
\section{\texorpdfstring{$g_{\text{jac}}$}{TEXT} calculation using SFT}\label{Appendix_A_gjac_calc}
In this appendix, we will calculate the contribution to the $g$ due to the collective mode field redefinition. This calculation was already done for the non-compact case in the appendix of \cite{Sen_202183} for non-zero momentum insertion. We will simply specialize to the case of zero momentum insertions along with the fact that the momentum is discrete now due to compactification. It will serve as a non-trivial sanity check because we would naively not expect a correlator of zero-momentum insertions(i.e. insertions not charged under the collective mode), to get corrections due to the collective mode redefinition. More details are in \cite{Sen_202183} and here, we will just produce the steps of the calculation that are changed in this case due to the zero-momentum insertion.

We will consider the C-C-O amplitude with the $\omega_1$ momentum insertion at $i$, zero momentum insertion at $iy$, and an open string insertion corresponding to $c\partial X$ (the collective mode) at $x$ on the boundary $\mathbb{R}$. The divergent part of the amplitude is formally given by, 
\begin{align}
\int_{-\infty}^{\infty}dx\int_0^1dy\frac{1}{2\pi}\left(-\frac{\omega_1}{x-i}+\frac{\omega_1}{x+i}\right)4g_s\left(\frac{2\sinh{\pi \omega_1}}{\pi \omega_1}\times 2\right)\left(\frac{y^{-2}}{2}\right)\nonumber\\
=-\frac{8ig_s}{\pi^2}\sinh{\pi \omega_1}\int_{-\infty}^{\infty}dx\int_{0}^1dy\frac{y^{-2}}{x^2+1}
\end{align}
where we have the disk one-point function normalized as follows,  
\begin{align}
    A_{\text{disk}}(\psi_c)=\frac{2\sinh{\pi \omega}}{\pi \omega}
\end{align}
so that it reduces to \eqref{Disk_1_pt_zeromom} for $\omega\rightarrow 0$ limit.
We will calculate the contribution from all the six regions of the moduli space characterized in appendix \ref{Vertices}. 
\begin{itemize}
    \item Region IV is characterized by $\tilde{\lambda}^{-2}-\tilde{\lambda}^{-4}\Delta(x)\leq y\leq 1$ where $\Delta(x)=F(x)^2+\frac{1}{4}$. This region can be further divided into two parts at the line $y=\tilde{\lambda}^{-2}$. We call the contribution from the region above and below this line respectively, $I_{\text{IV},1}$ and $I_{\text{IV},2}$. $I_{\text{IV},2}$ can be approximated simply by taking $y^{-2}=\tilde{\lambda}^{4}$ over the thin strip. We get,
    \begin{align}
        I_{\text{IV},1}=&-\frac{8ig_s}{\pi^2}\sinh{\pi \omega_1}\int_{-\infty}^{\infty}dx\int_{\tilde{\lambda}^{-2}}^1dy\frac{y^{-2}}{x^2+1}=-\frac{8ig_s}{\pi}\sinh{\pi \omega_1}(\tilde{\lambda}^2-1)\label{I1}\\
        I_{\text{IV},2}=&-\frac{8ig_s}{\pi^2}\sinh{\pi \omega_1}\int_{-\infty}^{\infty}dx\frac{\Delta(x)}{x^2+1}=-\frac{32ig_s}{\pi^2}\sinh{\pi \omega_1}\int_{\frac{1}{2\tilde{\lambda}}}^1\frac{d\beta}{1+\beta^2}\left(f(\beta)^2+\frac{1}{4}\right)\label{I2}
    \end{align}
    where we have used the fact that the width is $\Delta(x)\tilde{\lambda}^{-4}$ for the thin strip over which the $y$ integral is carried out in second line.  
    \item Regions III and III$'$ are characterized by \eqref{(x,y)_regionIII} and \eqref{u_beta_range}. It will be more convenient for us to introduce 
    \begin{align}\label{xidefn}
        \xi=\pm \frac{2\beta}{1-\beta^2}\text{ and }\phi(\xi):=f(\beta),
    \end{align}
    and work with $(u, \xi)$ instead of $(u,\beta)$. With this, \eqref{(x,y)_regionIII} will become
    \begin{align}
        y=\frac{u}{\tilde{\lambda}}(1+u^{2}\phi(\xi)^2)^{-1}, \qquad x=\xi-u^2\tilde{\lambda}^{-1}\phi(\xi)\label{x,y_u,xirelation}\\
        \Rightarrow dx\wedge dy=\tilde{\lambda}^{-1}d\xi\wedge du(1-3u^2\phi(\xi)^2),\label{measurechangeIII}
    \end{align}
    where we have kept terms up to leading order in $\tilde{\lambda}^{-1}$. Using \eqref{x,y_u,xirelation} and \eqref{measurechangeIII}, we get,
    \begin{align}
        I_{\text{III+III}'}=&-\frac{8ig_s}{\pi^2}\sinh{\pi \omega_1}\int_{\text{III+III'}}dxdy\frac{y^{-2}}{x^2+1}\nonumber\\
        =&-\frac{8ig_s}{\pi^2}\sinh{\pi \omega_1}\int_{\text{III+III'}}d\xi du\tilde{\lambda}^{-1}(1-3u^2\phi(\xi)^2)\frac{\tilde{\lambda}^2}{u^2}\frac{(1+u^2\phi(\xi))^2}{1+(\xi-u^2\tilde{\lambda}^{-1}\phi(\xi))^2}\nonumber\\
        =&\frac{8ig_s}{\pi^2}\sinh{\pi \omega_1}\int_{|\xi|\geq\frac{4\tilde{\lambda}}{4\tilde{\lambda}^2-1}}^{\infty}\frac{d\xi}{\xi^2+1}\left(\frac{4\tilde{\lambda}^2+1}{4}+ \phi(\xi) ^2\right),
    \end{align}
    where we have used the replacement rule \eqref{repl_rule_regionII} and ignored subleading terms in $\tilde{\lambda}^{-1}$. Now, changing the variable back to the $\beta$, we get,
    \begin{align}\label{I3}
        I_{\text{III+III}'}=\frac{32ig_s}{\pi^2}\sinh{\pi \omega_1}\int_{\frac{1}{2\tilde{\lambda}}}^{1}\frac{d\beta}{\beta^2+1}\left(\frac{4\tilde{\lambda}^2+1}{4}+ f(\beta) ^2\right) \ .
    \end{align}
    \item We can now consider the region II. Using the fact that region II (II$'$) is related to the region III (III$'$) by the coordinate transformation $y\rightarrow y$ and $x\rightarrow\tilde{x}=-y/x$,
    \begin{align}
        I_{\text{II+II}'}=&-\frac{8ig_s}{\pi^2}\sinh{\pi \omega_1}\int_{\text{II+II}'}dxdy\frac{y^{-2}}{x^2+1}\nonumber\\
        =&-\frac{8ig_s}{\pi^2}\sinh{\pi \omega_1}\int_{\text{III+III}'}d\tilde{x}dy\frac{y^{-1}}{\tilde{x}^2+y^2} \ .
    \end{align}
    We can again use the $\xi$ variable defined in \eqref{xidefn} and use \eqref{x,y_u,xirelation} (with $\tilde{x}$ instead of $x$) to get the, 
    \begin{align}
        I_{\text{II+II}'}=&-\frac{8ig_s}{\pi^2}\sinh{\pi \omega_1}\int_{\text{III+III}'}d\xi du\frac{\tilde{\lambda}^2\left(1-3 u^2 \phi (\xi )^2\right) \left(u^2 \phi (\xi )^2+1\right)}{u \left(\frac{u^2}{\left(1 +  u^2 \phi (\xi )^2\right)^2}+\left(\xi\tilde{\lambda} -u^2 \phi (\xi )\right)^2\right)} \ .
    \end{align}
    We can expand the integrand around $u=0$, the leading piece will be $u^{-1}$ which will simply vanish upon using the replacement rule \eqref{repl_rule_regionII}. All the subleading pieces will have negative powers of $\tilde{\lambda}$ and hence, have to be dropped. Therefore, 
    \begin{align}\label{I4}
        I_{\text{II+II}'}=0 \ .
    \end{align}
    \item We finally consider the region I. We will work with the $(u_1,u_2)$ variable. Using \eqref{y_region_I} and \eqref{x_region_I}, we get the change of measure,  
    \begin{align}
        dxdy=du_1du_2u_1\left(1+\frac{u_1^2}{2}-\frac{u_2^2}{4}\right) \ .
    \end{align}
    Hence, the integral becomes,
    \begin{align}
        I_{\text{I}+\text{I}'}=&-\frac{8ig_s}{\pi^2}\sinh{\pi \omega_1}\int_{\text{I+I}'}dxdy\frac{y^{-2}}{x^2+1}\nonumber\\
        =&-\frac{16ig_s}{\pi^2}\sinh{\pi \omega_1}\int_{0}^{\tilde{\lambda}^{-1}}\int_{0}^{\tilde{\lambda}^{-1}}du_1du_2u_1\left(1+\frac{u_1^2}{2}-\frac{u_2^2}{4}\right)\frac{y(u_1,u_2)^{-2}}{1+(x(u_1,u_2))^2}
    \end{align}
    where we have used $x\rightarrow-x$ symmetry of the integrand in the second step.
    The integrand can be expanded in small $u_1,u_2$, 
    \begin{align}
        \frac{1}{u_2^2}\left(\frac{1}{u_1}-\frac{u_1^3}{16}+...\right)+\left(\frac{1}{4u_1}+\frac{5u_1}{4}+...\right)+u_2^2\left(\frac{1}{16 u_1}+\frac{u_1}{8}+...\right).
    \end{align}
    Using the replacement rule \eqref{repl_rule_reg_I}, we can see that the leading contributions to this integral have negative powers of $\tilde{\lambda}$ and hence, we can set, 
    \begin{align}\label{I5}
        I_{\text{I+I}'}=0 \ .
    \end{align}
\end{itemize}
Adding \eqref{I1}, \eqref{I2}, \eqref{I3}, \eqref{I4} and \eqref{I5}, we get, 
\begin{align}
    -\frac{8ig_s}{\pi}\sinh{\pi \omega_1}(\tilde{\lambda}^2-1)+\frac{32ig_s}{\pi^2}\sinh{\pi \omega_1}\tilde{\lambda}^2\left(-\frac{1}{2 \tilde{\lambda} }+\frac{\pi }{4}\right)=\frac{8ig_s}{\pi}\sinh{\pi \omega_1}\left(1-\frac{2\tilde{\lambda}}{\pi}\right) \ .
\end{align}
The desired result for the coupling of the translation zero mode (or collective mode) $\phi$ is $-i\omega_1$ times the divergent part of the disc two-point amplitude with one $\omega_1$ and one zero momentum closed string insertion, 
\begin{align}
    -i\omega_1\times4g_s\left(\frac{2\sinh{\pi \omega_1}}{\pi \omega_1}\times 2\right)\times\left(-\frac{1}{2}\right)=\frac{8ig_s}{\pi}\sinh{\pi \omega_1}
\end{align}
where $-1/2$ is the divergent part of the disc two-point amplitude read from (5.7) of \cite{Sen_202183} (setting $\omega_2=0$). This means the difference is, 
\begin{align}
    -2ig_s\omega_1\frac{2\sinh{\pi \omega_1}}{\pi \omega_1}\times 2\times \frac{2\tilde{\lambda}}{\pi} \ .
\end{align}
The extra term in effective action that can account for this difference is of the form, 
\begin{align}\label{redef:intermediate_step-1}
    S_{\text{extra}}=-i\frac{g_s\phi}{2}\sum_{\omega_1\in \frac{1}{R}\mathbb{Z}}\int dP_1 dP_2\omega_1\frac{2\sinh{\pi \omega_1}}{\pi \omega_1}\times 2\times \frac{2\tilde{\lambda}}{\pi}\times \Phi_C(\omega_1,P_1)\Phi_C(0,P_2) \ .
\end{align}
Now, the disk one-point function with zero momentum insertion in our normalization is,
\begin{align}
    \frac{2\sinh \pi \omega}{\pi \omega}
\end{align}
and the CO amplitude with $\omega$ O insertion would be simply,
\begin{align}
    (-i \omega)\frac{2\sinh \pi \omega}{\pi \omega}=-i\frac{2\sinh \pi \omega}{\pi}
\end{align}
The corresponding term in the effective action will be of the form, 
\begin{align}\label{redef:intermediate_step-2}
    -i \phi\sum_{\omega\in \frac{\mathbb{Z}}{R}}\int dP\frac{2\sinh{\pi \omega}}{\pi} \Phi_C(\omega, P)
\end{align}
We now do a redefinition of the open string field $\phi$, 
\begin{align}
    \phi=\tilde{\phi}+4g_sA\tilde{\phi}\int dP'\Phi_C(0,P')
\end{align}
Putting this in \eqref{redef:intermediate_step-2}, we get, 
\begin{align}
    -i \tilde{\phi}\sum_{\omega\in \frac{\mathbb{Z}}{R}}\int dP\frac{2\sinh{\pi \omega}}{\pi} \Phi_C(\omega, P)-4ig_s A \tilde{\phi}\sum_{\omega\in \frac{\mathbb{Z}}{R}}\int dP'dP\frac{2\sinh{\pi \omega}}{\pi} \Phi_C(\omega, P)\Phi_C(0,P')
\end{align}
The extra term here should cancel the $S_{\text{extra}}$ we got in the \eqref{redef:intermediate_step-1} (with $\phi$ replaced with $\tilde{\phi}$) i.e., 
\begin{align}
    -8ig_s A \sum_{\omega_1\in \frac{\mathbb{Z}}{R}}\phi\int dP_2dP_1\frac{\sinh{\pi \omega_1}}{\pi} \Phi_C(\omega_1, P_1)\Phi_C(0,P_2)\nonumber\\
    -\frac{4ig_s\phi\tilde{\lambda}}{\pi}\sum_{\omega_1\in \frac{\mathbb{Z}}{R}}\int dP_1 dP_2\frac{\sinh{\pi \omega_1}}{\pi} \Phi_C(\omega_1,P_1)\Phi_C(0,P_2)=0
\end{align}
which just means that,
\begin{align}
    A=-\frac{\tilde{\lambda}}{2\pi}
\end{align}
Hence, the redefinition takes the following form, 
\begin{align}
    \phi=\tilde{\phi}\left(1-\frac{2g_s\tilde{\lambda}}{\pi}\int dP \Phi_C(0,P)\right)=\tilde{\phi}\exp\left(-\frac{2g_s\tilde{\lambda}}{\pi}\int dP \Phi_C(0,P)+\mathcal{O}(g_s^2)\right)
\end{align}
The jacobian associated with this field redefinition leads to the following term in the effective action, 
\begin{align}
    -\frac{g_s\tilde{\lambda}}{\pi}\int dP\times  2\times \Phi_C(0,P).
\end{align}
This means \textit{the zero momenta closed string one-point function will receive a contribution of the form},
\begin{align}
    -\frac{2g_s\tilde{\lambda}}{\pi}
\end{align}
Since this is proportional to $g_s$, this implies that the ratio of the annulus one-point function to the disk one-point function (which equals $2$ for the zero momentum insertion) will correspondingly get the following contribution,
\begin{align}\label{eq_gjac_1_result}
    g_{\text{jac}_1}=-\frac{\tilde{\lambda}}{\pi}
\end{align}
So far we have carried out the analysis for general $R$ but the result \eqref{eq_gjac_1_result} has finite $R\rightarrow 1$ limit and hence, can be used for the $R=1$ case. The same analysis can be carried out for the other two zero modes that appear at $R=1$ ($ce^{2iX}$ and $ce^{-2iX}$). We just have to choose the appropriate insertion at $i$ which is charged under these extra two generators. The answer would be the same using the symmetric nature of the $S^3$ (i.e. all directions are equivalent), this gives the total contribution due to Jacobian as follows,
\begin{align}
    g_{\text{jac}}=-\frac{3\tilde{\lambda}}{\pi}
\end{align}

\section{\texorpdfstring{$R\rightarrow1$}{TEXT} limit}\label{appendix_np_freeenergy}
Here, we elaborate a bit on the $R\rightarrow 1$ limit of the \eqref{freeenergyatR} and \eqref{npfreenergyatR}. Simply setting $R=1$ in \eqref{freeenergyatR}, we get the \eqref{freeenergyatR=1} and further extracting the imaginary part, we get, 
\begin{align}
    -i\text{Im}\int_{\frac{1}{\Lambda}}^{\infty}\frac{dt}{t}\frac{e^{i\mu t}}{4\sinh^2{\frac{t}{2}}}=&-\int_{\frac{1}{\Lambda}}^{\infty}\frac{dt}{t}\frac{e^{i\mu t}-e^{-i\mu t}}{8\sinh^2{\frac{t}{2}}}\nonumber\\
    =&-\bigg[\int_{-\infty}^{-\frac{1}{\Lambda}}+\int_{\frac{1}{\Lambda}}^{\infty}\bigg]\frac{dt}{t}\frac{e^{i\mu t}}{8\sinh^2{\frac{t}{2}}}\ .
\end{align}
Now, we can add the semicircle in UHP around the origin to join the contours and get a single contour $\tilde{\mathcal{C}}$. In the process, we have to subtract half the residue at the origin which will be just a polynomial in $\mu$ and won't affect the final non-perturbative contribution $\mathcal{F}_{\text{np}}$ we are interested in,
\begin{align}
    -i\text{Im}\int_{\frac{1}{\Lambda}}^{\infty}\frac{dt}{t}\frac{e^{i\mu t}}{4\sinh^2{\frac{t}{2}}}=&-\int_{\tilde{\mathcal{C}}}\frac{dt}{t}\frac{e^{i\mu t}}{8\sinh^2{\frac{t}{2}}}+\pi i\text{Res}_{t\rightarrow 0}\bigg(\frac{e^{i\mu t}}{8t\sinh^2{\frac{t}{2}}}\bigg)\nonumber\\
    =&-\int_{\tilde{\mathcal{C}}}\frac{dt}{t}\frac{e^{i\mu t}}{8\sinh^2{\frac{t}{2}}}-\frac{\pi i}{24}(1+6\mu^2)\ .
\end{align}
The contour $\tilde{\mathcal{C}}$ can be closed by adding a semicircular contour at infinity in UHP along which the integrand vanishes. For this closed contour, the integral would be just the sum of residues at the double poles on the positive imaginary $t$-axis parametrized as $t_n=2\pi in$ with $n\in \mathbb{Z}_+$, 
\begin{align}
    \mathcal{F}_{\text{np}}(\mu)\equiv-i\text{Im}\int_{\frac{1}{\Lambda}}^{\infty}\frac{dt}{t}\frac{e^{i\mu t}}{4\sinh^2{\frac{t}{2}}}=&-2\pi i\sum_{n\in \mathbb{Z}_+}\text{Res}_{t\rightarrow t_n}\bigg(\frac{e^{i\mu t}}{8t\sinh^2\frac{t}{2}}\bigg)\nonumber\\
    =&-\frac{\pi i}{4}\sum_{n\in \mathbb{Z}_+}\text{Res}_{t\rightarrow t_n}\bigg(\frac{e^{i\mu t}}{t\sinh^2\frac{t}{2}}\bigg)\ .
\end{align}
Since $t_n$ is a double pole, the residue is given by, 
\begin{align}
    \text{Res}_{t\rightarrow t_n}\frac{f(t)}{g(t)}=\bigg(\frac{2f'}{g''}-\frac{2fg'''}{3(g'')^2}\bigg)\bigg|_{t=t_n} \text{ where, } f(t)=\frac{e^{i\mu t}}{t} \text{ and }g(t)=\sinh^2{\frac{t}{2}} \ .
\end{align}
Because $g'''(t)=\sinh\frac{t}{2}\cosh\frac{t}{2}$, the second term will just vanish at $t=t_n$. Hence, we are just left with,
\begin{align}
    \text{Res}_{t\rightarrow t_n}\frac{f(t)}{g(t)}=\frac{2(\frac{i\mu e^{i\mu t}}{t}-\frac{ e^{i\mu t}}{t^2})}{\frac{1}{2}(\cosh^2{\frac{t}{2}}+\sinh^2{\frac{t}{2}})}\bigg|_{t=t_n}=\frac{4}{\pi}e^{-2\pi \mu n}\bigg(\frac{\mu}{2  n}+\frac{1}{4\pi n^2}\bigg)\ .
\end{align}
Hence, we end up with the following expression for $\mathcal{F}_{\text{np}}(\mu)$ as stated in \eqref{npfreeenergyatR=1},
\begin{align}\label{app_np_freeen_R=1}
    \mathcal{F}_{\text{np}}(\mu)=-i\sum_{n=1}^{\infty} e^{-2 \pi  \mu  n} \bigg(\frac{\mu }{2n}+\frac{1}{4\pi n^2}\bigg)
\end{align}
We could've lined it the other way as well by first extracting $\mathcal{F}_{\text{np},R}(\mu)$ \eqref{npfreenergyatR} from $\mathcal{F}_R(\mu)$ \eqref{freeenergyatR} and then taking the $R\rightarrow 1$ limit. We proceed similarly with free energy at general radius $R$ \eqref{freeenergyatR}. We can again add the small semicircular contour at the origin and subtract half the residue which would be just $-(1+R^2+12R^2\mu^2)/48R$. This is again a polynomial in $\mu$ and hence, the non-perturbative part of the free energy $\mathcal{F}_{\text{np},R}(\mu)$ stays unaffected. \textit{Hence, we can exclude the terms polynomial in $\mu$ for both $R=1$ and $R\neq 1$ cases while looking at the non-perturbative contributions}\footnote{However, these terms carry some meaning in the topological string context as indicated recently in \cite{hattab2024nonperturbativetopologicalstringtheory}.}. We can again close the contour in UHP. The non-perturbative part of the free energy will become a sum over residues at simple poles $t_n=2\pi in$ and $t'_n=2\pi inR$ for $n\in \mathbb{Z}_+$ i.e. 
\begin{align}
    \mathcal{F}_{\text{np},R}(\mu)=&-\frac{\pi i}{4}\sum_{n\in \mathbb{Z}_+}\bigg(\text{Res}_{t\rightarrow t_n}\bigg(\frac{e^{i\mu t}}{t\sinh{\frac{t}{2}}\sinh{\frac{t}{2R}}}\bigg)+\text{Res}_{t\rightarrow t_n'}\bigg(\frac{e^{i\mu t}}{t\sinh{\frac{t}{2}}\sinh{\frac{t}{2R}}}\bigg)\bigg)\nonumber\\
    =&i\sum_{n=1}^{\infty}\frac{1}{4n(-1)^n}\frac{e^{-2\pi n\mu}}{\sin\frac{\pi n}{R}}+i\sum_{n=1}^{\infty}\frac{1}{4n(-1)^n}\frac{e^{-2\pi n \mu R}}{\sin\pi Rn}
\end{align}
as expected. Now, we can take $R\rightarrow 1$ limit in the above expression and get $\mathcal{F}_{\text{np}}(\mu)$ in \eqref{app_np_freeen_R=1}. The two terms above are individually divergent with the divergence $\propto (R-1)^{-1}$ but the divergences from both the terms cancel and we end up with the finite result \eqref{app_np_freeen_R=1}. This completes the claim made in figure \ref{Commutingdiag}.

\section{Free boson annulus partition functions at \texorpdfstring{$R\neq 1$}{TEXT} and \texorpdfstring{$R=1$}{TEXT}}\label{app_freeboson_review}
We will review the boundary states of compactified free boson CFT ($X\sim X+2\pi R$) in this section. We will follow the discussion in \cite{Recknagel_Schomerus_2013}. We have the following mode expansion for the compactified case, 
\begin{align}
     X(z, \bar{z})=&X_L(z)+X_R(\bar{z})\\
     X_{L}(z)=&x_L-i\frac{\alpha'}{2}p_L\ln z+i\bigg(\frac{\alpha'}{2}\bigg)^{\frac{1}{2}}\sum_{m\in \mathbb{Z},m\neq 0}\frac{\alpha_m}{mz^m}\\
     X_{R}(\bar{z})=&x_R-i\frac{\alpha'}{2}p_R\ln \bar{z}+i\bigg(\frac{\alpha'}{2}\bigg)^{\frac{1}{2}}\sum_{m\in \mathbb{Z},m\neq 0}\frac{\tilde{\alpha}_m}{m\bar{z}^m}\\
     \text{where, } p_L\equiv &\bigg(\frac{2}{\alpha'}
     \bigg)^{\frac{1}{2}}\alpha_0=\frac{n}{R}+\frac{wR}{\alpha'}, \quad p_R\equiv \bigg(\frac{2}{\alpha'}
     \bigg)^{\frac{1}{2}}\tilde{\alpha}_0=\frac{n}{R}-\frac{wR}{\alpha'}
\end{align}
where $n,w\in \mathbb{Z}$ respectively represent the momentum and winding numbers. Now, there are two possibilities at the boundary, either Dirichlet or Neumann boundary condition. The corresponding gluing conditions are, 
\begin{align}
    (\alpha_n\pm \tilde{\alpha}_{-n})|(k,w)\rangle \rangle=0
\end{align}
where $+$ $(-)$ correspond to Neumann (Dirichlet). The solutions to the above constraints are the coherent states, 
\begin{align}
    |(0,w)\rangle\rangle_N=&\exp\bigg(-\sum_{n=1}^{\infty}\frac{\alpha_{-n}\tilde{\alpha}_{-n}}{n}\bigg)|(0,w)\rangle,\\
    |(k,0)\rangle\rangle_D=&\exp\bigg(\sum_{n=1}^{\infty}\frac{\alpha_{-n}\tilde{\alpha}_{-n}}{n}\bigg)|(k,0)\rangle \ .
\end{align}
These are the Ishibashi states and a general boundary state will be a linear combination of these states. However, only certain linear combinations lead to acceptable boundary conditions because we demand that the overlaps of these linear combinations produce the partition functions of the boundary theory (the so-called sewing constraints). Hence, we consider the following two linear combinations,
\begin{align}\label{eq: (E.8)}
    ||D(x_0)\rangle \rangle=&\bigg(\frac{\alpha'}{2R^2}\bigg)^{\frac{1}{4}}\sum_{k\in \mathbb{Z}}e^{ikx_0\sqrt{\alpha'}/R}|(k,0)\rangle\rangle_D, \nonumber\\
    ||N(\tilde{x}_0)\rangle \rangle=&\bigg(\frac{R^2}{2\alpha'}\bigg)^{\frac{1}{4}}\sum_{w\in \mathbb{Z}}e^{i Rw\tilde{x}_0/\sqrt{\alpha'}}|(0,w)\rangle\rangle_N
\end{align}
where these states respectively exhibit invariance under the transformation $x_0\rightarrow x_0+2l\pi R/\sqrt{\alpha'}$ and $\tilde{x}_0\rightarrow \tilde{x}_0+2l \pi \sqrt{\alpha'}/R$ for $l\in \mathbb{Z}$. From the overlaps between the Ishibashi states ($\tilde{v}:=e^{-\frac{2\pi}{t}}$), 
\begin{align}
    _D\langle \langle (k,0)|\tilde{v}^{L_0-\frac{c}{24}}| (k',0)\rangle \rangle_D&=\delta_{k,k'}\tilde{v}^{\frac{\alpha'k^2}{4R^2}}, \\
    _N\langle \langle (0,w)|\tilde{v}^{L_0-\frac{c}{24}}| (0,w')\rangle \rangle_N&=\delta_{w,w'}\tilde{v}^{\frac{w^2R^2}{4\alpha'}},\\
    _D\langle \langle (k,0)|\tilde{v}^{L_0-\frac{c}{24}}| (0,w)\rangle \rangle_N&=\delta_{k,0}\delta_{w,0}\tilde{v}^{-\frac{1}{24}}\prod_{n=1}^{\infty}(1+\tilde{v}^n)^{-1} \ .
\end{align}
We can calculate the annulus partition functions by calculating overlaps between boundary states \eqref{eq: (E.8)} using the Poisson resummation formula ($v$ $= e^{-2\pi t}$), 
\begin{align}
    Z_{N\tilde{x}_0, N\tilde{x}'_{0}}(v)=&\eta(it)^{-1}\sum_{w\in \mathbb{Z}}v^{\big(\frac{w\sqrt{\alpha'}}{R}+\frac{\tilde{x}_0-\tilde{x}_0'}{2\pi}\big)^2},\label{Neumannpartfunc}\\
    Z_{Dx_0, Dx_0'}(v)=&\eta(it)^{-1}\sum_{k\in \mathbb{Z}}v^{\big(\frac{kR}{\sqrt{\alpha'}}+\frac{x_0-x_0'}{2\pi}\big)^2},\label{Dirichlepartfunc}\\
    Z_{Dx_0, N\tilde{x}'_{0}}(v)=&v^{\frac{1}{48}}\prod_{n=1}^{\infty}(1-v^{n-\frac{1}{2}})^{-1}\ .\label{DNpartfunc}
\end{align}
\eqref{Dirichlepartfunc} with $\alpha'=1$ choice of normalization and $x_0=x_0'$ was used in \cite{alexandrov2023instantons} to calculate the exponentiated annulus. If we further set $R$ to the self-dual value i.e. $R=\sqrt{\alpha'}=1$, we end up with $Z_{X,g}(t)$ used in \eqref{cylpartfuncs} for any $g\in$ SU(2). Similarly, \eqref{Neumannpartfunc} also give $Z_{X,g}(t)$ upon setting $\tilde{x}_0=\tilde{x}_0'$ and $R=1$. The upshot is that at $R=1$, we can picturise the above partition functions as the special cases of the partition function for the mixed boundary condition we noted earlier in \eqref{mixedpartfunc}. In particular,
\begin{align}
    Z_{g_1,g_2;X}(v)\bigg|_{\alpha=0}=Z_{Dx_0,Dx_0'}(v)\bigg|_{R=1,x_0=x_0'}=Z_{Nx_0,Nx_0'}(v)\bigg|_{R=1,\tilde{x}_0=\tilde{x}_0'} \ .
\end{align}
 We can account for the cases $x_0\neq x_0'$ and $\tilde{x}_0=\tilde{x}_0'$ by taking non-zero $\alpha$. We can now finally show that $\alpha=\frac{\pi}{2}$ corresponds to the \eqref{DNpartfunc} as noted earlier. Consider \eqref{mixedpartfunc} with $\alpha=\frac{\pi}{2}$, 
\begin{align}
    Z_{g_1,g_2;X}(t)\bigg|_{\alpha=\frac{\pi}{2}}=\frac{1}{\eta(it)}\sum_{n\in \mathbb{Z}}e^{-2\pi t(n-\frac{1}{4})^2}=\frac{v^{\frac{1}{48}}\sum_{n\in \mathbb{Z}}v^{n^2-\frac{n}{2}}}{\prod_{n=1}^{\infty}(1-v^n)}\ .
\end{align}
Using the Jacobi triple product identity, we can convert the sum in the numerator into a product and then do some simple manipulations as follows\footnote{I thank Rajesh Gopakumar for pointing out these simple manipulations. It was earlier just a check using Mathematica.}, 
\begin{align}
    Z_{g_1,g_2;X}(t)\bigg|_{\alpha=\frac{\pi}{2}}=&v^{\frac{1}{48}}\prod_{n=1}^{\infty}\frac{(1-v^{2n})(1+v^{2n-1}v^{-\frac{1}{2}})(1+v^{2n-1}v^{\frac{1}{2}})}{(1-v^n)}\nonumber\\
    =&v^{\frac{1}{48}}\prod_{n=1}^{\infty}\frac{(1+v^{2n-1}v^{-\frac{1}{2}})(1+v^{2n-1}v^{\frac{1}{2}})}{(1-v^{2n-1})}\nonumber\\
    =&v^{\frac{1}{48}}\prod_{n=1}^{\infty}\frac{(1+v^{2n-1-\frac{1}{2}})(1+v^{2n-\frac{1}{2}})}{(1-v^{\frac{2n-1}{2}})(1+v^{\frac{2n-1}{2}})}\nonumber\\
    =&v^{\frac{1}{48}}\prod_{n=1}^{\infty}\frac{(1+v^{n-\frac{1}{2}})}{(1-v^{\frac{2n-1}{2}})(1+v^{n-\frac{1}{2}})}\nonumber\nonumber\\
    =&v^{\frac{1}{48}}\prod_{n=1}^{\infty}(1-v^{n-\frac{1}{2}})^{-1}\label{eq: app_equiv_R_1}
\end{align}
which is same as \eqref{DNpartfunc}.
%  We finally consider the $\log$ of the integrand in \eqref{eq: initial_int_R} excluding the three pieces in \eqref{eq: div_pieces_R}.
%  \begin{align}
%      -\sum_{n\neq  0,\pm 1}\log\left(1-\frac{1}{R^2(n+b)^2}\right)=&\sum_{k=1}^{\infty}\frac{1}{kR^{2k}}\sum_{n\neq 0,\pm 1}\frac{1}{(n+b)^{2k}}
%  \end{align}
%  It can be manipulated as follows using the Hurwitz Zeta Function, $\zeta(s,a):=\sum_{n=0}^{\infty}(n+a)^{-s}$, 
% \begin{align}
% -\sum_{n\neq  0,\pm 1}\log\left(1-\frac{1}{R^2(n+b)^2}\right)=&\sum_{k=1}^{\infty}\frac{1}{kR^{2k}}\left(\sum_{n=0}^{\infty}\frac{1}{(n+b+2)^{2k}}+\sum_{n=0}^{\infty}\frac{1}{(n+2-b)^{2k}}\right)\nonumber\\
%     =&\sum_{k=1}^{\infty}\frac{1}{kR^{2k}}\left(\zeta(2k,b+2)+\zeta(2k,2-b)\right)\nonumber\\
%     =&2\int_{0}^{\infty}\frac{dt}{t(1-e^{-t})}\sum_{k=1}^{\infty}\frac{t^{2k}}{2k\Gamma(k)R^{2k}}\left(e^{-(b+2)t}+e^{-(2-b)t}\right)\nonumber\\
%     =&2\int_{0}^{\infty}dt\frac{\left(\cosh(t/R)-1\right)\left(e^{-(b+2)t}+e^{-(2-b)t}\right)}{t(1-e^{-t})}\nonumber\\
%     =&4\int_{0}^{\infty}dt\frac{e^{-2t}\left(\cosh(t/R)-1\right)\cosh{bt}}{t(1-e^{-t})}\label{eq: app_conv_int}
% \end{align}
% where we have used the integral formula, 
% \begin{align}
%     \zeta(s,a)=\frac{1}{\Gamma(s)}\int_{0}^{\infty}dt\frac{t^{s-1}e^{-at}}{1-e^{-t}}, \quad \text{Re}(s)>1, \text{Re}(a)>0
% \end{align}
\section{Relation between the KM and IM model}\label{app: KMIMrel}
We will refer to the equation (64) of \cite{gopakumar2023derivingsimplestgaugestringduality}, which at a formal level establishes (choosing $V(x)=0$) (traces are implicit in the exponents), 
\begin{align}\label{eq: app_F_1}
    \frac{1}{(2\pi g_s)^{N^2}}\int dK dM e^{-\frac{1}{g_s}KM}=\frac{1}{(2\pi g_s)^{Q^2}}\int dAe^{\frac{A}{g_s}}(\text{det}_QA)^{-(N+Q)}\int dCe^{-\frac{C}{g_s}}(\text{det}_QC)^N
\end{align}
where, we have the KM model on the right hand side and a product of
IM with a Penner integral on the RHS.
We can exactly evaluate all three of these matrix integrals (using the orthogonal polynomial technique, for example, see \cite{marino2005chern}), 
\begin{align}
    \int dAe^{-\frac{A}{(-g_s)}}(\text{det}_QA)^{-(N+Q)}=&\frac{\text{Vol(U}(Q))}{(2\pi)^Q}\prod_{n=0}^{Q-1}\frac{\Gamma(n-N-Q+1)\Gamma(n+1)}{ (-g_s)^{N+Q-2n-1}}\nonumber\\
    \int dCe^{-\frac{C}{g_s}}(\text{det}_QC)^N=&\frac{\text{Vol(U}(Q))}{(2\pi)^Q}\prod_{n=0}^{Q-1}\frac{\Gamma(n+N+1)\Gamma(n+1)}{ g_s^{-N-2n-1}}
\end{align}
Using the above, we get the following for the right-hand side of \eqref{eq: app_F_1},
\begin{align}
    &\frac{1}{(2\pi)^{Q}(-1)^{NQ}}\prod_{n=0}^{Q-1}\Gamma(n+N+1)\Gamma(n-N-Q+1)\nonumber\\
    =&\frac{1}{(2\pi)^{Q}(-1)^{NQ}}\prod_{n=0}^{Q-1}\Gamma(n+N+1)\Gamma(-N-n)=\frac{1}{(2\pi)^{Q}(-1)^{NQ}}\prod_{n=0}^{Q-1}\frac{\pi}{\sin{((-N-n)\pi)}}\nonumber\\
    =&\frac{1}{(2\sinh{\mu\pi})^{Q}(-i)^Q(-1)^{-i\mu Q}}
\end{align}
where we have used $\text{Vol(U}(Q))=(2\pi)^{\frac{Q(Q+1)}{2}}/G_2(Q+1)$ and put $N=-i\mu$ as it is necessary for the convergence of it. Taking $\log$ of the RHS, we get, 
\begin{align}
    -Q\log(e^{\pi \mu}-e^{-\pi \mu})-Q\log(-i)+i\mu Q\log(-1)=-Q\log(-i(1-e^{-2\pi \mu}))
\end{align}
where we have chosen $\log(-1)=-i\pi$. Hence, we see that the convergence of the right-hand side of \eqref{eq: app_F_1}, requires the analytic continuation of $N$, which as a result, leads to a non-perturbative piece.

% BIBLIOGRAPHY
% use BIBTEX if you want
\bibliographystyle{JHEP}
\bibliography{main}


% \begin{thebibliography}{99}
% 
% \bibitem{a}
% Author, \emph{Title}, \emph{J. Abbrev.} {\bf vol} (year) pg.
% 
% \bibitem{b}
% Author, \emph{Title},
% arxiv:1234.5678.
% 
% \bibitem{c}
% Author, \emph{Title},
% Publisher (year).

% Please avoid comments such as "For a review'', "For some examples",
% "and references therein" or move them in the text. In general,
% please leave only references in the bibliography and move all
% accessory text in footnotes.

% Also, please have only one work for each \bibitem.

% \end{thebibliography}

\end{document}